\documentclass[longauth]{aa}

\usepackage{graphicx}
\usepackage{txfonts,textcomp}
\usepackage{upgreek}

\usepackage{hyperref}
\hypersetup{
    colorlinks=true,
    citecolor=blue,
    linkcolor=blue,
    filecolor=blue,
    urlcolor=blue
    }

\newcommand{\Rjup}{\ensuremath{\mathrm{R_{Jup}}}\xspace}
\newcommand{\MJup}{\ensuremath{\mathrm{M_{Jup}}}\xspace}
\newcommand{\Teff}{\ensuremath{\mathrm{T_{eff}}}\xspace}
\newcommand{\Mjup}{\MJup} 

\begin{document}

\title{Multimodal atmospheric \\ characterization of $\beta$ Pictoris b}
\subtitle{Adding high-resolution continuum spectra from GRAVITY}

\author{Matthieu~Ravet\inst{\ref{Lagrange}, \ref{IPAG}, \ref{MPIA}}
          \and
          M.~Bonnefoy\inst{\ref{IPAG}}
          \and
          G.~Chauvin\inst{\ref{MPIA}, \ref{Lagrange}}
          \and
          S.~Lacour\inst{\ref{LESIA}, \ref{ESO}, \ref{MPIEP}}
          \and
          M.~Nowak\inst{\ref{LESIA}, \ref{Cambridge}}
          \and
          B.~Charnay\inst{\ref{LESIA}}
          \and
          P.~Tremblin\inst{\ref{CEA}}
          \and
          D.~Homeier\inst{\ref{Aperio}, \ref{FPG}}
         \and
          C.~Morley\inst{\ref{DAUTA}}
          \and
          J.~Fortney\inst{\ref{DAAUC}}
           \and
          A.~Denis\inst{\ref{LAM}}
          \and
          S.~Petrus\inst{\ref{IFA}}
          \and
          P.~Palma-Bifani\inst{\ref{Lagrange}, \ref{LESIA}}
          \and
          R.~Landman\inst{\ref{Leiden}}
          \and
          L.~T.~Parker\inst{\ref{Oxford}}
          \and
          M.~Houllé\inst{\ref{Lagrange}}
          \and
          A.~Chomez\inst{\ref{LESIA}, \ref{IPAG}}
          \and
          K.~Worthen\inst{\ref{DPAJH}}
          \and
          F.~Kiefer\inst{\ref{LESIA}}
          \and
          G.-D.~Marleau\inst{\ref{MPIA},\ref{UDE},\ref{Bern}}
          \and
          Z.~Zhang\inst{\ref{DAAUC}, \ref{DPA}}
          \and
          J.~L.~Birkby\inst{\ref{Oxford}}
          \and
          F.~Millour\inst{\ref{Lagrange}}
          \and
          A.-M.~Lagrange\inst{\ref{LESIA}, \ref{IPAG}}
          \and
          A.~Vigan\inst{\ref{LAM}}
          \and
          G.P.P.L.~Otten\inst{\ref{Sinica}}
          \and
          J.~Shangguan\inst{\ref{Kavli}, \ref{MPIEP}}
}

\institute{Laboratoire J.-L. Lagrange, Université Côte d'Azur, Observatoire de la Côte d'Azur, CNRS, 06304 Nice, France\\\email{matthieu.ravet@oca.eu}\label{Lagrange}
        \and
             IPAG, Université Grenoble-Alpes, CNRS, F-38000 Grenoble, France\\\email{matthieu.ravet@univ-grenoble-alpes.fr}\label{IPAG}
        \and
            Max-Planck-Institut f\"ur Astronomie, K\"onigstuhl 17, D-69117 Heidelberg, Germany\\\email{maravet@mpia.de}\label{MPIA}
        \and
             LESIA, Observatoire de Paris, Université PSL, CNRS, 92195 Meudon, France\label{LESIA}
        \and
            European Southern Observatory, Karl-Schwarzschild-Straße 2, 85748 Garching, Germany\label{ESO}
        \and
            Max Planck Institute for Extraterrestrial Physics, Giessenbachstraße 1, 85748 Garching, Germany\label{MPIEP}
        \and
            Maison de la Simulation, CEA, CNRS, Univ. Paris-Sud, UVSQ, Université Paris-Saclay, 91191 Gif-sur-Yvette, France\label{CEA}
        \and
            Aperio Software Ltd., Insight House, Riverside Business Park, Stoney Common Road, Stansted, Essex, CM24 8PL, UK\label{Aperio}
        \and
            Förderkreis Planetarium Göttingen, Göttingen, Germany\label{FPG}
        \and
            Department of Astronomy, University of Texas at Austin, Austin, TX 78712, USA\label{DAUTA}
        \and
            Department of Astronomy and Astrophysics, University of California, Santa Cruz, CA 95064, USA\label{DAAUC}
        \and
            Department of Physics \& Astronomy, University of Rochester, Rochester, NY 14627, USA\label{DPA}
        \and
            Aix Marseille Univ, CNRS, CNES, LAM, Marseille, France\label{LAM}
        \and
            Instituto de Física y Astronomía, Facultad de Ciencias, Universidad de Valparaíso, 1111 Valparaíso, Chile\label{IFA}
        \and
            Institute of Astronomy, University of Cambridge, Madingley Road, Cambridge CB3 0HA, UK\label{Cambridge}
        \and
            Leiden Observatory, Leiden University, P.O. Box 9513, 2300 RA Leiden, The Netherlands\label{Leiden}
        \and
            Astrophysics, University of Oxford, Denys Wilkinson Building, Keble Road, Oxford, OX1 3RH, UK\label{Oxford}
        \and
            Department of Physics and Astronomy, Johns Hopkins University, 3400 N. Charles Street, Baltimore, MD 21218, USA\label{DPAJH}
        \and
            Fakultät für Physik, Universität Duisburg-Essen, Lotharstraße 1, 47057 Duisburg, Germany\label{UDE}
        \and
            Physikalisches Institut, Universit\"{a}t Bern, Gesellschaftsstr.~6, CH-3012 Bern, Switzerland\label{Bern}
        \and
            Academia Sinica, Institute of Astronomy and Astrophysics, 11F Astronomy-Mathematics Building, NTU/AS campus, No. 1, Section 4, Roosevelt Rd., Taipei 10617, Taiwan\label{Sinica}
        \and
            The Kavli Institute for Astronomy and Astrophysics, Peking University, Beijing 100871, China\label{Kavli}
}

\date{Received 24 January 2025; accepted 25 September 2025}

   \date{Received 24 January 2025; accepted 25 September 2025}

 
  \abstract
   {Characterizations of giant exoplanets such as $\beta$ Pictoris b (hereafter $\beta$~Pic~b)  are now routinely performed with multiple spectrographs and imagers  exploring different spectral bandwidths and resolutions, allowing for atmospheric retrieval of spectra with or without the conservation of the planet spectral continuum. The accounting of data multimodality in the analysis could provide a more comprehensive determination of the planets physical and chemical properties and inform on their formation history.} 
   {We present the first VLTI observations at R$_{\lambda}\sim$ 4,000 of $\beta$~Pic~b obtained for an exoplanet with GRAVITY at such a high resolution. We  upgraded the forward modelling code \texttt{\textit{ForMoSA}} to account for the data multimodality, including low-, medium-, and high-resolution spectroscopy  based on both a direct model-data comparison and an analysis of cross-correlation signals. We used the \texttt{\textit{ForMoSA}} code to refine the constraints on the atmospheric properties of the exoplanet and evaluated the sensitivity of the retrieved values to the input dataset.}
   {We obtained four high-signal-to-noise (S/N~$\sim$~20) spectra of $\beta$~Pic~b in the K band with GRAVITY at R$_{\lambda}\sim$~4,000 conserving both the pseudo-continuum and the pattern of molecular absorptions. We used \texttt{\textit{ForMoSA}} with four grids of self-consistent forward models (Exo-REM, ATMO, BT-Settl, and Sonora) to explore different \Teff, log(g), metallicity,  C/O, and $^{12}$CO/$^{13}$CO ratio values. We then combined the GRAVITY spectra with published 1--5 µm photometry (NaCo, VisAO, NICI, and SPHERE), low-to-medium-resolution ($R_{\lambda} \leq 700$ broadband, 0.9--7 µm) spectra, and echelle spectra covering narrower bandwidths ($R_{\lambda}\sim$  100,000, 2.1--5.2 µm).}
   {Sonora and Exo-REM are statistically preferred among all four models, regardless of the dataset used. Exo-REM predicts \Teff~$=1607.45^{+4.85}_{-6.20}$~K and log(g)~$=4.46^{+0.02}_{-0.04}$~dex when using only the GRAVITY epochs, whereas we have \Teff~$=1502.74^{+2.32}_{-2.14}$~K log(g)~$=4.00\pm0.01$~dex when incorporating all available datasets. The inclusion of archival data significantly affects all retrieved posteriors. When using all datasets, C/O mostly remains solar ($0.552^{+0.003}_{-0.002}$), while [M/H] reaches super-solar values (0.50~$\pm$~0.01). We report the first tentative constraint on the isotopic ratio log($^{12}$CO/$^{13}$CO)~=~1.12$^{+0.11}_{-0.08}$ in $\beta$~Pic~b's atmosphere; however, we note that this detection remains inconclusive due to telluric residuals affecting both the GRAVITY and SINFONI data. Additionally, we estimated the bolometric luminosity as log(L/L$\mathrm{_{\odot}}$)~=~-4.01$^{+0.04}_{-0.05}$~dex. Using a system age of 23~$\pm$~3~Myr, along with this bolometric luminosity and the constraints on the dynamical mass of $\beta$ Pic b, we were able to constrain the maximum of heavy element content of the planet to be on the order of 5\% (20--80~\ensuremath{\mathrm{M_{Earth}}}\xspace).} 
   {The joint access to the pseudo-continuum and molecular lines in the K band provided by GRAVITY have a significant impact on the retrieved metallicity, possibly owing to the collision-induced absorption driving the continuum shape of the K band. The echelle spectra do not dominate the final fit with respect to lower resolution data covering a broader portion of the spectral energy distribution and the latter keeps encapsulating more robust information on \Teff. Future multimodal frameworks should include a weighting scheme to account for the bandwidth and central wavelength of the observations.}

   
   \maketitle
%

\section{Introduction}

Direct imaging has  unearthed a few dozens of planetary-mass companions\footnote{\url{https://exoplanet.eu/}} across three decades in terms of the semi-major axis, found to be  orbiting brown dwarfs up to  massive stars hosts, as well as binaries, constituting hierarchical systems, or nested within belts of debris \citep[e.g.,][]{2004A&A...425L..29C,2010Natur.468.1080M, 2011AJ....141..119K, 2022A&A...664A...9S, 2023A&A...676L..10C}. This diverse population is likely the result of multiple formation pathways and a rich set of dynamical evolution mechanisms. 

The chemical composition of planet atmospheres has long been regarded as an additional way to constrain the formation and the dynamical evolution processes, starting from our Solar System \citep[e.g.,][]{1989oeps.book..487G, 1992mars.book..818O, 1996P&SS...44.1579L,  2003SSRv..106..121O}. Planets formed in disks are expected to display a wide range of bulk enrichments \citep{boley_heavy_2011, 2014A&A...566A.141M} and it remains to be determined whether these are detected in the object's spectra. The temperature of accreted solids (ice, dust, and pebbles) and gas in the disk at the planet formation radius is also expected to leave in addition an imprint on atomic and isotopic ratio \citep{2016ApJ...831L..19O, 2019ApJ...882L..29M, 2021A&A...654A..71S, 2021A&A...654A..72S, 2022ApJ...934...74M}. 

The direct imaging technique allows to capture high-quality spectra of giant exoplanets in a few hours of telescope time. Contrary to transit spectroscopy focusing on close-in and strongly irradiated objects, spectral features are formed over a broad range of pressure (several bars) and better account for the bulk composition of the atmosphere. Low-resolution (20 $\leq$ R$_{\lambda}\leq80$) integral field spectrographs fed by extreme adaptive-optics systems such as VLT/SPHERE \citep{beuzit_sphere_2019}, Gemini/GPI \citep{2014PNAS..11112661M}, Subaru/CHARIS \citep{2016SPIE.9908E..0OG}, or LMIRCam/ALES \citep{2015SPIE.9605E..1DS, 2018SPIE10702E..0CS} are collecting near-infrared (NIR:\ 1-5 µm) spectra of known imaged exoplanets below 2" around bright stars, revealing water features of M- and L- type companions \citep[e.g.,][]{2018AJ....155..226G, 2018AJ....156..291C,bonnefoy_first_2016, 2017A&A...605L...9C,2022AJ....163..217D} and methane features on a handful of cooler and least massive  T-type exoplanets  \citep{2017AJ....154...10R, 2023A&A...672A..93M, 2023A&A...672A..94D}. Such spectra offer to determine the bulk physical properties of the objects such as the effective temperature \Teff, radius R, and surface gravity log(g). These data have confirmed that young mid-L to early-T planets have redder under-luminous spectra than more massive and mature brown dwarfs. These properties can be explained by vigorous vertical mixing and reduced gravitational settling that change the relative abundance of methane and carbon monoxide \citep{2007ApJ...669.1248H, 2016ApJ...829...66M} and enhance the cloud opacity. Such clouds were recently detected for the first time in the atmosphere of the low-mass late-L companion VHS1256 b using the JWST \citep{2023ApJ...946L...6M}.

Ground and space-based medium-resolution (R$_{\lambda}\sim$ 2,000--5,000) integral field spectrographs (VLT/SINFONI and ERIS, Keck/OSIRIS, Gemini/NIFS, JWST/NIRSpec, and MIRI) can also provide high-fidelity infrared spectra (1--28 µm) of exoplanets  \citep{2013Sci...339.1398K, petrus_medium-resolution_2021, 2023AJ....166...85H, 2023ApJ...946L...6M, 2023A&A...670A..90P, kiefer_new_2024}. These instruments give access to the planet pseudo-continuum emission while de-blending partly the forests of ro-vibrational absorptions from key molecules (CO, H$_{2}$O, CH$_{4}$). They have been used to infer the C/O ratio \citep[and ref. therein]{2023AJ....166...85H} and even to estimate isotopic ratio \citep{zhang_13co-rich_2021, 2023ApJ...957L..36G}. Using interferometric techniques, GRAVITY and MATISSE on the VLTI have also demonstrated their capability to extract high S/N K-, L-, and M-band data at lower resolution (R$_{\lambda}$ = 150--700) spectra (2--5 µm) for exoplanets containing both the continuum and lines from which C/O and metallicity \citep{2019A&A...623L..11G, gravity_collaboration_peering_2020, blunt_first_2023, 2024arXiv240403776N} can be inferred.

Finally, more recently, single-mode fiber-fed echelle spectrographs coupled to adaptive optics (AO) systems  (HiRISE at VLT \citealp{2024A&A...686A.294C, Denis2025}, KPIC at Keck \citealp{2021JATIS...7c5006D}, REACH at Subaru \citealp{2020SPIE11448E..78K}) have started providing continuum-subtracted 1--5 µm spectra of exoplanets at R$_{\lambda}$~$\sim$~100,000 over a narrow spectral range. Cross-correlation techniques can then be applied for a joint detection and characterization of the objects, even those deeply buried into the stellar halo. The inversion of these data start to provide radial and rotational velocities measurements, as well as metallicity, C/O and isotopic ratio determinations for close-in planets \citep{2014Natur.509...63S, landman__2024, parker_into_2024, 2024arXiv240513125M, 2024A&A...682A..16V, 2024ApJ...970...71X, 2024arXiv240720952Z}.\\

For more than 40 years now since the discovery of \cite{smith_circumstellar_1984},
the young ($23\pm3$~Myr, \citealp{mamajek_age_2014}) and nearby ($\sim$19.8~pc, \citealp{gaia_collaboration_gaia_2023}) $\beta$ Pic system has been a unique laboratory for studying planetary formation, planet-disk interactions as well as the physics of young giant planets. The system is made up of a central star, a dust disk and two known giant planets, b and c (see Table \ref{BetaPicSys} for a more complete description). $\beta$ Pic b was the first planet discovered in the system \citep{lagrange_probable_2009} and it has since been thoroughly examined in numerous independent follow-up studies (e.g., \citealp{landman__2024, worthen_miri_2024, kiefer_new_2024, houlle_mathis_2025}, see Table \ref{litterature}). The planet has a semi-major axis of 9.93~$\pm$~0.03~au and dynamical mass of M~$=11.9^{+2.93}_{-3.04}$~\Mjup inferred from a monitoring of its orbital motion started following the initial discovery (2003 epoch published in 2009), high-precision astrometry obtained with GRAVITY, radial velocity measurements of the star, and, more recently, the joint analysis of GAIA and HIPPARCOS measurements \citep{lagrange_unveiling_2020, lacour_mass_2021, brandt_precise_2021}.

Following the firsts photometric measurements of $\beta$~Pic~b, \cite{bonnefoy_near-infrared_2013} used 1--5~µm photometry to determined the planet to be a young L-type object with an estimated luminosity of log(L/L$\mathrm{_{\odot}}$)= -3.87~$\pm$~0.08~dex. Their study and the following ones \citep{males_magellan_2014, chilcote_1-24_2017} used these data compared with evolutionary models to extract the first physical parameter constraints. Assuming an age of $23\pm3$~Myr, evolutionary models predicts \Teff between 1600~K and 1800~K, log(g)~$\sim$~4.0~dex, and a R~$\sim$~1.5~\Rjup.

Forward models predictions on the metallic content of $\beta$~Pic~b's atmosphere display a lot of variation in recent works. \cite{gravity_collaboration_peering_2020}, using GRAVITY R$_{\lambda}\sim$ 500 data, found both super-solar ([M/H]~$=0.68^{+0.11}_{-0.08}$) and sub-solar ([M/H]~$=-0.53^{+0.28}_{-0.34}$) metallicity when adding GPI data to their retrieval, while others have not. \cite{landman__2024} (using K-band CRIRES$_+$ data) obtained a range of metallicities between $-0.10$ and $0.72$ depending on the model they used; this result was confirmed by \cite{worthen_miri_2024} (using MIRI data), who also determined a large dispersion of values depending on the considered model ([M/H] between $-0.12$ and $0.90$). \cite{kiefer_new_2024} (using SINFONI data) found similar sub-solar values of $\sim-0.24$ using Exo-REM. The presence of the continuum in the data appears to be impactful on the retrieved [M/H] value. The treatment of clouds in models is also equally important for this object as they can lock up significant amount of metallic species (Fe, Mg, Si, and O). The main molecules predicted to condensate at chemical equilibrium are enstatite Mg$_2$Si$_2$O$_6$, forsterite Mg$_2$SiO$_4$, and corundum Al$_2$O$_3$.

The C/O measurements of $\beta$~Pic~b have also displayed a significant range in recent studies. The sub-solar value of C/O~$=0.43\pm0.5$ obtained by \cite{gravity_collaboration_peering_2020} suggested a formation through core-accretion, with a significant alteration of the envelop composition through the infall of planetesimals. Sub-solar values were also found by \cite{landman__2024} and \cite{worthen_miri_2024}. However, \cite{kiefer_new_2024} obtained a solar value of C/O~$=0.551\pm0.002$ using evolutionary-model informed priors but without continuum, a value that is better aligned with population studies \citep{nissen_carbon--oxygen_2013}. This value could still suggest a formation through core-accretion, but with a more moderate planetesimal accretion followed by an outward migration.
    
To our knowledge, no studies have been carried out on the isotopic content of the object's atmosphere. This is a difficult task because these absorptions are very subtle and the slightest systematic error in the models or issues with the data can hinder detectability.\\

In this paper, we propose to delve in the atmosphere of $\beta$~Pic~b using the full compilation of available data and adding the first VLTI/GRAVITY observations at R$_{\lambda}\sim$~4,000 of the planet (and the first ever obtained on an exoplanet with the instrument at such spectral resolution). The multimodality of the data in hand also motivated us to upgrade our Forward Modeling tool for Spectral Analysis (\texttt{\textit{ForMoSA}}\footnote{\url{https://formosa.readthedocs.io/en/latest/}}) code. Our different datasets are presented in Sect.~\ref{sec1}.  We present the multimodal version of our Bayesian inference tool \texttt{\textit{ForMoSA}} in Sect.~\ref{sec2}. We present the unified framework we used to analyse each individual epoch of GRAVITY spectra to account for the orbital motion of the planet in Sect.~\ref{sec3}. We describe our  inversion of the whole set of existing data in Sects.~\ref{sec4} and \ref{sec5}. We discuss our findings and conclusions in Sects.~\ref{sec6} and \ref{sec7}. Additional material is provided in the appendices.

\begin{table}[ht]
    \setlength{\tabcolsep}{6pt}
    \renewcommand{\arraystretch}{1}
    \centering
    \caption{Physical properties of the $\beta$ Pictoris system.}
    \begin{tabular}{lll}
    \hline
    \hline
    \multicolumn{2}{c}{$\beta$ Pictoris A} & Refs. \\
    \hline
    Spectral type & A6V & (a) \\
    \Teff (K) & $7890^{+13}_{-17}$ & (h) \\
    log(g) (dex) & $3.83^{+0.03}_{-0.02}$ & (h) \\
    \text{[M/H]} (dex) & $-0.21^{+0.03}_{-0.02}$ & (h) \\
    Distance (pc) & $19.8294\pm0.0571$ & (j) \\
    Age (Myr) & $23\pm3$ & (d) \\
    Mass (M$_{\odot}$) & $1.75^{+0.03}_{-0.02}$ & (i) \\
    \hline
    \\
    \multicolumn{2}{c}{$\beta$ Pictoris b} & \\
    \hline
    Spectral type & L2~$\pm$~2 & (b) \\
    \Teff (K) & $1742\pm10$ & (e) \\
    log(g) (dex) & $4.34^{+0.08}_{-0.09}$ & (e) \\
    \text{[M/H]} (dex) & $0.68^{+0.11}_{-0.08}$ & (e) \\
    C/O & $0.43^{+0.04}_{-0.03}$ & (e) \\
    Semi-major axis (au) & $9.93\pm0.03$ & (i) \\
    Eccentricity & $0.103\pm0.003$ & (i) \\
    Inclination (°) & $89.00\pm0.01$ & (i) \\
    Period (days) & $8634\pm84$ & (i) \\
    Mass (\Mjup) & $11.9^{+2.93}_{-3.04}$ & (i) \\
    \hline
    \\
    \multicolumn{2}{c}{$\beta$ Pictoris c} & \\
    \hline
    \Teff (K) & $1250\pm50$ & (f) \\
    log(g) (dex) & $\sim4.0$ & (f) \\
    Semi-major axis (au) & $2.68\pm0.02$ & (i) \\
    Eccentricity & $0.32\pm0.02$ & (i) \\
    Inclination (°) & $88.95^{+0.09}_{-0.10}$ & (i) \\
    Period (days) & $1210\pm17$ & (i) \\
    Mass (\Mjup) & $8.89\pm0.75$ & (i) \\
    \hline
    \\
    \multicolumn{2}{c}{Dust disk} & \\
    \hline
    Location (au) & $20-1000$ & (g) \\
    Inclination (°) & $85.27^{+0.26}_{-0.19}$ & (c) \\
    Position angle (°) & $30.35^{+0.14}_{-0.13}$ & (c) \\
    \end{tabular}
    \tablefoot{The atmospheric parameters given here are taken from the two GRAVITY collaboration papers using atmospheric models (refer to Table \ref{litterature} for evolutionary models predictions).}
    \tablefoottext{a}{\cite{gray_2006_contribution},}\tablefoottext{b}{\cite{bonnefoy_near-infrared_2013},}\tablefoottext{c}{\cite{millar-blanchaer_beta_2015},}\tablefoottext{d}{\cite{mamajek_age_2014},}\tablefoottext{e}{\cite{gravity_collaboration_peering_2020},}\tablefoottext{f}{\cite{nowak_direct_2020},}\tablefoottext{g}{\cite{2021A&A...646A.132J},}\tablefoottext{h}{\cite{swastik_host_2021},}\tablefoottext{i}{\cite{lacour_mass_2021}}and \tablefoottext{j}{\cite{gaia_collaboration_gaia_2023}}.
    \label{BetaPicSys}
\end{table}


\section{Observations and data reduction}\label{sec1}

This section presents the various spectroscopic and photometric data we use for our analysis, all combined and summarized in Table \ref{table_archival}. We  briefly describe the GRAVITY instrument, the observing conditions, and the additional post-processing we applied in the following.

\subsection{New GRAVITY observations at high-spectral resolution}\label{sec1.1}

GRAVITY \citep{2017A&A...602A..94G} is an interferometric spectro-imager that combines light coming from the four 8m Unit Telescopes (UTs) of the VLT. It operates in the K band between 1.97-2.4 µm. The four new epochs of GRAVITY observations of $\beta$~Pic~b where obtained in the span of three years (2019/11/09, 2019/11/11, 2021/08/27, and 2022/01/24; respective ESO program IDs : 1104.C-0651(A), 1104.C-0651(D), and 1104.C-0651(B)) using the high spectral resolution mode of the instrument, resulting in an effective spectral resolving power of R$_{\lambda}\sim$ 4,000. They represent the first observations of an exoplanet at this resolution with the instrument.

Each epoch has been reduced following the GRAVITY reduction pipeline \citep{nowak_2017_2019}. We created a BT-NextGen stellar model spectra \citep{hauschildt_nextgen_1999}, using an effective temperature of \Teff~=~7890~K, a surface gravity of log(g) = 3.83~dex and a solar metallicity of [M/H]~=~0 \citep{swastik_host_2021}, to calibrate the final flux. This was done by scaling the synthetic stellar model with a 2MASS K-band magnitude of 3.526~mag. The resulting merged and corrected spectrum can be found in Fig.~\ref{GRAVITY_spec_ExoREMvsATMO}.
The pipeline allows covariance matrices to be empirically estimated from the raw interferometric data (see detailed description in the appendix of \citealp{gravity_collaboration_peering_2020}).

\subsection{Archival data}

In addition to GRAVITY data, we used archival spectroscopic and photometric data. Table \ref{table_archival} offers more details on the characteristics of each observations. Data reduction and post-processing analysis can be found in the referenced paper of each study. The specific analysis of high-resolution data is presented in more details in Sect.~\ref{sec5}. We did not include the six new JWST/NIRCam photometric points presented in \cite{kammerer_2024_JWST}, as the two PSF subtraction methods used in that work produced significant scaling differences in the final flux values. Moreover, we did not expect these additional points to contribute meaningfully to the fit in our current weighted scheme.

\begingroup

\setlength{\tabcolsep}{6pt}
\renewcommand{\arraystretch}{1.2}
\begin{table*}[ht!]
    \centering
    \caption{Summary of the new and archival data used.}
    \begin{tabular}{ccccccc}
    \hline
    \hline
    Instrument & Type & Spectral coverage & Spectral resolution & S/N & Continuum ? & Refs. \\
    \hline
    VLT/NaCo & \multicolumn{1}{l}{J-band photometry} & \multicolumn{1}{l}{1.255~$\pm$~0.193 µm} & <30 & $\sim$3 & yes & (a) \\
    & \multicolumn{1}{l}{H-band photometry} & \multicolumn{1}{l}{1.655~$\pm$~0.307 µm} & <30 & $\sim$4 & yes \\
    & \multicolumn{1}{l}{K$_s$-band photometry} & \multicolumn{1}{l}{2.159~$\pm$~0.324 µm} & <30 & $\sim$7 & yes \\
    & \multicolumn{1}{l}{L'-band photometry} & \multicolumn{1}{l}{3.813~$\pm$~0.573 µm} & <30 & $\sim$9 & yes \\
    & \multicolumn{1}{l}{[4.05] photometry} & \multicolumn{1}{l}{4.056~$\pm$~0.0.060 µm} & <30 & $\sim$4 & yes \\
    & \multicolumn{1}{l}{M'-band photometry} & \multicolumn{1}{l}{4.818~$\pm$~0.416 µm} & <30 & $\sim$7 & yes \\
    \hline
    Magellan/VisAO & \multicolumn{1}{l}{Y$_s$-band photometry} & \multicolumn{1}{l}{0.985~$\pm$~0.084 µm} & <30 & $\sim$9 & yes & (a) \\
    \hline
    Magellan/Clio & \multicolumn{1}{l}{[3.1] photometry} & \multicolumn{1}{l}{3.102~$\pm$~0.090 µm} & <30 & $\sim$15 & yes & (a) \\
    & \multicolumn{1}{l}{[3.3] photometry} & \multicolumn{1}{l}{3.345~$\pm$~0.300 µm} & <30 & $\sim$11 & yes \\
    & \multicolumn{1}{l}{L'-band photometry} & \multicolumn{1}{l}{3.761~$\pm$~0.619 µm} & <30 & $\sim$15 & yes \\
    & \multicolumn{1}{l}{M'-band photometry} & \multicolumn{1}{l}{4.687~$\pm$~0.185 µm} & <30 & $\sim$7 & yes \\
    \hline
    VLT/SPHERE & \multicolumn{1}{l}{YJ-band spectroscopy} & \multicolumn{1}{l}{0.95--1.35 µm} & $\sim$50 & $\sim$22.5 & yes & (*) \\
    & \multicolumn{1}{l}{H-band photometry} & \multicolumn{1}{l}{1.59 µm and 1.667 µm} & <30 & $\sim$39 & yes \\
    \hline
    Gemini/GPI & \multicolumn{1}{l}{YJHK-band spectroscopy} & \multicolumn{1}{l}{0.9--2.4 µm} & $\sim$45 & $\sim$10 & yes & (b) \\
    \hline
    VLTI/GRAVITY & \multicolumn{1}{l}{K-band spectroscopy} & \multicolumn{1}{l}{1.97--2.4 µm} & $\sim$4,000 & $\sim$20 & yes & (**) \\
    \hline
    VLT/SINFONI & \multicolumn{1}{l}{K-band spectroscopy} & \multicolumn{1}{l}{1.929--2.472 µm} & 4,120 & $\sim$20 & no & (c) \\
    \hline
    VLT/CRIRES$_+$ & \multicolumn{1}{l}{K-band spectroscopy} & \multicolumn{1}{l}{2.06--2.47 µm} & $\sim$100,000 & >5 & no & (d, e) \\
    & \multicolumn{1}{l}{M-band spectroscopy} & \multicolumn{1}{l}{3.51--5.22 µm} & $\sim$100,000 & >5 & no \\
    \hline
    VLTI/MATISSE & \multicolumn{1}{l}{LM-band spectroscopy} & \multicolumn{1}{l}{2.76--4.99 µm} & 387--700 & $\sim$35 & yes & (f) \\
    \hline
    JWST/MIRI & \multicolumn{1}{l}{M-band spectroscopy} & \multicolumn{1}{l}{4.90--6.8 µm} & 30--40 & $\sim$15 & yes & (g) \\
    \hline
    \end{tabular}
    \tablefoot{The first column contains the name of the instrument, the second the type (spectroscopy or photometry), the third the wavelength coverage, the forth the (effective) spectral resolution R$_{\lambda}$, the fifth the average S/N, and the last indicates whether the data has continuum information or not.}
    \tablefoottext{a}{\cite{morzinski_magellan_2015},}\tablefoottext{b}{\cite{chilcote_1-24_2017},}\tablefoottext{c}{\cite{kiefer_new_2024},}\tablefoottext{d}{\cite{landman__2024},}\tablefoottext{e}{\cite{parker_into_2024},}\tablefoottext{f}{\cite{houlle_mathis_2025},}\tablefoottext{g}{\cite{worthen_miri_2024} obtainable from \href{https://archive.stsci.edu/doi/resolve/resolve.html?doi=10.17909/7xb7-hh14}{DOI},}\tablefoottext{*}{unpublished, and}\tablefoottext{**}{this work.}
    \label{table_archival}
\end{table*}

\endgroup

We divided our analysis in tree main parts. In Sect.~\ref{sec3}, we focus on the analysis of the new GRAVITY spectra. In Sects.~\ref{sec4} and \ref{sec5}, we describe the combination of this spectrum with the available datasets presented in Sect.~\ref{sec1}. All of the observational data, together with the complete set of models and results is available on Zenodo\footnote{\url{https://zenodo.org/records/17609314}}.

\section{Methods}\label{sec2}

In this section, we present our fitting strategy based on forward modeling, as well as our Bayesian  \texttt{\textit{ForMoSA}} code. Sects.~\ref{sec2.1} and \ref{sec2.2}, respectively, code with the forward models we used and the new additions made to \texttt{\textit{ForMoSA}} for the purposes of this work.
\\

To extract information from exoplanet spectra, a common approach is to compare them with synthetic spectra generated from models of atmosphere. The comparison heavily relies on Bayesian inversion methods that can be divided into two categories: forward modeling and retrieval. The forward modeling approach (\citealp{samland_spectral_2017}) works "from the model to the data" by using a pre-computed grid of several thousand of spectra and comparing them to the data using Bayesian inference methods. This approach has the advantage of being "self-consistent," meaning that it faithfully maintains the physical interpretation of the observed spectrum, as each synthetic spectrum in the grid is computed with strong physical priors. It is also relatively fast since you do not need to generate a new model at each step, but only need to re-interpolate the grid. The main drawbacks are the dependencies on both the quality and the size of the parameter space proposed by the model grid, as well as the systematic deviations from the models that can arise from a lack of physics or hidden fixed free-parameters in the forward model.
\\

Each model is compared iteratively using Bayesian inversion techniques. The goal is to estimate the posterior probability p($\theta|d$) of the model parameters $\theta$ given the data $d$ and prior knowledge on the parameters p$(\theta)=\pi(\theta)$d$\theta$, using Bayes theorem,

\begin{equation}
    \text{p}(\theta|d) = \frac{p(d|\theta).p(\theta)}{p(d)}
.\end{equation}

Two terms are worth mentioning here:

\begin{itemize}
    \item p$(d|\theta)=\mathcal{L}$ is the probability of the data given a parameter set, also known as the likelihood function. This is the function we want to maximize during the inversion.
    \item p$(d)=\mathcal{Z}$ is the Bayesian evidence, computed by marginalizing $\mathcal{L}$ over the parameter space. It can be written as $\mathcal{Z}=\int\mathcal{L}\pi(\theta)$d$\theta=\int\mathcal{L}$dX, with X as the prior volume.
\end{itemize}

The Bayesian evidence is especially useful when comparing two competing models, $m_1$ and $m_2$. Unlike the likelihood $\mathcal{L}$, it is independent of any single set of parameters as it integrates over all parameters (and is therefore more general in that sense). Concretely, if we label $\mathcal{Z}_1$ and $\mathcal{Z}_2$ as the Bayesian evidence of the two competing models, we can compute the Bayes factor using

\begin{equation*}
    B_m = \frac{\mathcal{Z}_1}{\mathcal{Z}_2} \iff \ln B_m = \ln\mathcal{Z}_1 - \ln\mathcal{Z}_2.
\end{equation*}

Typically, a Bayes factor above 12 (or $\ln B_m$ above 2.5) can be interpreted as a "moderate need" for the first model to describe the data instead of the second ($2.7\sigma$ significance, see Table 4 of  \citealp{benneke_how_2013}). detection. Recently, \citet{kipping_exoplaneteers_2025} demonstrated that this commonly used correspondence between $\ln B_m$ and $\sigma$ can lead to overestimations of the actual detection significance. For completeness, we report both metrics, but we rely primarily on the Bayesian evidence as the main diagnostic tool for the model comparison. This approach will be used in Sects.~\ref{sec3.2} and \ref{sec4} to assess the significance of the $^{13}$CO detection. For a more detailed description of the Bayesian inference methods, we refer to \cite{knuth_bayesian_2015}.
\\

For this work, we chose to use the forward modeling approach with our Bayesian inversion tool, \texttt{\textit{ForMoSA}}. A detailed description of the code can be found in \cite{petrus_new_2020}, \cite{petrus_medium-resolution_2021}, \cite{petrus_x-shyne_2023}, and \cite{petrus_jwst_2024}. In brief, \texttt{\textit{ForMoSA}} uses a grid of forward models (see Sect.~\ref{sec2.1}) along with spectroscopic and/or photometric data. The tool then proceeds to adapt these models to the data (in terms of wavelength range coverage, resolution, continuum, etc.) to make them comparable. The inversion is performed by maximizing the likelihood function $\mathcal{L}$ using the nested sampling method (see \citealp{skilling_nested_2004}). The grid is interpolated linearly in the N dimension of the parameter space using the python packaged \texttt{\textit{xarray}}\footnote{\url{https://docs.xarray.dev/en/stable/}}.

\subsection{Forward models}\label{sec2.1}

\begin{figure}[ht!]
\centering
    \includegraphics[scale=0.75]{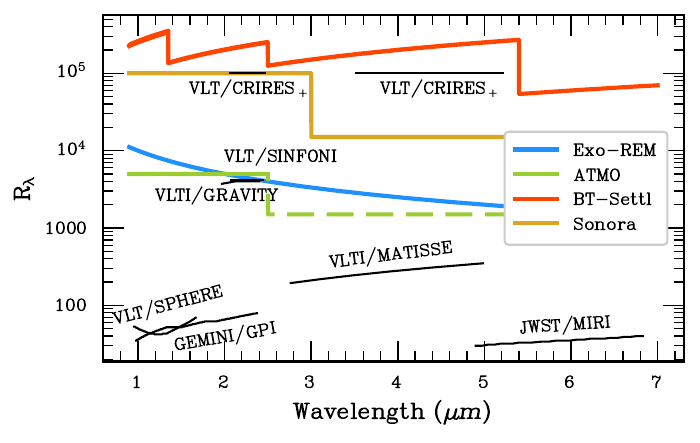}
    \caption{Spectral resolution, R$_{\lambda}$, as a function of the wavelength coverage for each forward modeling used in this study. The observed data are represented by the black lines while the forward models are in shades of color. Dotted lines represent the extensions (both in wavelength and resolution) used in Sects.~\ref{sec4} and \ref{sec5}.}
    \label{model_data_res}
\end{figure}

Atmospheric models aim to replicate the emission spectra of exoplanets. To do this, they typically compute a blackbody-like emission spectrum from an interior model, which is then modified by a radiative-convective atmosphere composed of various opacity sources (and sometimes clouds), following a parameterized pressure-temperature profile; all within a simplified 1D framework. These models cover different wavelength ranges and spectral resolutions, as illustrated in Fig.~\ref{model_data_res}. We give  a brief summary of the grids we used below, previously presented in \cite{petrus_jwst_2024}.\\

Exo-REM \citep{charnay_self-consistent_2018} is our reference model. It includes non-equilibrium chemistry of CO--CH$_4$ and N$_2$--NH$_3$ and takes into account multiple opacity sources: H$_2$--H$_2$ and H$_2$--He collisions induced absorptions, H$_2$O, CH$_4$, CO, CO$_2$, NH$_3$, PH$_3$, TiO, VO, and FeH ro-vibrationnal transitions as well as Na and K resonant lines. Clouds formation based on Si, Fe, Na$_2$S, KCl, and H$_2$O species was considered, but with a simplified cloud micro-physics (i.e., only the fastest chemical processes were taken into account). We used a custom version of the grid to explore the abundance of the isotope $^{13}$C resulting in these five free parameters: \Teff $\in$ [1200~K; 2000~K], log(g) $\in$ [3.5~dex; 4.5~dex], [M/H] $\in$ [-0.5~dex; 1~dex], C/O $\in$ [0.40; 0.65], and log($^{12}$C/$^{13}$C) $\in$ [1.0; 2.6]. Exo-REM is known to produce deviant spectra so we performed a simple cleanup by computing the integrated fluxes across the grids and removing the spectra that did not follow the approximate Stefan–Boltzmann law. Furthermore, we chose to remove all nodes at \Teff = 1500~K from the grid, as they appeared to have some deviant spectra driving the solutions.

ATMO \citep{tremblin_fingering_2015} are a family of cloudless models. They explore numerous opacity sources : H$_2$--H$_2$ and H$_2$--He collision induced absorptions, NH$_3$, H$_2$O, CH$_4$, CO, TiO, and VO  ro-vibrationnal transitions as well as Na and K resonant lines. These models include two non-equilibrium chemical reactions: CO--CH$_4$ and N$_2$--NH$_3$ that will drive fingering convection in the atmosphere and change the temperature gradient. This will uniquely emulate the reddening effect of clouds. We also utilized a custom version of this grid exploring $^{13}$C. The parameter space we explored is: \Teff $\in$ [1600~K; 1900~K], log(g) $\in$ [3.5~dex; 4.5~dex], [M/H] $\in$ [-0.6; 0.6], $\gamma\in$ [1.01; 1.05], and log($^{12}$C/$^{13}$C) $\in$ [1.0; 2.6] ($\gamma$ is the adiabatic index driving the temperature gradient). The grids were set with C/O = 0.55. In Sect.~\ref{sec3}, we use an extended grid in \Teff and log(g) that explored the following parameters : \Teff $\in$ [800~K; 3000~K], log(g) $\in$ [2.5~dex; 4.5~dex], [M/H] $\in$ [-0.6; 0.6], and C/O $\in$ [0.3; 0.7].

BT-Settl \citep{allard_models_2012} model grids explore more complex cloud microphysics than Exo-REM and include non-equilibrium chemistry (for CO, CH$_4$, N$_2$, NH$_3$, and CO$_2$) as well as vertical mixing. To properly take into account the cloud micro-physics, BT-Settl divides the atmosphere into multiple layers and computes the distribution and size of grains by comparing their characteristic times of condensation, coalescence, dispersion, and sedimentation in the atmosphere. These clouds are Fe and Si based. We used a custom version of this grid exploring the C/O, giving the following parameter space: \Teff $\in$ [1400~K; 2200~K], log(g) $\in$ [2.5; 5.0], and C/O $\in$ [0.2754; 1.096]. The metallicity is solar ($[M/H]=0$) for this grid.

Sonora \citep{morley_sonora_2024} models include vertical mixing, multiple opacity sources and parameterized clouds. We used the custom Diamondback sub-model grid that contains, in total, 31 molecules, atoms, and ions as opacity sources (hydrocarbons, CO, CO$_2$, OCS, Li, K, Na, LiH, LiF), as well as multiple collision induced absorptions (H$_2$--H$_2$, H$_2$--He, H$_2$--N$_2$, H$_2$--CH$_4$). Clouds were parameterized using \cite{ackerman_precipitating_2001} approach, meaning that their opacities were computed for Mie diffusion on MgSiO$_3$, Mg$_2$SiO$_4$, and Al$_2$O$_3$. In the end, the model grid effectively explores the following free parameters: \Teff $\in$ [900~K; 2400~K], log(g) $\in$ [3.5~dex; 5.5~dex], [M/H] $\in$ [-0.5; 0.5], and f$_\text{sed}$ $\in$ [1; 8] (f$_\text{sed}$ is the sedimentation efficiency parameter). The grids were set with C/O~=~0.55.

\subsection{\texttt{\textit{ForMoSA}} 2.0}\label{sec2.2}

Atmospheric retrievals and forward modeling approaches both deal with a number of issues when combining different observations into a single Bayesian framework. A non-exhaustive list of theses limitations include: data that spectraly overlap (i.e., GPI and GRAVITY); data that have different noise estimates (i.e., spectral covariances or not); data that are not calibrated in flux or have inaccurate flux calibration; data that are taken at different epochs and spectra for which the spectral continuum can not be trusted or has been removed at previous analysis steps (Doppler spectroscopy, etc). Inspired by the work of \cite{hayoz_crocodile_2023}, we incorporated a Multimodal Option for Simultaneous Analysis and Improved Constrains (MOSAIC) in \texttt{\textit{ForMoSA}} aim to reduce theses limitation by adapting and inverting the different datasets in parallel.

\begin{figure}[ht!]
\centering
    \includegraphics[scale=0.35]{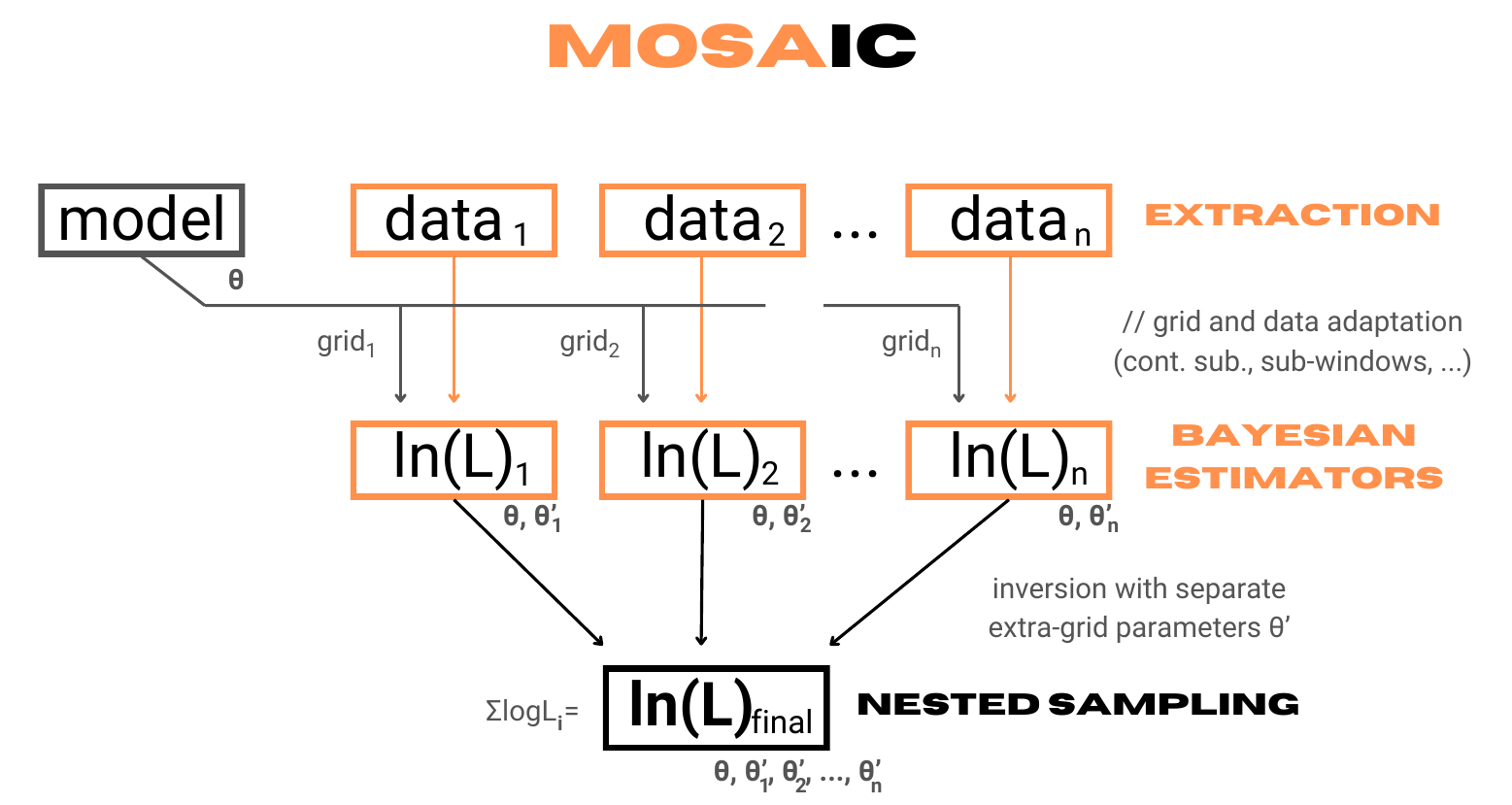}
    \caption{Schematic of MOSAIC. Here, "data" refer to spectroscopic and/or photometric observations and the "model" is a pre-computed grid of forward model spectra. $\theta$ represents the grid parameters (i.e \Teff, log(g), [M/H]), and $\theta'$ the extra-grids parameters that can be defined separately (or not) for each sub-inversions (i.e., rv, $v\sin i$, ...). $\ln\mathcal{L}_{final}$ is the final log-likelihood function that goes in the nested-sampling algorithm and is computed with the assumption of independent observations, i.e $\ln\mathcal{L}_{final}=\sum_i \ln\mathcal{L}_i$.}
    \label{MOSAIC_schem}
\end{figure}

The general idea is summarized on Fig.~\ref{MOSAIC_schem}. Each dataset was first imported and adapted alongside a grid of synthetic spectra taken from a single atmospheric model. This means that the continuum subtraction and/or wavelength windows can be applied separately for each data-grid pair and then stored for the inversion. Secondly, each dataset was compared with the interpolated grid using separate likelihood functions $\ln \mathcal{Z}_i$ during the inversion. The grids were modified by both interpolating them with the original model parameters $\theta$ and then adding transformations with extra-grid parameters, $\theta'_i$, that can vary from dataset to dataset (i.e., different scaling factors, c$_k$, radial velocities, rv, projected rotational velocities, $v\sin i$, etc.). In this way, we were able to control the transformations we want to apply on a given  data+grid pair. The stored sub-grids are usually much smaller than the original grids. Therefore, this step is relatively fast.

New likelihood functions are now available to account for differences in the data reduction scheme and, in particular, noise processing. Given a set of data (spectrum or photometry), $d$, its associated error bars, $\sigma,$ or covariance matrix, C, and the model vector, m, the classical $\chi^2$ distribution can be defined as

\begin{equation}
    \ln\mathcal{L}\propto-\frac{\chi^2}{2}=-\frac{1}{2}\sum\left(\frac{d-m}{\sigma}\right)^2,
\end{equation}

where the  observational data points are independent. In addition, we implemented

\begin{equation}
    \ln\mathcal{L}\propto-\frac{\chi^2}{2}=-\frac{1}{2}\sum\left(d-m\right)^T C^{-1} \left(d-m\right),
\end{equation}

when the data covariances are provided as input of the code. We also allowed for various proposed mappings \citep{zucker_cross-correlation_2003, brogi_retrieving_2019, ruffio_radial_2019} of  cross-correlation signals to likelihood (CCF-to-log$\mathcal{L}$), which can be used when Doppler spectroscopy (e.g., Keck/KPIC, VLT/HiRISE, etc) was performed solely or in addition to regular spectrophotometric observations (e.g., JWST, VLT/SPHERE, Gemini/GPI, etc.).

We computed the joint likelihood function $\ln\mathcal{Z}_{final}=\sum_i \ln\mathcal{Z}_i$, summing over the likelihoods corresponding to each dataset taken as input. This means that this upgraded version of ForMoSA does not account for correlations either between datasets from the same instrument taken at different epochs or between datasets from different instruments. As is common practice in the field, each dataset was treated as statistically independent.

\section{Inversion of GRAVITY spectra}\label{sec3}
\begingroup

\setlength{\tabcolsep}{4.5pt}
\renewcommand{\arraystretch}{1.5}
\begin{table*}[ht!]
\tiny
    \centering
    \caption{Grid priors and posteriors for each forward model are presented in this section.}
    \begin{tabular}{lllllllll}
    \hline
    \hline
    \textbf{Parameter} & \Teff & log(g) & [M/H] & C/O & $\gamma$ & f$_\text{sed}$ & log($^{12}$C/$^{13}$C) & log(L/L$_{\odot}$) \\
    & (K) & (dex) & & & & & & (dex) \\
    \hline
    Exo-REM priors & $U(1200, 2000)$ & $U(3.5, 4.5)$ & $U(-0.5, 1)$ & $U(0.4, 0.65)$ & & & $U(1.0, 2.6)$ & \\
    Exo-REM posteriors & $B(1551.89, 1604.44)$ & $B(4.18, 4.44)$ & $B(0.65, 0.83)$ & $0.55\pm0.01$ & & & $<1.08$ & $B(-3.723, -3.719)$ \\
    strongest node & $1607.45^{+4.85}_{-6.20}$ & $4.46^{+0.02}_{-0.04}$ & $0.82\pm0.02$ & $0.55\pm0.01$ & & & $<1.08$ & $-3.723\pm0.002$ \\
    \hline
    ATMO priors & $U(1400, 1900)$ & $U(3.5, 4.5)$ & $U(-0.6, 0.6)$ & & $U(1.01, 1.05)$ & & $U(1.0, 2.6)$ & \\
    ATMO posteriors & $>1894.93$ & $<3.51$ & $-0.17\pm0.02$ & & $>1.05$ & & $<1.55$ & $-3.780\pm0.001$ \\
    \hline
    BT-Settl priors & $U(1400, 2000)$ & $U(3.5, 5.0)$ & & $U(0.2754, 1.096)$ & & & & \\
    BT-Settl posteriors & $1668.43^{+5.62}_{-5.55}$ & $3.60\pm0.04$ & & $0.65\pm0.02$ & & & & $-3.745\pm0.006$ \\
    \hline
    Sonora priors & $U(900, 2400)$ & $U(3.5, 5.5)$ & $U(-0.5, 0.5)$ & & & $U(1, 8)$ & & \\
    Sonora posteriors & $1625.67^{+11.48}_{-11.21}$ & $4.56^{+0.06}_{-0.07}$ & $>0.49$ & & & $<1.03$ & & $-3.791\pm0.001$ \\
    \hline
    \end{tabular}
    \tablefoot{Hereafter $U(a,b)$ refers to a uniform distribution and $B(a,b)$ to a bimodal distribution between $a$ and $b$. The error bars correspond to the lower and upper bonds in the parameter space encompassing 68\% of the retrieved solutions around the best fit. The C/O ratio is fixed at 0.55 for ATMO, 0.458 for Sonora and the metallicity is fixed at 0.0 for the BT-Settl.}
    \label{priors_post_GRAVITY}
\end{table*}

\setlength{\tabcolsep}{21pt}
\renewcommand{\arraystretch}{1.5}
\begin{table*}[ht!]
\tiny
    \centering
    \caption{Extra-grid priors and posteriors for each forward modeling presented in this section.}
    \begin{tabular}{lllllll}
    \hline
    \hline
    Parameter & R & rv$_{2019/11/09}$ & rv$_{2019/11/11}$ & rv$_{2021/08/27}$ & rv$_{2022/01/24}$ \\
    & (\Rjup) & (km.s$^{-1}$) & (km.s$^{-1}$) & (km.s$^{-1}$) & (km.s$^{-1}$) \\
    \hline
    Exo-REM priors & $U(1, 2)$ & $U(-50, 50)$ & $U(-50, 50)$ & $U(-50, 50)$ & $U(-50, 50)$ \\
    Exo-REM posteriors & $B(1.74, 1.87)$ & $B(29.39, 49.97)$ & $24.12^{+2.18}_{-2.19}$ & $24.79^{+2.18}_{-2.19}$ & $40.32^{+1.03}_{-1.06}$ \\
    strongest node & $1.73\pm0.01$ & $B(29.39, 49.97)$ & $24.12^{+2.18}_{-2.19}$ & $24.79^{+2.18}_{-2.19}$ & $40.32^{+1.03}_{-1.06}$ \\
    \hline
    ATMO priors & $U(1, 2)$ & $U(-50, 50)$ & $U(-50, 50)$ & $U(-50, 50)$ & $U(-50, 50)$ \\
    ATMO posteriors & $1.160^{+0.005}_{-0.002}$ & $B(30.66, 50.00)$ & $26.12^{+1.68}_{-1.75}$ & $26.63^{+2.31}_{-2.30}$ & $42.28^{+1.08}_{-1.10}$ \\
    \hline
    BT-Settl priors & $U(1, 2)$ & $U(-50, 50)$ & $U(-50, 50)$ & $U(-50, 50)$ & $U(-50, 50)$ \\
    BT-Settl posteriors & $1.56\pm0.01$ & $42.18^{+1.89}_{-2.10}$ & $28.52^{+3.29}_{-3.35}$ & $26.88^{+4.58}_{-4.48}$ & $34.93^{+2.08}_{-2.02}$ \\
    \hline
    Sonora priors & $U(1, 2)$ & $U(-50, 50)$ & $U(-50, 50)$ & $U(-50, 50)$ & $U(-50, 50)$\\
    Sonora posteriors & $1.56\pm0.02$ & $40.02^{+2.28}_{-2.89}$ & $25.94^{+3.17}_{-3.23}$ & $26.02^{+4.45}_{-4.53}$ & $32.61^{+2.11}_{-2.24}$ \\ 
    \hline
    \end{tabular}
    \\~\\
    \tablefoot{The radial velocities display hereafter are in the reference frame of the sun (e.g., rv~=~rv$_{\beta~\text{Pic~A}}$~+~rv$_{\beta~\text{Pic~b}}$).}
    \label{priors_post_GRAVITY_extra}
\end{table*}

\endgroup

This section summarizes the forward modeling of the four GRAVITY spectra of $\beta$~Pic~b using the forward models introduced in Sect.~\ref{sec2.1}. It describes below the bulk properties derived for the atmosphere of $\beta$ Pictoris b, its atmospheric structure and the attempt to constrain the presence of CO isotopologues.  

\begin{figure*}[t]
    \centering
    \includegraphics[scale=0.483]{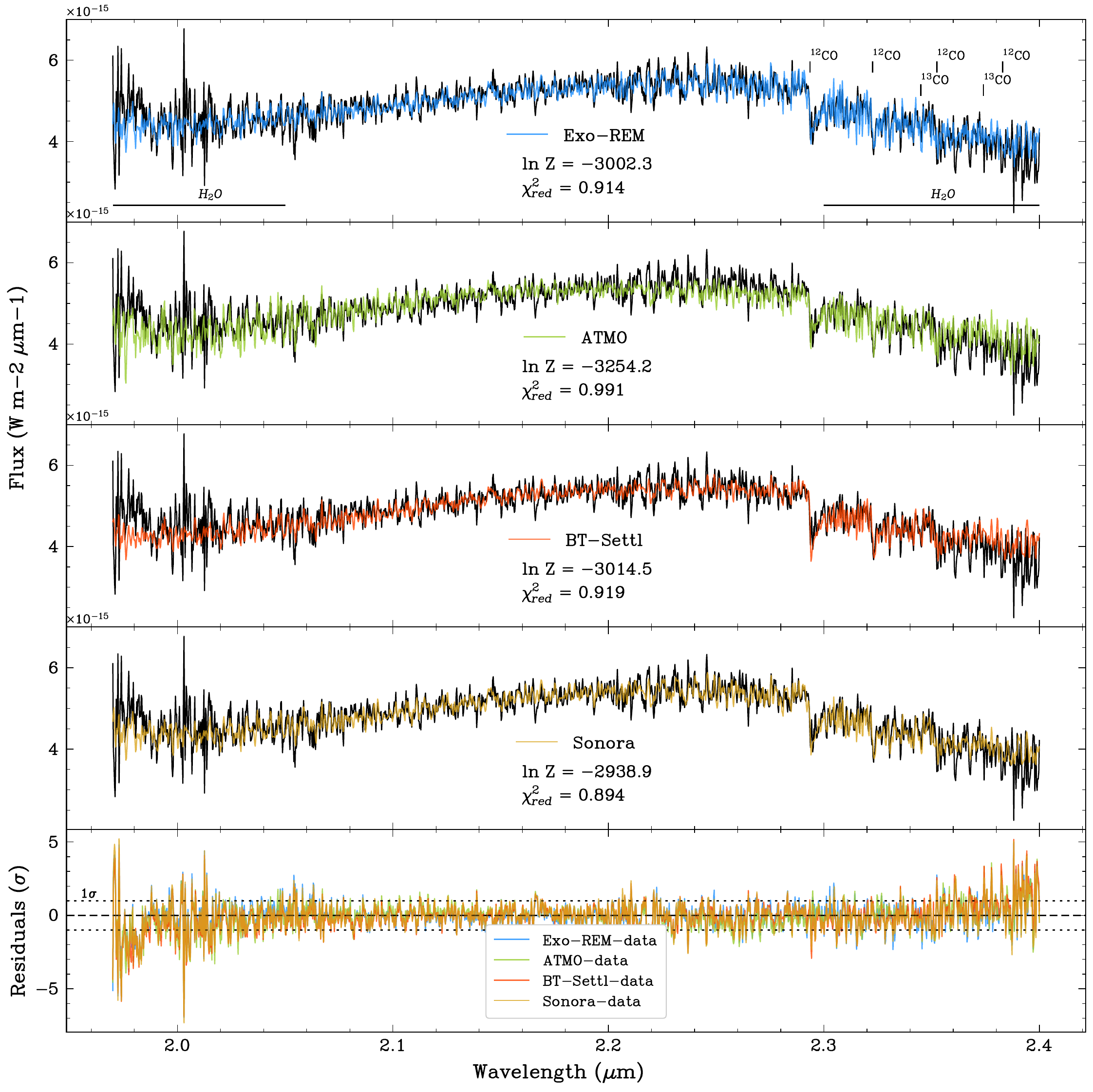}
    \caption{Results of the forward modeling of the $\beta$~Pic~b GRAVITY spectrum. The combined GRAVITY spectrum (in black) compared data from the other panels in this figure. \textit{First panel:}  Best-fit Exo-REM (in blue). \textit{Second panel:} Best-fit ATMO (in green). \textit{Third panel:} Best-fit BT-Settl (in red). \textit{Fourth panel:} Best-fit Sonora (in yellow). The two $^{13}$CO absorption lines are annotated at 2.345 µm and 2.374 µm; the four $^{12}$CO lines at 2.294 µm, 2.323 µm, 2.352 µm and 2.383 µm as well as the two H$_2$O band heads between 1.75--2.05 µm and 2.3--3.2 µm. \textit{Bottom panel:} Residuals for each fit. Dotted lines represents the $\pm\sigma$ (68$\%$) confidence interval}
    \label{GRAVITY_spec_ExoREMvsATMO}
\end{figure*}

For this inversion, the likelihood function was computed using the generalized $\chi^2$ distribution (see Sect.~\ref{sec2.2}) taking as input the covariance matrixes provided by the GRAVITY pipeline. The inversion was performed across the entire spectral coverage of GRAVITY (1.97--2.4~µm) using 1~000 live points. We also explored the additional parameters provided by \texttt{\textit{ForMoSA}}: the radius (R), radial velocity (rv), and the projected rotational velocity ($v\sin i$). The priors and their associated posteriors  are presented in Tables \ref{priors_post_GRAVITY} and \ref{priors_post_GRAVITY_extra} for the different models. Figure~\ref{GRAVITY_spec_ExoREMvsATMO} shows the best fit for the four family of atmospheric models. Posterior distributions can be found in the Appendices. 

\begingroup
\setlength{\tabcolsep}{11.5pt}
\renewcommand{\arraystretch}{1.5}
\begin{table}[ht]
\tiny
\centering
    \caption{Compilation of the Bayesian evidences ($\ln\mathcal{Z}$) and reduced $\chi^2$ ($\chi_{red}^2$) for each forward model are presented in this section.}
    \begin{tabular}{lllll}
    \hline
    \hline
    Model & Exo-REM & ATMO & BT-Settl & Sonora \\
    \hline
    $\ln\mathcal{Z}$ & $-3002.3$ & $-3254.2$ & $-3014.5$ & $-2938.9$  \\
    $\chi_{red}^2$ & $0.914$ & $0.991$ & $0.919$ & $0.894$  \\
    \hline
    \end{tabular}
    \tablefoot{For Exo-REM and ATMO, the values are those of the complete model (i.e., with the exploration of $^{12}$CO/$^{13}$CO).}
    \label{bayesian_ev_chi}
\end{table}
\endgroup

All models faithfully reproduce the observations, accurately capturing the shape of the pseudo-continuum, the pattern of H$_2$O and CO absorptions, and the absolute flux level. Residuals are relatively low ($\leq1\,\sigma$) except at the edges of the wavelength interval ($<2.02\mu$m; $>2.35\mu$m) where telluric residuals are more prominent (see Fig.~\ref{GRAVITY_post_comp}). Comparing the Bayesian evidence (see Table \ref{bayesian_ev_chi} below), Exo-REM and Sonora are statistically preferred among the four models (see Sect.~\ref{sec2} for more details). The Bayesian evidence for the ATMO model fit is significantly lower. This model proved to be the least able to give physical constraints on the planet's physical parameters, with parameters converging at the border of the grid. 

\subsection{Physical and atmospheric parameters}\label{sec3.1}

For the effective temperatures, except ATMO, all models give consistent values between $1607.45^{+4.85}_{-6.20}$ and $1668.43^{+5.62}_{-5.55}$~K. These values are slightly lower than the ones obtained by \cite{gravity_collaboration_peering_2020} using a lower resolution K-band spectrum ($1700\pm50$~K for Exo-REM and $1847\pm55$~K for \texttt{\textit{petitRADTRANS}}\footnote{\url{https://petitradtrans.readthedocs.io/en/latest/}} \citealp{2019A&A...627A..67M}). They are also lower than those predicted by evolutionary models (\citealp{chilcote_1-24_2017}, $1724\pm15$~K).

For surface gravity, evolutionary models predict values for $\beta$~Pic~b of $4.18\pm0.01$~dex \citep{chilcote_1-24_2017}. Exo-REM and Sonora retrieve larger values of $4.46^{+0.02}_{-0.04}$~dex and $4.56^{+0.06}_{-0.07}$~dex, respectively, while BT-Settl is noticeably lower at $3.60\pm0.04$~dex. ATMO also converges at the border of the considered grids, meaning that the model likely has difficulties to properly represent the pseudocontinuum in the K band.

In \texttt{\textit{ForMoSA}}, the radius is constrained by the dilution factor c$_k$~=~(R/D)$^2$ when scaling the model spectra. We use the GAIA DR3 prediction for the distance, D~$=19.8294\pm0.0571$~pc. For Exo-REM, we obtained a radius of $1.73\pm0.01$~\Rjup. A similar value of $\sim1.7$~\Rjup was found by \cite{gravity_collaboration_peering_2020}, who also used Exo-REM on LRS GRAVITY data, but with a slightly lower prior distance of $19.4401\pm0.0454$~pc. For ATMO, the radius converged to $1.160^{+0.005}_{-0.002}$~\Rjup, while BT-Settl and Sonora retrieved similar radii of $1.56\pm0.01$~\Rjup and $1.56\pm0.02$~\Rjup respectively. 

For metallicity, Exo-REM and ATMO models retrieve highly super-solar values Surprisingly, \cite{gravity_collaboration_peering_2020} obtained a sub-solar value when using the LRS from GRAVITY and only obtained super-solar values when adding the GPI data ($\sim-0.5$ and $\sim0.5$ with Exo-REM respectively).

The carbon-to-oxygen ratio obtained with Exo-REM is solar ($0.55\pm0.01$) and super-solar for BT-Settl ($0.65\pm0.02$). These values differ significantly from those adopted by \cite{gravity_collaboration_peering_2020} ($0.43\pm0.05$), \cite{landman__2024} ($0.41\pm0.04$), and \cite{worthen_miri_2024} ($0.39^{+0.10}_{-0.06}$), but is in agreement with the recent result of \cite{kiefer_new_2024} ($0.551\pm0.002$) and \cite{houlle_mathis_2025} ($0.539\pm0.003$), both using Exo-REM.

The cloud sedimentation efficiency parameter, f$_\text{sed}$, is explored only by the Sonora model grids. This parameter ranges from 1 (fully cloudy atmosphere) to 8 (no clouds). In our analysis, we obtain an upper constraint of f$_{sed}<1.03$, indicating a cloud-rich atmosphere for this object, probably form by silicate and/or metal-oxide condensates as expected from L/T objects surveys \citep{petrus_fulllib_2025}. This result is somewhat surprising, as Sonora predicts high metallicity, with metals typically trapped in clouds for such objects. Additionally, the adiabatic index $\gamma$, explored only by ATMO, converged towards the upper edge of the grid with $\gamma>1.05$. A very high adiabatic index suggests a strong temperature gradient in the atmosphere, which could simulate the presence of clouds. This finding is consistent with the low sedimentation efficiency obtained using Sonora.

\subsection{Atmospheric structure}\label{sec3.4}

\begin{figure}[th]
\centering
    \includegraphics[scale=0.68]{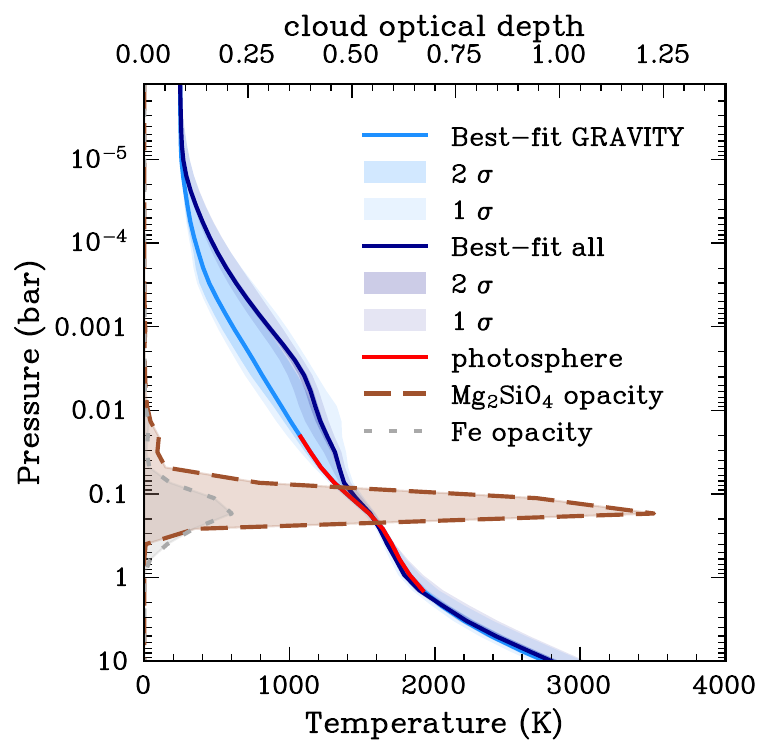}
    \caption{Retrieved best-fit pressure-temperature profiles when using GRAVITY (light blue) and all datasets except CRIRES$_+$ (dark blue) for Exo-REM forward modeling. Shaded regions represent the 1 and 2 $\sigma$ confidence interval, respectively. They are compared with the cloud optical depths of Mg$_2$SiO$_4$ (dashed brown) and Fe (dashed gray) defined as d$\tau=\sigma_c$\,n\,dz with $\sigma_c$ the effective extinction cross-section of the cloud (in m$^2$), n its concentration (in particles/m$^3$) and dz the thickness of the atmospheric layer (in m). The red curve represents the photosphere (region from which most thermal emission originates). In practice, we used a similar approach as \cite{charnay_self-consistent_2018} and constrained the pressure-temperature between the minimum and maximum of the brightness temperature computed between 0.625 to 10~µm.}
    \label{Free_MOSAIC_PT}
\end{figure}

Exo-REM grids predict the equilibrium temperature profile and the mixing ratio profiles of the most important gases. Figure~\ref{Free_MOSAIC_PT} shows the retrieved temperature profile for our best fit using the full model (in light blue)
compared with the opacities of two cloud species, Fe and Mg$_2$SiO$_4$ (in gray and brown respectively), as a function of pressure. We observe a well-constrained pressure-temperature (P-T) profile, with a cloud deck forming around P~$\sim10$~bar, where the temperature becomes high enough for clouds to be optically thick and impact the observed spectrum. The implications in terms of variability will have to be explored with 3D simulations. Stellar irradiation is not accounted for in these models, as seen in the almost isothermal top of the atmosphere. However, for $\beta$~Pic~b, a planet relatively close to its star ($9.93\pm0.03$~au), irradiation is likely to have an impact. Using predictions for $\beta$~Pic~A (see Table \ref{BetaPicSys} and \citealp{zwintz_revisiting_2019}) for the stellar radius), we estimate the stellar irradiation received by $\beta$~Pic~b to be F~$\sim108.2$~W.m$^{-2}$, which is about twice the amount Jupiter receives today ($\sim50.5$~W.m$^{-2}$). This increased irradiation could affect the planet's atmospheric chemistry and, consequently, the retrieved parameters. The coupling between self-consistent atmospheric models and irradiation models is an area of atmospheric modeling that requires further exploration.

\subsection{CO Isotopologue}\label{sec3.2}

\begin{figure}[ht!]
\centering
    \includegraphics[scale=0.75]{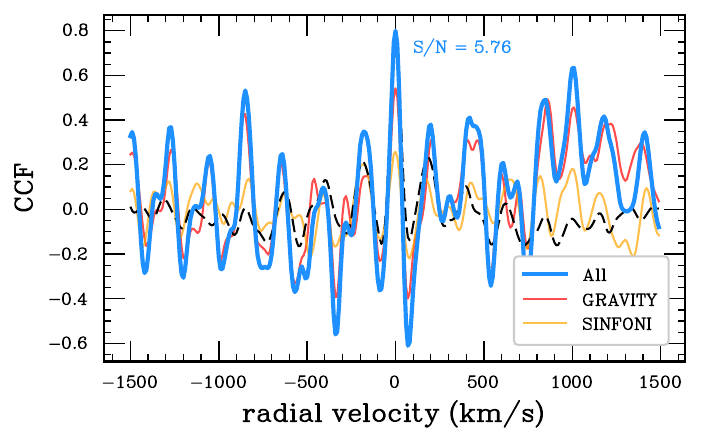}
    \caption{Cross-correlation function (CCF) between the observational residuals (i.e., data minus model without $^{13}$CO) and the $^{13}$CO model template (i.e., model with $^{13}$CO minus model without $^{13}$CO) for the two highest resolution data sets (GRAVITY and SINFONI). The sum of all CCFs is shown in blue, while the CCFs for GRAVITY and SINFONI are shown in red and orange, respectively. The auto-correlation function (dashed-black) has been re-normalized to the peak of the CCF. The CCF has been computed between 2.33--2.40 µm with a skipedge of 50 bins to properly cover the two $^{13}$CO band heads. The S/N corresponds to that of the central peak and was  computed using equation (1) of \cite{Houlle_2020} for each individual CCF, then propagated.}
    \label{Free_MOSAIC_CCF}
\end{figure}

Given the S/N of the GRAVITY spectrum ($\sim20$) and the relatively high spectral resolution, we investigated whether we could retrieve some constraint on the isotopic ratio $^{12}$CO/$^{13}$CO. With Exo-REM, we only manage to obtain a loose upper constraint of log($^{12}$C/$^{13}$C)~<~1.08. Similarly, ATMO only constrained it up to log($^{12}$C/$^{13}$C)~<~1.55. Since both grids does not explore the parameter space below log($^{12}$C/$^{13}$C)~=~1.0, it is hard to conclude on the exact upper bound, because we could just be seeing the end tail of a much larger posterior distribution. We used a similar strategy as \cite{zhang_13co-rich_2021} to assess the detection and its statistical significance. The statistical significance is directly obtain by comparing the Bayesian evidences, computing the Bayes factor,

\begin{equation*}
    \ln B_m=\ln\mathcal{Z}\,\text{(model with $^{13}$CO)} - \ln \mathcal{Z}\,\text{(model without $^{13}$CO)},
\end{equation*}

for two runs using respectively the model with and without $^{13}$CO. The model without $^{13}$CO is the same model as the one exploring the $^{13}$CO abundance, but with a fixed log($^{12}$C/$^{13}$C)=100. Focusing on Exo-REM, we obtained $ln(B_m)=29.3\pm0.2$; meaning that the observation favor a full model with $^{13}$CO at a significance level $7.9\sigma$. To support this value, we computed the CCF between the observational residuals (i.e., data minus model without $^{13}$CO) and a $^{13}$CO model template (i.e., model with $^{13}$CO minus model without $^{13}$CO). This CCF, performed around the $^{13}$CO band heads (2.34--2.4 µm) compared with the re-normalized auto-correlation, is shown on Fig.~\ref{Free_MOSAIC_CCF}. We see a tentative peak at 0 km.s$^{-1}$ (S/N~=~4.54) which could indicate a $^{13}$CO detection. However, the presence of strong residuals of the order of the variation induced by the $^{13}$CO makes this statement very dubious. Indeed, analyzing each GRAVITY observation separately revealed that the first night (2019/11/09) had significantly more telluric residuals. Inverting the data for each of the four nights individually provided only upper limits for the first night, while no constraints (i.e., a flat posterior) were obtained for the other three nights. We further investigate these detection limits and their significance using synthetic spectra in Appendix \ref{sec8.2}.

\subsection{Radial velocity}\label{sec3.3}

Radial velocities presented in Table \ref{priors_post_GRAVITY_extra} have been corrected from the Earth barycentric velocity. We note that Exo-REM's and ATMO's predictions are significantly shifted compared to those of BT-Settl and Sonora for the 2019/11/09 and 2022/01/24 epochs. Coincidentally, Exo-REM and ATMO are the only two grids with spectral sampling lower than GRAVITY beyond $\sim$2.3~µm, which may partially explain these discrepancies. Accounting for the systemic velocity of the host star ($v_{sys}=20\pm0.7$~km.s$^{-1}$, \citealp{gontcharov_pulkovo_2006}) and propagating its error bars, we show on Fig.~\ref{all_rvs} the retrieved velocities for the BT-Settl model together with the orbital predictions from the astrometric measurements \footnote{\url{https://github.com/semaphoreP/whereistheplanet}} (\citealp{wang_asclnet_whereistheplanet}). They are relatively consistent with the predictions except the 2019/11/09 epoch. 

\section{Combining low and medium spectral resolution}\label{sec4}

This section focuses on the joint datasets described in Table \ref{table_archival},  excluding the CRIRES$+$ data. Given the multimodality of the datasets, two main approaches are followed for the inversion, described below.

\begin{itemize}
    \item \textbf{Free:} we assume here that none of the datasets are perfectly calibrated in flux. \texttt{\textit{ForMoSA}} applies a dilution factor c$_k$ to scale the model to the data flux, which is calculated analytically \citep{cushing_atmospheric_2008}:
    
    \begin{equation}
        \text{c}_k=\frac{\sum d \times m / \sigma^2}{\sum m^2/\sigma^2},
    \end{equation}
    
    where $m$ is the model, $d$ the data, and $\sigma$ their associated uncertainties.
    \item \textbf{Physically informed:} In contrast to the  approach above, we introduce constraints on the planet's radius, $R$, and use a new scaling parameter, $\alpha_i$, for each dataset, $i$. This parameter is a multiplicative factor defined as
    \begin{equation}
        \text{c}_{k,i}=\alpha_i\,(\text{R}/\text{D})^{2}.
    \end{equation}
    
    We imposed a broad Gaussian prior on the radius ($R=1.4\pm0.1$~\Rjup) extended from \cite{chilcote_1-24_2017}, and use informed priors on parameters sensitive to flux calibration, such as log(g)~=~$4.18\pm0.13$~dex from \cite{lacour_mass_2021}'s dynamical mass measurements. To account for degeneracies, we further constrained $\alpha_i$ with a Gaussian prior ($\alpha_i=1\pm1$), assuming the data are reasonably well-calibrated and allowing only small variations around 1.
\end{itemize}

\begin{figure*}[t]
    \centering
    \includegraphics[scale=0.47]{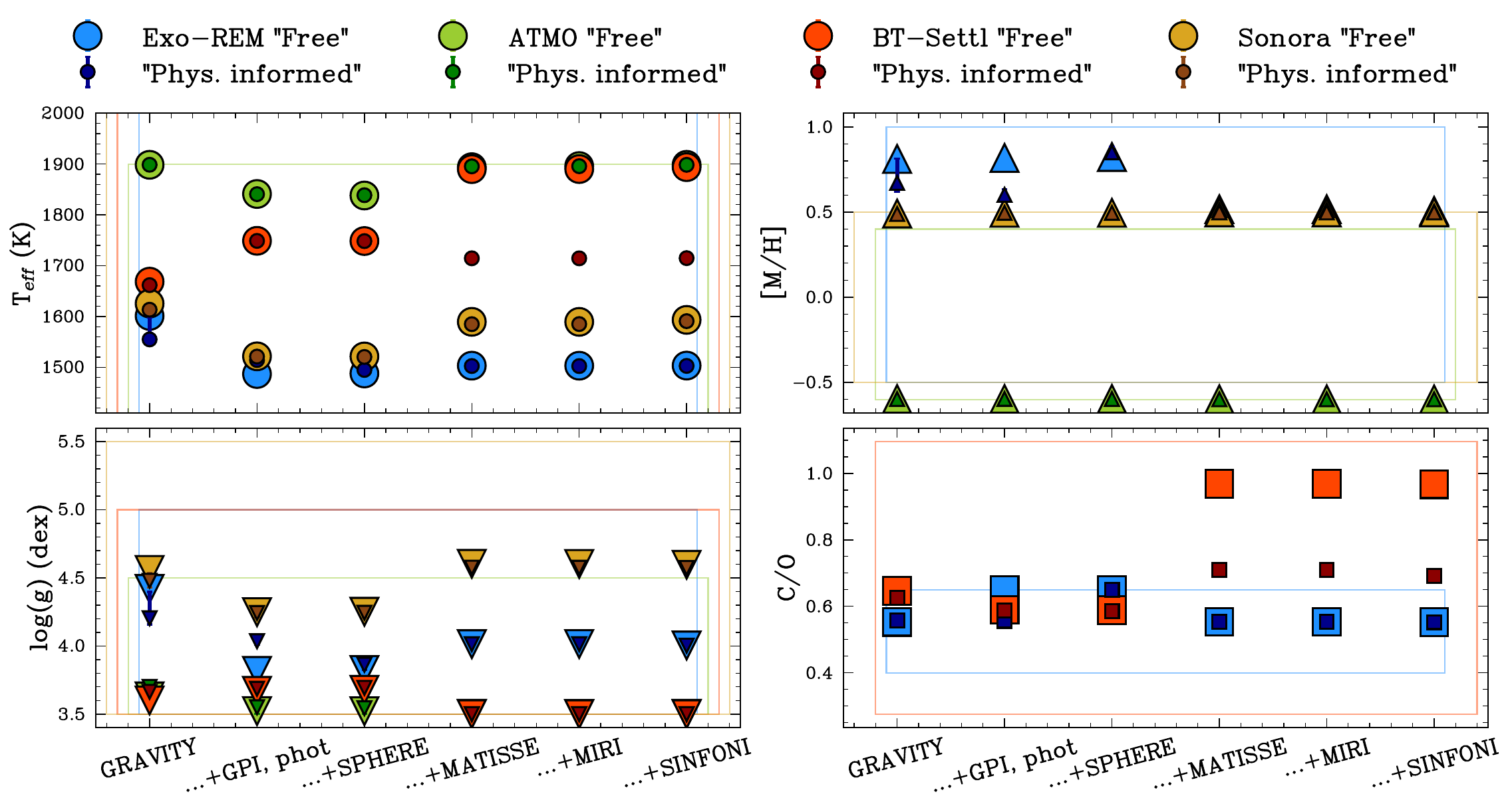}
    \caption{Evolution of the grid parameters when adding observations iteratively from only GRAVITY (left) all except CRIRES$_+$ (right). Each color represents one of the model tested: blue for Exo-REM, green for ATMO, red for BT-Settl and orange for Sonora. Colored boxes indicate the boundaries of the respective model grids. Large markers represent the inversions with the "free" approach, while small markers represent those with the "physically informed" approach. The error bars correspond to the lower and upper bonds in the parameter space encompassing 68\% of the retrieved solutions around the best fit.}
    \label{Free_MOSAIC_post}
\end{figure*}

We report in Tables \ref{priors_post_MOSAIC_Free}, \ref{priors_post_MOSAIC_Phys}, and \ref{priors_post_extra_MOSAIC_All} the main characteristics of the posterior distributions when datasets are  added iteratively. The main free parameters for each model are shown in Fig.~\ref{Free_MOSAIC_post}. Both the Exo-REM and Sonora models remain statistically preferred over ATMO and BT-Settl, as supported by the Bayesian evidences and reduced  $\chi^2$ (Table \ref{bayesian_ev_MOSAIC_Free}). We note that only the Exo-REM grid did not go off the edge of the grids when we added all the data, which, in our view, makes it the most robust in terms of derived physical parameters.

The best fits for each model are shown in Fig.~\ref{GRAVITY_MOSAIC_fullspec}. The overall pseudo-continuum shape is reproduced by all models but the residuals reveal noticeable differences. We interpret this as a consequence of the preponderance of some datasets containing the largest number of datapoints or with the largest S/N (e.g., GRAVITY and SINFONI) in the likelihood computation. As a consequence, datasets  at lower resolutions exhibit stronger residuals, but which remain within the 1$\sigma$ envelope.

\begingroup
\setlength{\tabcolsep}{6pt}
\renewcommand{\arraystretch}{1.5}
\begin{table}[ht]
\tiny
\centering
    \caption{Comparison of the Bayesian evidence ($\ln\mathcal{Z}$) and reduced chi-square ($\chi_{red}^2$) associated with each fits presented in this section ("free" approach).}
    \begin{tabular}{lllll}
    \hline
    \hline
    Model & Exo-REM & ATMO & BT-Settl & Sonora \\
    \hline
    $\ln\mathcal{Z}\,(\text{GRAVITY})$ & $-2968.7$ & $-3198.8$ & $-2981.5$ & $-2905.9$ \\
    $\ln\mathcal{Z}\,(\text{...+GPI,phot})$ & $-3078.9$ & $-3348.9$ & $-3419.2$ & $-3027.8$ \\
    $\ln\mathcal{Z}\,(\text{...+SPHERE})$ & $-3089.9$ & $-3363.3$ & $-3437.3$ & $-3040.4$ \\
    $\ln\mathcal{Z}\,(\text{...+MATISSE})$ & $-3639.3$ & $-4592.7$ & $-5163.8$ & $-3564.7$ \\
    $\ln\mathcal{Z}\,(\text{...+MIRI})$ & $-3645.3$ & $-4597.8$ & $-5171.4$ & $-3570.5$ \\
    $\ln\mathcal{Z}\,(\text{...+SINFONI})$ & $-5039.7$ & $-6187.0$ & $-6532.6$ & $-4650.0$ \\
    \hline
    $\chi_{red}^2\,(\text{GRAVITY})$ & $0.905$ & $0.975$ & $0.911$ & $0.886$ \\
    $\chi_{red}^2\,(\text{...+GPI,phot})$ & $0.915$ & $0.987$ & $1.013$ & $0.898$ \\
    $\chi_{red}^2\,(\text{...+SPHERE})$ & $0.913$ & $0.985$ & $1.012$ & $0.896$ \\
    $\chi_{red}^2\,(\text{...+MATISSE})$ & $1.038$ & $1.303$ & $1.471$ & $1.013$ \\
    $\chi_{red}^2\,(\text{...+MIRI})$ & $1.035$ & $1.300$ & $1.467$ & $1.011$ \\
    $\chi_{red}^2\,(\text{...+SINFONI})$ & $1.186$ & $1.450$ & $1.534$ & $1.090$ \\
    \hline
    \end{tabular}
    \tablefoot{For Exo-REM, the values are those of the complete model (i.e., with the exploration of $^{12}$CO/$^{13}$CO).}
    \label{bayesian_ev_MOSAIC_Free}
\end{table}
\endgroup

Compared to the previous inversion on the GRAVITY datasets and considering mainly the Exo-REM and Sonora solutions, we note the following points.

\begin{itemize}

\item Since the effective temperature is strongly influenced by the shape of the continuum, the inclusion of low-resolution data with broad spectral coverage from GPI and photometry significantly impacts its value. There is a significant decrease by typically $\sim100$\,K for Exo-REM that shifts from $1607.45^{+4.85}_{-6.20}$~K to $1503.38^{+2.30}_{-2.31}$~K when all datasets are considered. For Sonora, \Teff decreases but remains compatible within 2$\sigma$.

\item For the surface gravity, the posteriors remain fairly centered around log(g)~=~$4.00\pm0.01$ dex with Exo-REM. With Sonora, it remains fairly constant around $4.60\pm0.02$~dex, illustrating the model-to-model systematics on the determination of that parameter. 

\item The radius was fitted separately using the "physically~informed" runs using a relatively strong prior of 1.4~$\pm$~0.1~\Rjup. Results are stored in the first column of Table \ref{priors_post_extra_MOSAIC_All}. Exo-REM and Sonora end up finding more consistent radii of $1.43\pm0.07$~\Rjup and $1.37\pm0.08$~\Rjup, respectively. The scaling parameters, $\alpha_i$, for the four models can be found in Fig.~\ref{alpah_evo}.

\item The metallicity remains super-solar when more datasets are added. For Exo-REM, we observe a shift with the inclusion of MATISSE data, where the metallicity decreases from $0.82\pm0.03$ to $0.52\pm0.02$. Sonora shows no changes compared to the GRAVITY-only inversion, with [M/H]~$>0.49$ at the grid border, suggesting very high metallicity. 

\item the carbon-oxygen ratio briefly converges to a super-solar value ($>0.646$) when only the photometry, and the GPI and SPHERE data are considered in the fit for Exo-REM. However, with the addition of MATISSE, it returns to a near-solar value of $0.554\pm0.003$, which remains consistent with the value determined using the SINFONI data alone. 

\item Figure \ref{Free_MOSAIC_PT} shows the evolution of the retrieved Exo-REM pressure-temperature profile when we  iteratively add observations. We see that adding datasets has a noticeable effect on the profile, constraining  the 1$\sigma$ contour significantly. The overall shape of the pressure-temperature profile is maintained. The two dominant cloud species (Fe and Mg$_2$SiO$_4$) are overlaid as a function of pressure when using all datasets. They form in the photospheric region of the atmosphere. 

\item For the detection of the CO isotopologue, when adding the SINFONI datasets to the GRAVITY one, we note that the Exo-REM models constrain a sub-solar log($^{12}$C/$^{13}$C)~$=1.12^{+0.11}_{-0.08}$, although a significant part of the tail of the distribution remains at the grid edge, as seen in Fig.~\ref{Free_MOSAIC_13CO}. The Bayes factor between the model with and without $^{13}$CO is $\ln B_m =(-5039.7)-(5107.9) = 68.2$ indicates that the model model with  $^{13}$CO is preferred at $11.9\,\sigma$. To support this, we also computed the CCF between the two residuals as done previously. The results are shown in Fig.~\ref{Free_MOSAIC_CCF} with the CCFs obtained using GRAVITY-alone, SINFONI-alone, and their combination. As in Sec.\,\ref{sec3.2}, we observe that the GRAVITY signal is heavily affected by residual telluric absorption. The SINFONI spectrum is less affected by this issue and the $^{13}$CO signal appears more pronounced. This weak and systematics-prone constraint does not support a robust $^{13}$CO detection at this stage.

\item Finally, similarly to what is described in Sect.~\ref{sec3.3}, the radial velocities were fitted separately for the four GRAVITY epochs and SINFONI. Results are stored in Table \ref{priors_post_extra_MOSAIC_All} in Appendix \ref{sec8.3}.
For GRAVITY, all models yield similar radial velocities, regardless of the added datasets, with values ranging between 5~km.s$^{-1}$ and 15~km.s$^{-1}$; this is in good agreement with predictions from the most recent orbital elements  (\citealp{wang_asclnet_whereistheplanet} and Fig.~\ref{all_rvs}).

\end{itemize}

\section{Adding CRIRES+ spectra}\label{sec5}

\subsection{Modeling approach}\label{sec5.1}

This section presents the results obtained when adding the two CRIRES$_+$ datasets. These data differ from the rest, as they contain both the companion signal and the host star light. This required us to implement a dedicated analysis scheme within our framework, following the approach described in \cite{landman__2024} for high resolution, while still incorporating the low- and medium-resolution observations.

\subsection{Results}\label{sec5.2}

For the combined inversion on the CRIRES$_+$ data, we used the likelihood mapping from \cite{ruffio_radial_2019} on the A and B nods, with the two nights processed separately to avoid re-interpolation. Each order was also fitted separately to avoid border issues when computing the continuum. We constrained the exploration of the radial velocity using the priors from \cite{landman__2024}, with rv $\in$ [20 km.s$^{-1}$, 40 km.s$^{-1}$]. For the M-band data, we avoided the B nod of the OB2 (second night) as well as the third order in all frames. All contribution where fitted in parallel for the K-band data while we only use the residual flux of the M-band data. We use the BT-Settl model since it is the only one capable of reaching both K- and M-band resolution requirement of R$_{\lambda}$~$\sim$100,000 as shown on Fig.~\ref{model_data_res}.

\begin{figure*}[ht]
    \centering
    \includegraphics[scale=0.4]{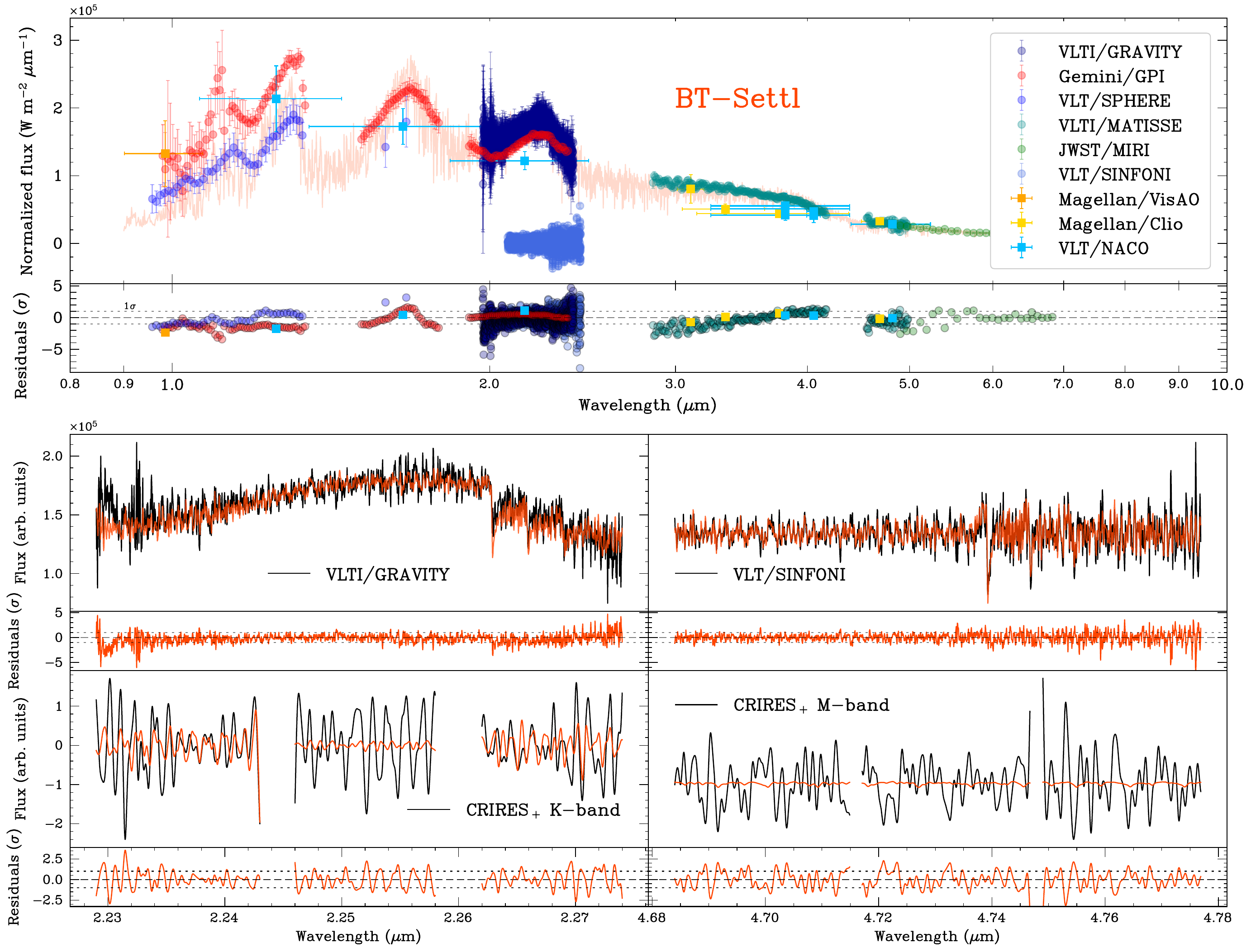}
    \caption{Results of the forward modeling of $\beta$~Pic~b using the most recent spectro/photometric observations of the planet (including CRIRES$_+$) using BT-Settl. \textit{Top panel:} Spectra (points) and photometry (squares) of $\beta$~Pic~b alongside a R$_{\lambda}\sim$ 4,000 spectrum extracted using the best fit parameters for the inversion. Observations have been re-normalized using the analytic scaling factors computed during the inversion. Associated residuals are shown below. Dotted lines represents the $\pm\sigma$ (68$\%$) confidence interval. \textit{Bottom sub-panels:} Zoom versions of the top panel on GRAVITY (upper left), SINFONI (upper right), CRIRES$_+$-K (lower left) and CRIRES$_+$-M (lower right). Similarly to \cite{landman__2024}, each CRIRES$_+$ spectra are shown smoothed by their retrieved broadening to show the planet features more clearly.}
    \label{ALL_MOSAIC_CRIRES_spec}
\end{figure*}

Posteriors are provided in Table \ref{priors_post_MOSAIC_CRIRES}, and the corresponding best fit is shown in Fig.~\ref{ALL_MOSAIC_CRIRES_spec}. In the first lower panel of Fig.~\ref{ALL_MOSAIC_CRIRES_spec}, we observe that BT-Settl struggles to reproduce the the GPI YJ bands, a behavior noted earlier in Sect.~\ref{sec4} when fitting for the $\alpha$ parameter. The model also shows deeper $^{12}$CO absorption lines, seen in both the GRAVITY and SINFONI data which is probably correlated to the high C/O retrieved. The last two lower panels display one of the spectral orders alongside the fitted model for the two CRIRES$_+$ observations. The thermal contribution in the M band is particularly pronounced, with the fitted planetary signal appearing almost flat, while in the K band, we retrieve some broad planetary features. 

We do not observe any major changes when comparing the atmospheric parameters obtained using the "physically informed" approach from Sect.~\ref{sec4} with those retrieved here, even after adding the CRIRES$_+$ data. This suggests that CRIRES$_+$ does not provide any strong additional information on the atmospheric parameters. As expected, the absence of the planet's continuum in the CRIRES$_+$ datasets makes it difficult to constrain \Teff and log(g), and the values retrieved align with those predicted using only low- and medium-resolution datasets. However, changes in the posterior distributions of the C/O is more complex, as both the K band (2.34--2.38~µm) and M band (4.25--5.22~µm) feature strong CO absorption lines, which should influence the retrieved C/O. This is evident in the significantly lower C/O values obtained when inverting each CRIRES$_+$ dataset individually, C/O~=~$0.35^{+0.06}_{-0.05}$ for the K band and C/O~=~$0.46^{+0.20}_{-0.12}$ for the M band (using priors of G(1700~K, 100~K) for \Teff and G(4.18~dex, 0.13~dex) for log(g)). In this simple framework we are fitting all possible orders which makes it likely that a significant amount of noise is introduced into the CCF signal used to compute the likelihood functions. This may reduce the prominence of the CO signal compared to that found in the GRAVITY, SINFONI, and MATISSE datasets. In particular in the M band, the large error-bars associated to the thermal noise render the likelihood function almost flat.

\begin{figure}[ht]
\centering
    \includegraphics[scale=0.47]{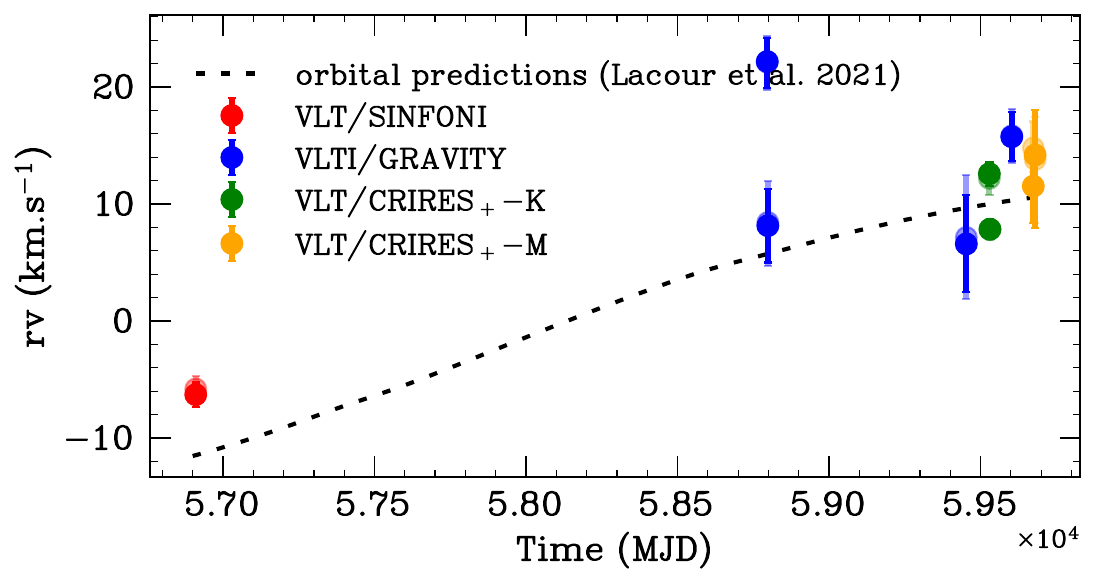}
    \caption{Orbital predictions from \cite{lacour_mass_2021} (dotted line) vs. retrieved radial velocities (colored bars) for each epoch, using forward modeling (here BT-Settl). The four GRAVITY epochs were fitted in parallel using MOSAIC, and we can see that the observation on the 2019/11/09 is significantly shifted.}
    \label{all_rvs}
\end{figure}

The retrieved radial velocities for each CRIRES$_+$ epoch agreed relatively well with orbital predictions although we do observe a significant spread around the expected value for the two epochs in the K band. Figure~\ref{all_rvs} compiles all the retrieved radial velocities for each epochs using BT-Settl. For the rotational velocity, we use the broadening function \textit{FastRotBroad} from \texttt{\textit{PyAstronomy}}\footnote{\url{https://pyastronomy.readthedocs.io/en/latest/}} jointly on the K- and M-band CRIRES$_+$ epochs. We obtained a value of $v\sin i$~$=27.20^{+1.29}_{-1.16}$~km.s$^{-1}$, consistent at 1$\sigma$ with the values of \cite{snellen_fast_2014} (25~$\pm$~3~km.s$^{-1}$) and \cite{kiefer_new_2024} (25$^{+5}_{-6}$~km.s$^{-1}$) and at 2$\sigma$ with the values of \cite{landman__2024} (19.9~$\pm$~1.0~km.s$^{-1}$) and \cite{parker_into_2024} (22~$\pm$~2~km.s$^{-1}$). One possible explanation for the discrepancy between the value retrieved here and those reported by \cite{landman__2024} and \cite{parker_into_2024} is that the forward model grids used in this study do not have a well-defined spectral resolution. Additional sources of uncertainty arise from the fact that, for computational efficiency, the grid is resampled and truncated in wavelength before the inversion. This inevitably affects both the retrieved radial and rotational velocities. Consequently, both the retrieved $v\sin i$ should be interpreted as a broadening coefficient that accounts for these uncertainties and for the planet's true rotational velocity.

\setlength{\tabcolsep}{5pt}
\renewcommand{\arraystretch}{1.5}
\begin{table}[ht!]
\tiny
    \centering
    \caption{Grid and extra-grids priors and posteriors for the inversion including CRIRES$_+$ (Sect.~\ref{sec5}).}
        \begin{tabular}{lll}
    \hline
    \hline
    Parameter & BT-Settl priors & BT-Settl posteriors\\
    \hline
    \Teff (K) & $U(1400, 2000)$ & $1710.49^{+2.73}_{-2.67}$ \\
    log(g) (dex) & $U(3.5, 5.0)$ &  $<3.50$ \\
    C/O & $U(0.2754, 1.096)$ & $0.701\pm0.005$ \\
    rv$_\text{GRAVITY~(2019/11/09)}$ (km.s$^{-1}$) & $U(-50, 50)$ & $42.18^{+1.91}_{-2.16}$ \\
    rv$_\text{GRAVITY~(2019/11/11)}$ (km.s$^{-1}$) & $U(-50, 50)$ & $28.15^{+3.03}_{-3.07}$ \\
    rv$_\text{GRAVITY~(2021/08/27)}$ (km.s$^{-1}$) & $U(-50, 50)$ & $26.59^{+4.10}_{-4.06}$ \\
    rv$_\text{GRAVITY~(2022/01/24)}$ (km.s$^{-1}$) & $U(-50, 50)$ & $35.75^{+1.98}_{-1.95}$ \\
    rv$_\text{SINFONI}$ (km.s$^{-1}$) & $U(-50, 50)$ & $13.71^{+0.77}_{-0.80}$ \\
    rv$_{\text{CRIRES}_+\text{-K~(2021/11/11)}}$ (km.s$^{-1}$) & $U(20, 40)$ & $32.59^{+0.73}_{-0.79}$ \\
    rv$_{\text{CRIRES}_+\text{-K~(2021/11/13)}}$ (km.s$^{-1}$) & $U(20, 40)$ & $27.85^{+0.46}_{-0.44}$ \\
    rv$_{\text{CRIRES}_+\text{-M~(2022/04/05)}}$ (km.s$^{-1}$) & $U(20, 40)$ & $31.53^{+2.63}_{-3.06}$ \\
    rv$_{\text{CRIRES}_+\text{-M~(2022/04/11)}}$ (km.s$^{-1}$) & $U(20, 40)$ & $34.19^{+3.78}_{-6.20}$ \\
    $v\sin i_{\text{CRIRES}_+}$ (km.s$^{-1}$) & $U(1, 100)$ & $27.20^{+1.29}_{-1.16}$ \\
    \hline
    \end{tabular}
    \label{priors_post_MOSAIC_CRIRES}
\end{table}

\section{Discussion}\label{sec6}

\subsection{The multimodal analysis: Advantages and limitations}\label{sec6.1}

Using multiple datasets in a single and comprehensive framework is both powerful, but also poses several challenges when wanting to infer on the physical properties of an exoplanet. Photometry and low-resolution spectra primarily constrain \Teff and log(g), while high-resolution data lack the information on the emission continuum but capture the molecular features which are set by molecular abundances in the gas phase. When relying on a limited set of observations, the resulting posteriors can be degenerate. Combining multiple datasets within a joint likelihood framework can help break these degeneracies. This is illustrated in Fig.~\ref{bimodal_exemple}, where the bimodal solution obtained with Exo-REM (without $^{13}$CO) disappears once the MATISSE observation is included.

Known sources of error from each observation are properly propagated into the final posteriors; either through observational error bars or, when available, covariance matrices. Nevertheless, unaccounted noise and/or bias may still remain. This is clearly illustrated by the nonphysical "uncertainties" retrieved on the atmospheric parameters in all our inversions. These uncertainties result from the propagation of observational errors (or covariances) but do not account for model systematics \citep{petrus_jwst_2024}. This issue is particularly evident in Figure~\ref{Free_MOSAIC_post}, where the retrieved uncertainties are significantly smaller than the differences between models; even between the two best-fitting ones, Exo-REM and Sonora.

Several studies \citep{2015ApJ...812..128C,kawahara_2025_autodiff,De-Regt_2024_ESO,De-Regt_2025_ESO,rotman_2025_enable} have demonstrated the strength of Gaussian processes (GPs) in modeling unaccounted noise and spectral correlations, and in propagating them into the final posteriors, in the context of atmospheric modeling of exoplanets and brown dwarfs. Thus, we argue that adapting these developments into our forward modeling framework is the next logical step in \texttt{\textit{ForMoSA}} if we want to increase the robustness of our physical interpretations.

\begin{figure}[ht!]
\centering
    \includegraphics[scale=0.42]{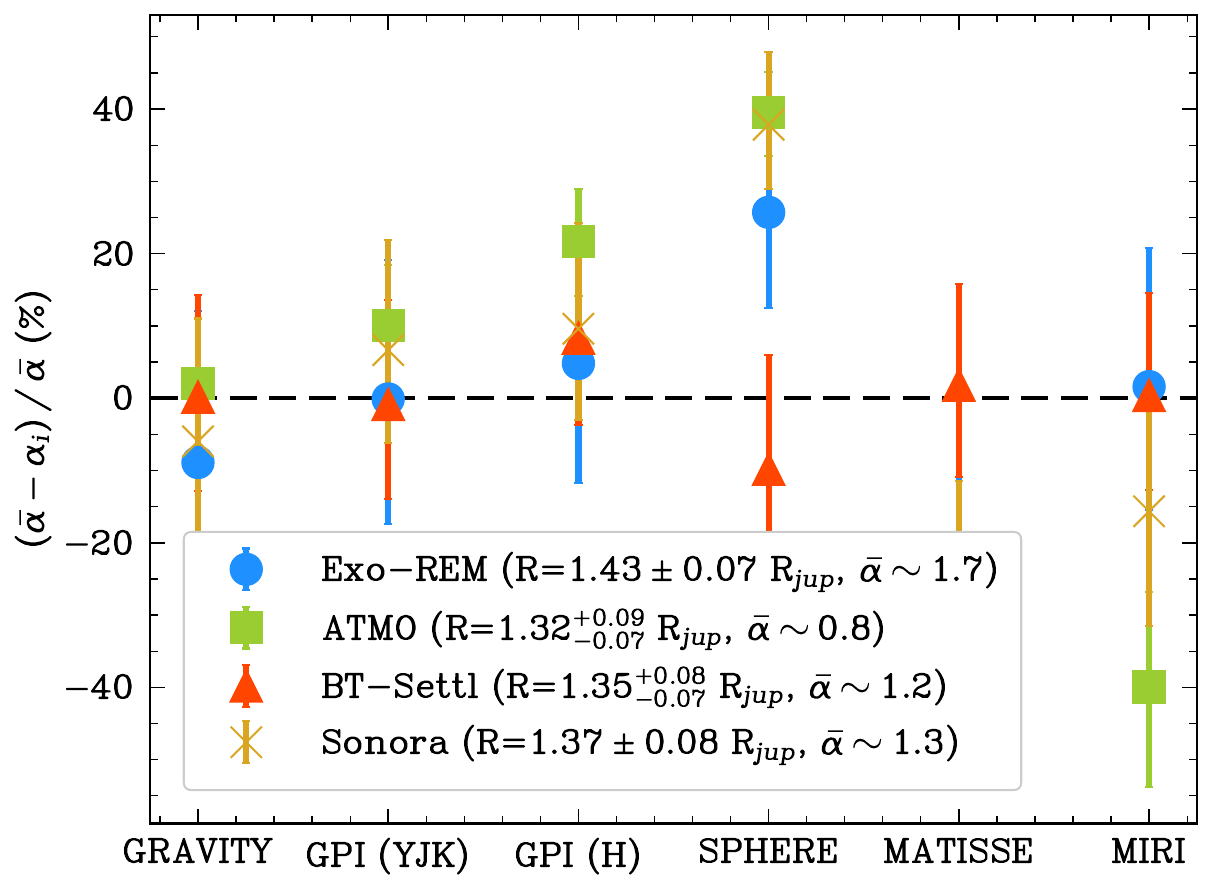}
    \caption{Evolution of the relative scaling parameters $(\bar{\alpha}-\alpha_{i})\,/\,\bar{\alpha}$ with respect to the dataset considered. $\alpha_{i}$ is the scaling parameter for the $i$-th observation, while $\bar{\alpha}$ is the mean of $\alpha_{i}$ taken over all $i$. $\bar{\alpha}$ values for each model are reported in the legend. Only datasets with continuum information were rescaled. Each color represents one of the model tested: blue for Exo-REM, green for ATMO, red for BT-Settl, and orange for Sonora. The error bars correspond to the lower and upper bonds in the parameter space encompassing 68\% of the retrieved solutions around the best fit. }
    \label{alpah_evo}
\end{figure}

Except for BT-Settl, we find no major differences between the posteriors retrieved with the "free" and "physically informed" inversions (Tables~\ref{priors_post_MOSAIC_Free} and \ref{priors_post_MOSAIC_Phys}, Appendix~\ref{sec8.3}). Yet, a closer look at the retrieved scaling parameters ($\alpha_i$) reveals significant inter-dataset discrepancies; up to 40\% in some cases (Fig.\ref{alpah_evo}). BT-Settl shows the smallest scaling adjustments ($<10$\%), yet still fails to reproduce the spectral shape (see residuals in Fig.\ref{ALL_MOSAIC_CRIRES_spec}). Exo-REM also applies limited inter-dataset scaling, but with a higher average scaling factor ($\bar{\alpha} \approx 1.7$), suggesting that this model systematically underestimates the absolute flux at a given radius. Overall, the interpretation of the scaling parameters is non-trivial, as they reflect a combination of both model deficiencies and data-related systematics.

Fitting for separate velocity shifts and broadenings for different instruments or observations is a powerful way to avoid biasing the estimated atmospheric parameters. This approach helps account for potential biases within and between instruments and models, preventing their propagation into the retrieved parameters. This is particularly important for parameters sensitive to line profiles such as log(g), C/O and [M/H].
\\

Finally, we would like to underline that, in its current form, the MOSAIC framework is constructed under the assumption of independent datasets/likelihood (see Sect. \ref{sec2.2} Fig. \ref{MOSAIC_schem}). Temporal independence is generally valid when combining data from different instruments, as they do not share the same instrumental systematics. Furthermore, it is standard practice to assume that photon noise is uncorrelated across time.
However, when multiple datasets originate from the same instrument at different epochs, correlated systematics can be present across spectra, which may violate the assumption of independence. In the present study, this applies to GRAVITY and the two CRIRES$_+$ observations. For GRAVITY, we argue that the poor wavelength calibration (see Fig. \ref{all_rvs}) and the limited number of epochs (only four) may hinder our ability to accurately estimate and propagate these correlations.
A more in-depth data-merging analysis would be required, but this is beyond the scope of the present work.

\subsection[On the atmospheric properties of beta Pictoris b]{Exploring the atmospheric properties of $\beta$ Pictoris b}

Across our study, Exo-REM proved to be the most suitable grid for modeling $\beta$~Pic~b's data. Using only GRAVITY in Sect.~\ref{sec3}, the two preferred models, Exo-REM and Sonora, predict similar values for \Teff ($1607.45^{+4.85}_{-6.20}$ and $1625.67^{+11.48}_{-11.21}$K) and log(g) ($4.46^{+0.02}_{-0.04}$ and $4.56^{+0.06}_{-0.07}$~dex). These results are consistent with previous atmospheric and evolutionary model predictions (see Table \ref{litterature}). However, Exo-REM estimates a radius that is about 10\% larger than Sonora's: $1.73\pm0.01$~\Rjup compared to $1.56\pm0.01$~\Rjup. By combining these radii and surface gravities, it is possible to estimate the planet's mass. Exo-REM provides a mass estimate of $34.82\pm0.40$~\Mjup, while Sonora gives $35.65\pm0.92$~\Mjup; both of which are significantly higher than the more robust dynamical measurements from \cite{lacour_mass_2021} of $11.9^{+2.93}_{-3.04}$~\Mjup. This tendency of forward models to under or overestimate both radius and surface gravity has been observed in other recent works \citep{2023A&A...670A..90P} and could stem from incorrect predictions of the emergent flux by the models after rescaling.

The retrieved metallicity using only GRAVITY is super-solar for both Exo-REM ($0.82\pm0.02$) and Sonora ($>0.49$). A high metallicity, combined with thick cloud coverage as indicated by the very low sedimentation efficiency in Sonora (f$_{\text{sed}}\sim1$) and previous photometric surveys \citep{currie_combined_2013, bonnefoy_near-infrared_2013, morzinski_magellan_2015, chilcote_1-24_2017, gravity_collaboration_peering_2020, 2020A&A...635A.182S}, is a surprising result, considering that clouds typically sequester a significant amount of metallic species (e.g., Mg, Si, Fe, O), which tends to lower the observed metallicity. In the case of Exo-REM, the metallicity is mainly driven by H$_2$O absorption features while Mg, Si and Fe condense. The likely explanation for this issue (clouds + high metallicity) is likely to come from an incorrect or imperfect reproduction of the continuum by the model. To demonstrate this, we repeated the inversion presented in Sect.~\ref{sec3}, this time removing the continuum from the GRAVITY data. The results, shown in Table \ref{priors_post_ExoREM_nocont} below, show a noticeable change in every parameter, particularly in [M/H], which shifts from $0.82\pm0.02$ to $0.24\pm0.12$. This lower metallicity is in fact more aligned with most planet formation model predictions, as shown on the top panel of Fig.~\ref{Schneider_pop}. This imperfect reproduction of the continuum may stems from and incomplete and/or imperfect cloud and chemistry modeling \citep{2017AJ....154...10R, 2020A&A...640A.131M}: Firstly, it is possible that the collision-induced absorption of H$_2$, which is the dominant absorber at the pressure levels probed by the K band, is not properly modeled. Conversely, it is possible that for some spectra in the grid, the emitted flux passing through the cloud deck is not properly spectrally redistributed which can strongly modify the shape of the K band. In both cases, the impact on the K-band shape can be sufficient enough to explain this discrepancy. Super-solar metallicities were also found by \cite{gravity_collaboration_peering_2020}, \cite{worthen_miri_2024} and \cite{houlle_mathis_2025} using Exo-REM on the combined GRAVITY and available LRS data. More recently, \cite{parker_into_2024} detected an SiO signal in the M band using CRIRES$_+$, further complicating the tension between clouds and metallicity, as SiO is expected to be readily sequestered into clouds (see Sect.~\ref{sec3.4} and Fig.~\ref{Free_MOSAIC_PT}). 

\begingroup
\setlength{\tabcolsep}{5pt}
\renewcommand{\arraystretch}{1.5}
\begin{table}[ht]
\tiny
    \centering
    \caption{Grid priors and posteriors obtain on the continuum-removed GRAVITY spectrum using the Exo-REM model.}
    \tiny
    \begin{tabular}{lllll}
    \hline
    \hline
    Exo-REM & \Teff & log(g) & [M/H] & C/O \\
    & (K) & (dex) & & \\
    \hline
    priors & $U(1200, 2000)$ & $U(3.5, 4.5)$ & $U(-0.5, 1.0)$ & $U(0.40, 0.65)$ \\
    posteriors & $1905.54^{+67.13}_{-76.07}$ & $4.26^{+0.13}_{-0.16}$ & $0.24\pm0.12$ & $0.63^{+0.01}_{-0.02}$ \\
    \hline
    \end{tabular}
    \label{priors_post_ExoREM_nocont}
\end{table}
\endgroup

\begin{figure}[ht!]
\centering
    \includegraphics[scale=0.5]{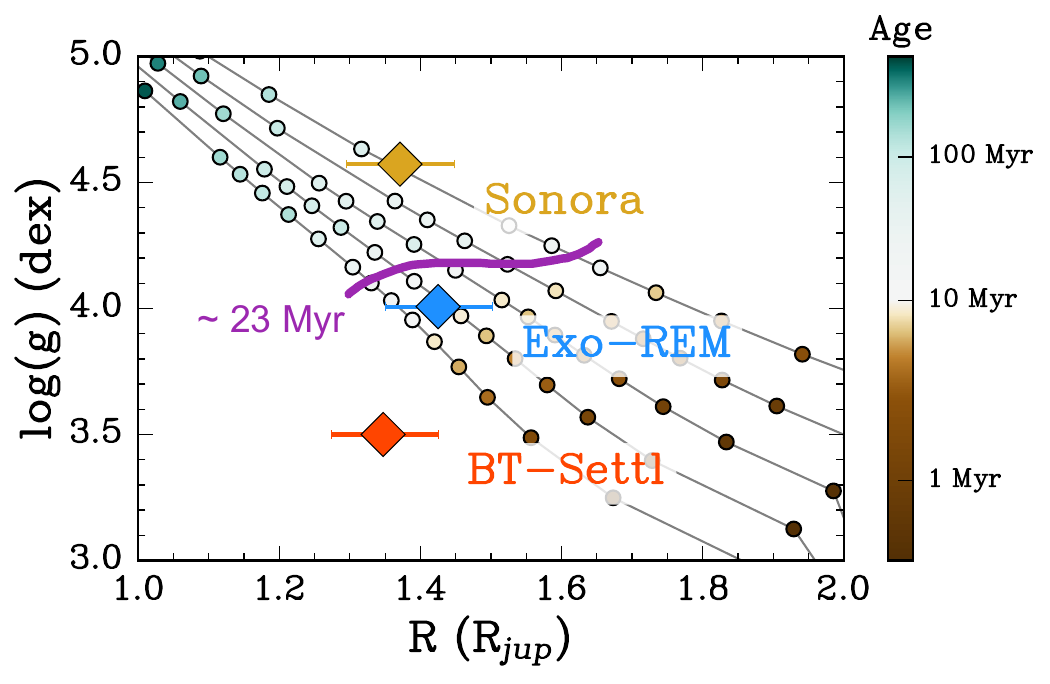}
    \caption{Radius vs. surface gravity as a function of age and effective temperature, based on the BEX Hottest cond03 evolutionary models. The color scale on the right represents the age in Myr. Each colored diamond corresponds to predictions using all datasets (Sect.~\ref{sec4}) from each atmospheric model: blue for Exo-REM, green for ATMO (not visible here), red for BT-Settl, and orange for Sonora. The purple line indicates the approximate formation age of the $\beta$ Pic system.}
    \label{mod_evo_MOSAIC}
\end{figure}

\begin{figure}[ht!]
\centering
    \includegraphics[scale=0.40]{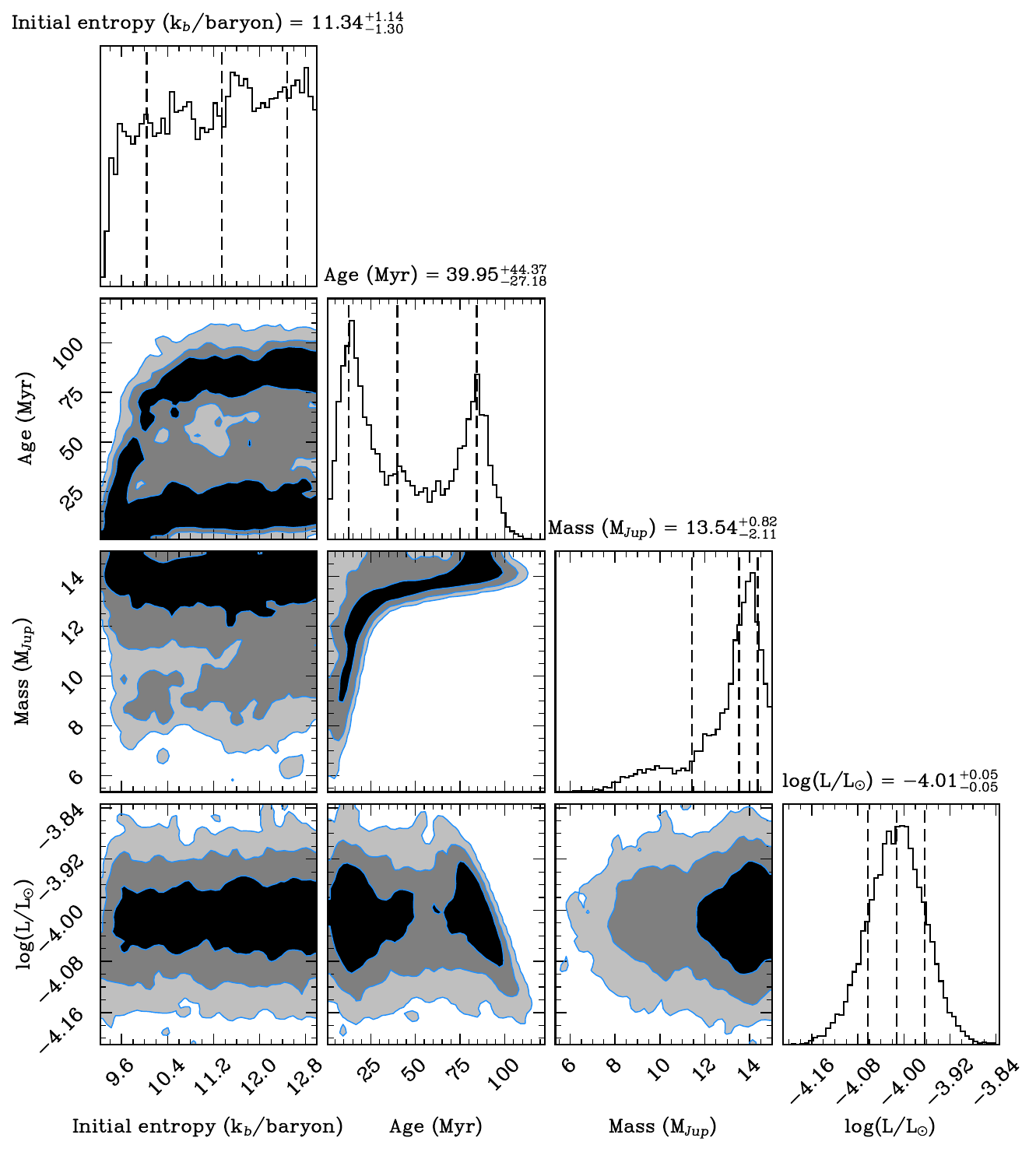}
    \caption{Corner plot illustrating the predictions for initial entropy, age, mass, and bolometric luminosity based on the models from \cite{2012ApJ...745..174S}. The dotted lines indicate the first (16\%), second (50\%), and third (84\%) quantiles. These results were obtained using the method outlined in \cite{2024RNAAS...8..114Z}, employing the nested sampling approach with 4,000 live points.}
    \label{Zhang_corner}
\end{figure}

Conversely, Sect.~\ref{sec4} and \ref{sec5} revealed a significantly more complex picture of $\beta$ Pic b's atmosphere using a self-consistent and multi-wavelength modeling of its spectra. As previously discussed, all atmospheric parameters are extremely sensitive to the considered data, but two main regimes seem to emerge for both \Teff and log(g): a "low-\Teff / high-log(g)" regime (\Teff $\in$ 1500--1600~K / log(g) $\in$ 4.0--5.0~dex) as predicted by Exo-REM and Sonora, and a "high-\Teff / low-log(g)" regime (\Teff~>~1800~K / log(g)~<~3.5~dex) as predicted by ATMO and BT-Settl. This trend is reflected in the Bayesian evidences (see Table \ref{bayesian_ev_MOSAIC_Free}), the scaling parameters $\alpha_i$ (see Fig.~\ref{alpah_evo}) and the residuals (see Figs.~\ref{GRAVITY_MOSAIC_fullspec} and \ref{ALL_MOSAIC_CRIRES_spec}), indicating that ATMO and BT-Settl generally struggle more to reproduce the combined dataset shape. In the "physically informed" setup, Exo-REM retrieved the following values: \Teff~$=1502.74^{+2.32}_{-2.14}$~K, log(g)~$=4.00\pm0.01$~dex, [M/H]~$=0.50\pm0.01$ and C/O~$=0.552^{+0.003}_{-0.002}$. Exo-REM also remains the only model capable of retrieving a dynamically consistent mass of M~$=8.25\pm0.81$~\Mjup. This radius-log(g) pair is in relatively good agreement with hot-start predictions from evolutionary tracks, which estimate the formation age of $\beta$~Pic~b to be $\sim$9.6~Myr, as shown on Fig.~\ref{mod_evo_MOSAIC}. However, this simple comparison suggests a much later formation for the planet, given that its host star formed $23\pm3$~Myr ago  \citep{mamajek_age_2014} which does not align with the typical protoplanetary disk dispersal timescales (2--3~Myr, \citealp{2009AIPC.1158....3M}) if we believe $\beta$ Pic to have been formed by core accretion. A more robust way to estimate the age of $\beta$ Pic b is to use our derived bolometric luminosity (log(L/L$\mathrm{_{\odot}}$)~$=-4.01^{+0.04}_{-0.05}$~dex) alongside dynamical mass predictions (M~$=11.9^{+2.93}_{-3.04}$~\Mjup, \citealp{lacour_mass_2021}) to explore the parameter space provided by evolutionary models. Using the approach outlined in \cite{2024RNAAS...8..114Z} combined with the models from \cite{2012ApJ...745..174S}, we obtained a bimodal posterior distribution for the age, with one peak near 13~Myr and a secondary peak around 84~Myr (see Fig.~\ref{Zhang_corner}). While the first solution is physically plausible, it is important to acknowledge that both the initial entropy and age remain poorly constrained. This behavior, also noted in \cite{2024RNAAS...8..114Z}, mainly stems from the large dynamical mass uncertainties. Detailed orbital monitoring is necessary if we want to push for independent planet age estimations. A more complete discussion on formation scenarios can be found in the next section.

\subsection{Interpreting formation tracers}

\begin{figure}[t!]
\centering
    \includegraphics[scale=0.65]{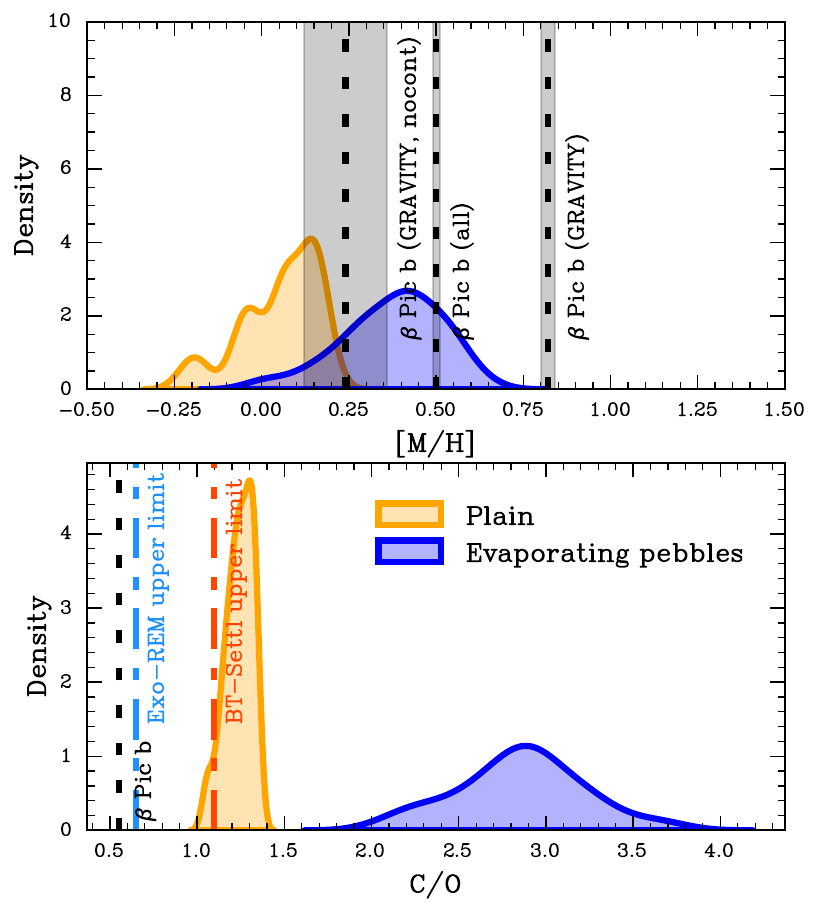}
    \caption{\textit{Top panel:} Exo-REM [M/H] predictions (dashed black lines) based on three inversions: GRAVITY with continuum (0.82~$\pm$~0.02), GRAVITY without continuum (0.24~$\pm$~0.12), and all datasets combined (0.50~$\pm$~0.01). Colored regions show simulated population predictions from \citet{2021A&A...654A..71S} center around the predicted mass (5~\Mjup $\leq$ M $\leq$ 15~\Mjup) and semi-major axis (5~au $\leq$ a $\leq$ 15~au) for $\beta$ Pic b: in yellow are planets form if the disk is plain (i.e., no evaporation and condensation), while in blue are planets form when the full model is included. \textit{Bottom panel:} Similar to the top panel, but for C/O predictions. All Exo-REM predictions cluster near C/O~$\sim$~0.55, so only one dashed line is shown. Dashed blue and red lines mark the maximum C/O values explored by Exo-REM and BT-Settl, respectively.}
    \label{Schneider_pop}
\end{figure}

 Exo-REM and Sonora both predicts super-solar metallicities regardless of the dataset added, apart when using GRAVITY data alone with Sonora (Sect.~\ref{sec4}). The challenges mentioned above regarding the reproduction of the continuum still holds here. We can add that, for the Y to K bands, the treatment of collisionally broadened K I doublet is also essential. In particular, \cite{2020A&A...637A..38P} showed that a simple broadening model has a strong impact on the Y-, J-, H- (SPHERE, GPI), and K-band (GRAVITY) shapes. The unprecedented wavelength coverage (0.9--7~µm) provided by all these observations also enable us to retrieve robust measurements of the bolometric luminosity (log(L/L$\mathrm{_{\odot}}$)~$=-4.01^{+0.04}_{-0.05}$~dex, Exo-REM), which could be used to constrain the maximum metallic content of the atmosphere using simple interior models. We used the methodology presented in \cite{2014MNRAS.437.1378M, 2019A&A...624A..20M} to compare cooling models constraints on planet’s mass and post-formation entropy with the results of population synthesis from \cite{2021A&A...656A..69E, 2021A&A...656A..70E}. Assuming the metallic core is completely diluted (i.e., metallic content = atmospheric metallicity), we get an optimistic upper limit on [M/H] of 0.05. This means that the super-solar metallicity we retrieve ($0.50\pm0.01$) is hard to reach with this setup. Interaction with the disk, as discussed in the next section (top panel of Fig.~\ref{Schneider_pop}) might help reaching a super-solar value.

 We obtain a solar value for the C/O using Exo-REM, which remains robust with the addition of datasets, yielding a final value of C/O~$=0.552^{+0.003}_{-0.002}$. 
 The solar value retrieve by Exo-REM is in tension with \cite{gravity_collaboration_peering_2020} ($0.43\pm0.05$), \cite{landman__2024} ($0.48\pm0.03$) and \cite{worthen_miri_2024} ($0.36^{+0.13}_{-0.05}$) but in good agreement with more recent studies from \cite{kiefer_new_2024} ($0.551\pm0.002$) and \cite{houlle_mathis_2025} ($0.524\pm0.004$). All of these values (except those from \citealp{landman__2024}) were also obtained using Exo-REM, ruling out the simpler cloud prescription in the retrieval as a possible explanation for the observed discrepancies.
\cite{kiefer_new_2024} report C/O values similar to ours and their conclusions on the formation history of $\beta$~Pic~b therefore still hold based on simple assumptions on the disk composition.
We propose here the additional use of planet population  models, assuming the pebble accretion scenario and encompassing the core+envelope formation of the planets as well as drift and evaporations from pebbles \citep{2021A&A...654A..71S, 2021A&A...654A..72S} to further interpret our results.  Figure~\ref{Schneider_pop} compares planets formed with this model that fall within the same mass and semi-major axis range as $\beta$ Pic b with the different Exo-REM predictions presented in this work. While this model could potentially explain the higher metallicity we observe (top panel), the situation is reversed for the C/O ratio. Both the full model (evaporating pebbles) and the reduced model (plain) predict C/O values greater than 1. We can notice as well that our C/O and M/H estimates might be biased because atmospheric models rarely explore values beyond this threshold. This is shown by the blue and red vertical lines in the bottom panel, representing the upper limits of Exo-REM and BT-Settl respectively.
 
\section{Conclusion}\label{sec7}

In this work, we present the first high-resolution spectro-interferometric observation of $\beta$ Pic b using VLTI/GRAVITY at R$_\lambda\sim$~4,000. This new dataset brings insights onto the atmosphere of this object. First, the continuum in the K band is primarily shaped by H$_2$O absorption, allowing us to place strong constraints on the effective temperature and surface gravity of the object. Although statistical metrics rank Sonora as the best overall grid, half of its intrinsic parameters reach the grid's edges. Therefore, we adopted the second-best statistically preferred grid, Exo-REM, as our best-fit model. It yields \Teff$=1607.45^{+4.85}_{-6.20}$~K and log(g)$=4.46^{+0.02}_{-0.04}$. Collision-induced absorption of H$_2$ which is sensitive to atmospheric pressure, also introduce metallicity-related information into this continuum. However, discrepancies between the retrieved metallicities from this spectrum and its continuum-subtracted counterpart suggest an issue in the modeling; either due to inaccuracies in the collision-induced absorption, deviant spectra in the grids, or other systematics that prevents us from placing robust constraints on [M/H].

We found a near-solar C/O value of $0.55\pm0.01$, but we were only able to place a loose upper limit on log($^{12}$C/$^{13}$C), likely due to the combined effects of telluric residuals, wavelength calibration issues, and/or limitations in grid sampling, as the $^{13}$C molecular features are very narrow. We also investigated the combination of previously obtained spectro-photometric observations using our newly developed framework, MOSAIC. This framework allowed us to simultaneously invert multiple observations and explore their impact on the retrieved parameters when adding them iteratively. We found that all atmospheric parameters are significantly influenced by the dataset used, and that the models typically separate into two categories: "low-\Teff / high-log(g)" (\Teff $\in$ 1500--1600~K / log(g) $\in$ 4.0--5.0~dex) as predicted by Exo-REM and Sonora, and a "high-\Teff / low-log(g)" (\Teff~>~1800~K / log(g)~<~3.5~dex) as predicted by ATMO and BT-Settl. The first category appears to be statistically favored and more aligned with previous studies on the target.

Adding high-resolution spectroscopy from CRIRES$_+$ in this framework proved to be challenging (forward models resolution, inversion time, etc.). The shorter coverage of both the K- and M-band observations, as well as their noise, limits their impact on the final fitted parameters. Finally, for the future, new data from the improved GRAVITY$_+$ instrument could better mitigate telluric contamination in the K band and enhance the retrieval of the $^{13}$CO signal. Meanwhile, $\beta$~Pic~b will be a compelling target for short-term monitoring with JWST, allowing us to explore potential variability and gain new insights into its cloud coverage and properties. Additionally, the advanced capabilities of the upcoming Extremely Large Telescope (ELT) by $\sim$2030 will enable more detailed analyses, including the use of Doppler imaging to map temperature and chemical inhomogeneities and clouds.

\begin{acknowledgements}
      This work is heavily supported by a large collaboration of international people: Atmospheric modeling groups represented by B. Charnay (Exo-REM), P. Tremblin (ATMO), D. Homeier, F. Allard (deceased) (BT-Settl), J. Fortney, C. Morley (Sonora). Collaborators who provided us with the data are represented by the GRAVITY collab (GRAVITY high-res.), A. Chomez (SPHERE), F. Kiefer (SINFONI), M. Houllé (MATISSE), R. Landman, L T. Parker (CRIRES+) and K. Worthen (MIRI). This work is based on observations collected at the European Southern Observatory under ESO Large programme ExoGRAVITY: ID 1104.C-0651. S.~Lacour\ acknowledges the support of the French \emph{Agence Nationale de la Recherche} (ANR), under grant ANR-21-CE31-0017 (project ExoVLTI) and grant ANR-23-CE31-0006-01 (MIRAGES). G.-D. Marleau acknowledges support from the European Research Council under the Horizon 2020 Framework Program via the ERC Advanced Grant Origins 83 24 28 (PI: Th. Henning). This work is based [in part] on observations made with the NASA/ESA/CSA James Webb Space Telescope. The data were obtained from the Mikulski Archive for Space Telescopes at the Space Telescope Science Institute, which is operated by the Association of Universities for Research in Astronomy, Inc., under NASA contract NAS 5-03127 for JWST. These observations are associated with GTO program 1294. 
\end{acknowledgements}

%
%

\bibliographystyle{aa}
\bibliography{refs}

\begin{thebibliography}{124}
\expandafter\ifx\csname natexlab\endcsname\relax\def\natexlab#1{#1}\fi

\bibitem[{{Ackerman} \& {Marley}(2001)}]{ackerman_precipitating_2001}
{Ackerman}, A.~S. \& {Marley}, M.~S. 2001, \apj, 556, 872

\bibitem[{{Allard} {et~al.}(2012){Allard}, {Homeier}, \&
  {Freytag}}]{allard_models_2012}
{Allard}, F., {Homeier}, D., \& {Freytag}, B. 2012, Philosophical Transactions
  of the Royal Society of London Series A, 370, 2765

\bibitem[{{Baudino} {et~al.}(2015){Baudino}, {B{\'e}zard}, {Boccaletti},
  {Bonnefoy}, {Lagrange}, \& {Galicher}}]{baudino_interpreting_2015}
{Baudino}, J.~L., {B{\'e}zard}, B., {Boccaletti}, A., {et~al.} 2015, \aap, 582,
  A83

\bibitem[{{Benneke} \& {Seager}(2013)}]{benneke_how_2013}
{Benneke}, B. \& {Seager}, S. 2013, \apj, 778, 153

\bibitem[{{Beuzit} {et~al.}(2019){Beuzit}, {Vigan}, {Mouillet}, {Dohlen},
  {Gratton}, {Boccaletti}, {Sauvage}, {Schmid}, {Langlois}, {Petit},
  {Baruffolo}, {Feldt}, {Milli}, {Wahhaj}, {Abe}, {Anselmi}, {Antichi},
  {Barette}, {Baudrand}, {Baudoz}, {Bazzon}, {Bernardi}, {Blanchard}, {Brast},
  {Bruno}, {Buey}, {Carbillet}, {Carle}, {Cascone}, {Chapron}, {Charton},
  {Chauvin}, {Claudi}, {Costille}, {De Caprio}, {de Boer}, {Delboulb{\'e}},
  {Desidera}, {Dominik}, {Downing}, {Dupuis}, {Fabron}, {Fantinel}, {Farisato},
  {Feautrier}, {Fedrigo}, {Fusco}, {Gigan}, {Ginski}, {Girard}, {Giro},
  {Gisler}, {Gluck}, {Gry}, {Henning}, {Hubin}, {Hugot}, {Incorvaia}, {Jaquet},
  {Kasper}, {Lagadec}, {Lagrange}, {Le Coroller}, {Le Mignant}, {Le Ruyet},
  {Lessio}, {Lizon}, {Llored}, {Lundin}, {Madec}, {Magnard}, {Marteaud},
  {Martinez}, {Maurel}, {M{\'e}nard}, {Mesa}, {M{\"o}ller-Nilsson}, {Moulin},
  {Moutou}, {Orign{\'e}}, {Parisot}, {Pavlov}, {Perret}, {Pragt}, {Puget},
  {Rabou}, {Ramos}, {Reess}, {Rigal}, {Rochat}, {Roelfsema}, {Rousset}, {Roux},
  {Saisse}, {Salasnich}, {Santambrogio}, {Scuderi}, {Segransan}, {Sevin},
  {Siebenmorgen}, {Soenke}, {Stadler}, {Suarez}, {Tiph{\`e}ne}, {Turatto},
  {Udry}, {Vakili}, {Waters}, {Weber}, {Wildi}, {Zins}, \&
  {Zurlo}}]{beuzit_sphere_2019}
{Beuzit}, J.~L., {Vigan}, A., {Mouillet}, D., {et~al.} 2019, \aap, 631, A155

\bibitem[{{Blunt} {et~al.}(2023){Blunt}, {Balmer}, {Wang}, {Lacour}, {Petrus},
  {Bourdarot}, {Kammerer}, {Pourr{\'e}}, {Rickman}, {Shangguan},
  {Winterhalder}, {Abuter}, {Amorim}, {Asensio-Torres}, {Benisty}, {Berger},
  {Beust}, {Boccaletti}, {Bohn}, {Bonnefoy}, {Bonnet}, {Brandner},
  {Cantalloube}, {Caselli}, {Charnay}, {Chauvin}, {Chavez}, {Choquet},
  {Christiaens}, {Cl{\'e}net}, {Du Foresto}, {Cridland}, {Dembet}, {Drescher},
  {Duvert}, {Eckart}, {Eisenhauer}, {Feuchtgruber}, {Garcia}, {Garcia Lopez},
  {Gendron}, {Genzel}, {Gillessen}, {Girard}, {Haubois}, {Hei{\ss}el},
  {Henning}, {Hinkley}, {Hippler}, {Horrobin}, {Houll{\'e}}, {Hubert}, {Jocou},
  {Keppler}, {Kervella}, {Kreidberg}, {Lagrange}, {Lapeyr{\`e}re}, {Le
  Bouquin}, {L{\'e}na}, {Lutz}, {Maire}, {Mang}, {Marleau}, {M{\'e}rand},
  {Molli{\`e}re}, {Monnier}, {Mordasini}, {Mouillet}, {Nasedkin}, {Nowak},
  {Ott}, {Otten}, {Paladini}, {Paumard}, {Perraut}, {Perrin}, {Pfuhl}, {Pueyo},
  {Rameau}, {Rodet}, {Rustamkulov}, {Shimizu}, {Sing}, {Stolker},
  {Straubmeier}, {Sturm}, {Tacconi}, {van Dishoeck}, {Vigan}, {Vincent},
  {Ward-Duong}, {Widmann}, {Wieprecht}, {Wiezorrek}, {Woillez}, {Yazici},
  {Young}, \& {Exogravity Collaboration}}]{blunt_first_2023}
{Blunt}, S., {Balmer}, W.~O., {Wang}, J.~J., {et~al.} 2023, \aj, 166, 257

\bibitem[{{Boley} {et~al.}(2011){Boley}, {Helled}, \&
  {Payne}}]{boley_heavy_2011}
{Boley}, A.~C., {Helled}, R., \& {Payne}, M.~J. 2011, \apj, 735, 30

\bibitem[{{Bonnefoy} {et~al.}(2013){Bonnefoy}, {Boccaletti}, {Lagrange},
  {Allard}, {Mordasini}, {Beust}, {Chauvin}, {Girard}, {Homeier}, {Apai},
  {Lacour}, \& {Rouan}}]{bonnefoy_near-infrared_2013}
{Bonnefoy}, M., {Boccaletti}, A., {Lagrange}, A.~M., {et~al.} 2013, \aap, 555,
  A107

\bibitem[{{Bonnefoy} {et~al.}(2011){Bonnefoy}, {Lagrange}, {Boccaletti},
  {Chauvin}, {Apai}, {Allard}, {Ehrenreich}, {Girard}, {Mouillet}, {Rouan},
  {Gratadour}, \& {Kasper}}]{bonnefoy_high_2011}
{Bonnefoy}, M., {Lagrange}, A.~M., {Boccaletti}, A., {et~al.} 2011, \aap, 528,
  L15

\bibitem[{{Bonnefoy} {et~al.}(2014){Bonnefoy}, {Marleau}, {Galicher}, {Beust},
  {Lagrange}, {Baudino}, {Chauvin}, {Borgniet}, {Meunier}, {Rameau},
  {Boccaletti}, {Cumming}, {Helling}, {Homeier}, {Allard}, \&
  {Delorme}}]{bonnefoy_physical_2014}
{Bonnefoy}, M., {Marleau}, G.~D., {Galicher}, R., {et~al.} 2014, \aap, 567, L9

\bibitem[{{Bonnefoy} {et~al.}(2016){Bonnefoy}, {Zurlo}, {Baudino}, {Lucas},
  {Mesa}, {Maire}, {Vigan}, {Galicher}, {Homeier}, {Marocco}, {Gratton},
  {Chauvin}, {Allard}, {Desidera}, {Kasper}, {Moutou}, {Lagrange}, {Antichi},
  {Baruffolo}, {Baudrand}, {Beuzit}, {Boccaletti}, {Cantalloube}, {Carbillet},
  {Charton}, {Claudi}, {Costille}, {Dohlen}, {Dominik}, {Fantinel},
  {Feautrier}, {Feldt}, {Fusco}, {Gigan}, {Girard}, {Gluck}, {Gry}, {Henning},
  {Janson}, {Langlois}, {Madec}, {Magnard}, {Maurel}, {Mawet}, {Meyer},
  {Milli}, {Moeller-Nilsson}, {Mouillet}, {Pavlov}, {Perret}, {Pujet}, {Quanz},
  {Rochat}, {Rousset}, {Roux}, {Salasnich}, {Salter}, {Sauvage}, {Schmid},
  {Sevin}, {Soenke}, {Stadler}, {Turatto}, {Udry}, {Vakili}, {Wahhaj}, \&
  {Wildi}}]{bonnefoy_first_2016}
{Bonnefoy}, M., {Zurlo}, A., {Baudino}, J.~L., {et~al.} 2016, \aap, 587, A58

\bibitem[{{Brandt} {et~al.}(2021){Brandt}, {Brandt}, {Dupuy}, {Li}, \&
  {Michalik}}]{brandt_precise_2021}
{Brandt}, G.~M., {Brandt}, T.~D., {Dupuy}, T.~J., {Li}, Y., \& {Michalik}, D.
  2021, \aj, 161, 179

\bibitem[{{Brogi} \& {Line}(2019)}]{brogi_retrieving_2019}
{Brogi}, M. \& {Line}, M.~R. 2019, \aj, 157, 114

\bibitem[{{Charnay} {et~al.}(2018){Charnay}, {B{\'e}zard}, {Baudino},
  {Bonnefoy}, {Boccaletti}, \& {Galicher}}]{charnay_self-consistent_2018}
{Charnay}, B., {B{\'e}zard}, B., {Baudino}, J.~L., {et~al.} 2018, \apj, 854,
  172

\bibitem[{{Chauvin} {et~al.}(2017){Chauvin}, {Desidera}, {Lagrange}, {Vigan},
  {Gratton}, {Langlois}, {Bonnefoy}, {Beuzit}, {Feldt}, {Mouillet}, {Meyer},
  {Cheetham}, {Biller}, {Boccaletti}, {D'Orazi}, {Galicher}, {Hagelberg},
  {Maire}, {Mesa}, {Olofsson}, {Samland}, {Schmidt}, {Sissa}, {Bonavita},
  {Charnay}, {Cudel}, {Daemgen}, {Delorme}, {Janin-Potiron}, {Janson},
  {Keppler}, {Le Coroller}, {Ligi}, {Marleau}, {Messina}, {Molli{\`e}re},
  {Mordasini}, {M{\"u}ller}, {Peretti}, {Perrot}, {Rodet}, {Rouan}, {Zurlo},
  {Dominik}, {Henning}, {Menard}, {Schmid}, {Turatto}, {Udry}, {Vakili}, {Abe},
  {Antichi}, {Baruffolo}, {Baudoz}, {Baudrand}, {Blanchard}, {Bazzon}, {Buey},
  {Carbillet}, {Carle}, {Charton}, {Cascone}, {Claudi}, {Costille}, {Deboulbe},
  {De Caprio}, {Dohlen}, {Fantinel}, {Feautrier}, {Fusco}, {Gigan}, {Giro},
  {Gisler}, {Gluck}, {Hubin}, {Hugot}, {Jaquet}, {Kasper}, {Madec}, {Magnard},
  {Martinez}, {Maurel}, {Le Mignant}, {M{\"o}ller-Nilsson}, {Llored}, {Moulin},
  {Orign{\'e}}, {Pavlov}, {Perret}, {Petit}, {Pragt}, {Puget}, {Rabou},
  {Ramos}, {Rigal}, {Rochat}, {Roelfsema}, {Rousset}, {Roux}, {Salasnich},
  {Sauvage}, {Sevin}, {Soenke}, {Stadler}, {Suarez}, {Weber}, {Wildi},
  {Antoniucci}, {Augereau}, {Baudino}, {Brandner}, {Engler}, {Girard}, {Gry},
  {Kral}, {Kopytova}, {Lagadec}, {Milli}, {Moutou}, {Schlieder},
  {Szul{\'a}gyi}, {Thalmann}, \& {Wahhaj}}]{2017A&A...605L...9C}
{Chauvin}, G., {Desidera}, S., {Lagrange}, A.~M., {et~al.} 2017, \aap, 605, L9

\bibitem[{{Chauvin} {et~al.}(2004){Chauvin}, {Lagrange}, {Dumas}, {Zuckerman},
  {Mouillet}, {Song}, {Beuzit}, \& {Lowrance}}]{2004A&A...425L..29C}
{Chauvin}, G., {Lagrange}, A.~M., {Dumas}, C., {et~al.} 2004, \aap, 425, L29

\bibitem[{{Chilcote} {et~al.}(2015){Chilcote}, {Barman}, {Fitzgerald},
  {Graham}, {Larkin}, {Macintosh}, {Bauman}, {Burrows}, {Cardwell}, {De Rosa},
  {Dillon}, {Doyon}, {Dunn}, {Erikson}, {Gavel}, {Goodsell}, {Hartung},
  {Hibon}, {Ingraham}, {Kalas}, {Konopacky}, {Maire}, {Marchis}, {Marley},
  {Marois}, {Millar-Blanchaer}, {Morzinski}, {Norton}, {Oppenheimer}, {Palmer},
  {Patience}, {Perrin}, {Poyneer}, {Pueyo}, {Rantakyr{\"o}}, {Sadakuni},
  {Saddlemyer}, {Savransky}, {Serio}, {Sivaramakrishnan}, {Song}, {Soummer},
  {Thomas}, {Wallace}, {Wiktorowicz}, \& {Wolff}}]{chilcote_first_2015}
{Chilcote}, J., {Barman}, T., {Fitzgerald}, M.~P., {et~al.} 2015, \apjl, 798,
  L3

\bibitem[{{Chilcote} {et~al.}(2017){Chilcote}, {Pueyo}, {De Rosa}, {Vargas},
  {Macintosh}, {Bailey}, {Barman}, {Bauman}, {Bruzzone}, {Bulger}, {Burrows},
  {Cardwell}, {Chen}, {Cotten}, {Dillon}, {Doyon}, {Draper}, {Duch{\^e}ne},
  {Dunn}, {Erikson}, {Fitzgerald}, {Follette}, {Gavel}, {Goodsell}, {Graham},
  {Greenbaum}, {Hartung}, {Hibon}, {Hung}, {Ingraham}, {Kalas}, {Konopacky},
  {Larkin}, {Maire}, {Marchis}, {Marley}, {Marois}, {Metchev},
  {Millar-Blanchaer}, {Morzinski}, {Nielsen}, {Norton}, {Oppenheimer},
  {Palmer}, {Patience}, {Perrin}, {Poyneer}, {Rajan}, {Rameau},
  {Rantakyr{\"o}}, {Sadakuni}, {Saddlemyer}, {Savransky}, {Schneider}, {Serio},
  {Sivaramakrishnan}, {Song}, {Soummer}, {Thomas}, {Wallace}, {Wang},
  {Ward-Duong}, {Wiktorowicz}, \& {Wolff}}]{chilcote_1-24_2017}
{Chilcote}, J., {Pueyo}, L., {De Rosa}, R.~J., {et~al.} 2017, \aj, 153, 182

\bibitem[{{Chomez} {et~al.}(2023){Chomez}, {Squicciarini}, {Lagrange},
  {Delorme}, {Viswanath}, {Janson}, {Flasseur}, {Chauvin}, {Langlois},
  {Rubini}, {Bergeon}, {Albert}, {Bonnefoy}, {Desidera}, {Engler}, {Gratton},
  {Henning}, {Mamajek}, {Marleau}, {Meyer}, {Reffert}, {Ringqvist}, \&
  {Samland}}]{2023A&A...676L..10C}
{Chomez}, A., {Squicciarini}, V., {Lagrange}, A.~M., {et~al.} 2023, \aap, 676,
  L10

\bibitem[{{Costes} {et~al.}(2024){Costes}, {Xuan}, {Vigan}, {Wang}, {D'Orazi},
  {Molli{\`e}re}, {Baker}, {Bartos}, {Blake}, {Calvin}, {Cetre}, {Delorme},
  {Doppmann}, {Echeveri}, {Finnerty}, {Fitzgerald}, {Hsu}, {Jovanovic},
  {Lopez}, {Mawet}, {Morris}, {Pezzato}, {Phillips}, {Ruffio}, {Sappey},
  {Schneeberger}, {Schofield}, {Skemer}, {Wallace}, \&
  {Wang}}]{2024A&A...686A.294C}
{Costes}, J.~C., {Xuan}, J.~W., {Vigan}, A., {et~al.} 2024, \aap, 686, A294

\bibitem[{{Currie} {et~al.}(2018){Currie}, {Brandt}, {Uyama}, {Nielsen},
  {Blunt}, {Guyon}, {Tamura}, {Marois}, {Mede}, {Kuzuhara}, {Groff},
  {Jovanovic}, {Kasdin}, {Lozi}, {Hodapp}, {Chilcote}, {Carson}, {Martinache},
  {Goebel}, {Grady}, {McElwain}, {Akiyama}, {Asensio-Torres}, {Hayashi},
  {Janson}, {Knapp}, {Kwon}, {Nishikawa}, {Oh}, {Schlieder}, {Serabyn},
  {Sitko}, \& {Skaf}}]{2018AJ....156..291C}
{Currie}, T., {Brandt}, T.~D., {Uyama}, T., {et~al.} 2018, \aj, 156, 291

\bibitem[{{Currie} {et~al.}(2013){Currie}, {Burrows}, {Madhusudhan},
  {Fukagawa}, {Girard}, {Dawson}, {Murray-Clay}, {Kenyon}, {Kuchner},
  {Matsumura}, {Jayawardhana}, {Chambers}, \& {Bromley}}]{currie_combined_2013}
{Currie}, T., {Burrows}, A., {Madhusudhan}, N., {et~al.} 2013, \apj, 776, 15

\bibitem[{{Cushing} {et~al.}(2008){Cushing}, {Marley}, {Saumon}, {Kelly},
  {Vacca}, {Rayner}, {Freedman}, {Lodders}, \&
  {Roellig}}]{cushing_atmospheric_2008}
{Cushing}, M.~C., {Marley}, M.~S., {Saumon}, D., {et~al.} 2008, \apj, 678, 1372

\bibitem[{{Czekala} {et~al.}(2015){Czekala}, {Andrews}, {Mandel}, {Hogg}, \&
  {Green}}]{2015ApJ...812..128C}
{Czekala}, I., {Andrews}, S.~M., {Mandel}, K.~S., {Hogg}, D.~W., \& {Green},
  G.~M. 2015, \apj, 812, 128

\bibitem[{{de Regt} {et~al.}(2024){de Regt}, {Gandhi}, {Snellen}, {Zhang},
  {Ginski}, {Gonz{\'a}lez Picos}, {Kesseli}, {Landman}, {Molli{\`e}re},
  {Nasedkin}, {S{\'a}nchez-L{\'o}pez}, \& {Stolker}}]{De-Regt_2024_ESO}
{de Regt}, S., {Gandhi}, S., {Snellen}, I.~A.~G., {et~al.} 2024, \aap, 688,
  A116

\bibitem[{{de Regt} {et~al.}(2025){de Regt}, {Snellen}, {Allard}, {Gonz{\'a}lez
  Picos}, {Gandhi}, {Grasser}, {Landman}, {Molli{\`e}re}, {Nasedkin},
  {Stolker}, \& {Zhang}}]{De-Regt_2025_ESO}
{de Regt}, S., {Snellen}, I.~A.~G., {Allard}, N.~F., {et~al.} 2025, \aap, 696,
  A225

\bibitem[{{De Rosa} {et~al.}(2023){De Rosa}, {Nielsen}, {Wahhaj}, {Ruffio},
  {Kalas}, {Peck}, {Hirsch}, \& {Roberson}}]{2023A&A...672A..94D}
{De Rosa}, R.~J., {Nielsen}, E.~L., {Wahhaj}, Z., {et~al.} 2023, \aap, 672, A94

\bibitem[{{Delorme} {et~al.}(2021){Delorme}, {Jovanovic}, {Echeverri}, {Mawet},
  {Kent Wallace}, {Bartos}, {Cetre}, {Wizinowich}, {Ragland}, {Lilley},
  {Wetherell}, {Doppmann}, {Wang}, {Morris}, {Ruffio}, {Martin}, {Fitzgerald},
  {Ruane}, {Schofield}, {Suominen}, {Calvin}, {Wang}, {Magnone}, {Johnson},
  {Sohn}, {L{\'o}pez}, {Bond}, {Pezzato}, {Sayson}, {Chun}, \&
  {Skemer}}]{2021JATIS...7c5006D}
{Delorme}, J.-R., {Jovanovic}, N., {Echeverri}, D., {et~al.} 2021, Journal of
  Astronomical Telescopes, Instruments, and Systems, 7, 035006

\bibitem[{{Denis} {et~al.}(2025){Denis}, {Vigan}, {Costes}, {Chauvin},
  {Radcliffe}, {Ravet}, {Balmer}, {Palma-Bifani}, {Petrus}, {Parmentier},
  {Martos}, {Simonnin}, {Bonnefoy}, {Cadet}, {Forveille}, {Charnay}, {Kiefer},
  {Lagrange}, {Chiavassa}, {Stolker}, {Lavail}, {Godoy}, {Janson}, {Pourcelot},
  {Delorme}, {Rickman}, {Cont}, {Reiners}, {De Rosa}, {Anwand-Heerwart},
  {Charles}, {Costille}, {El Morsy}, {Garcia}, {Houll{\'e}}, {Lopez}, {Murray},
  {Muslimov}, {Otten}, {Paufique}, {Phillips}, {Seemann}, {Viret}, \&
  {Zins}}]{Denis2025}
{Denis}, A., {Vigan}, A., {Costes}, J., {et~al.} 2025, \aap, 696, A6

\bibitem[{{Doelman} {et~al.}(2022){Doelman}, {Stone}, {Briesemeister},
  {Skemer}, {Barman}, {Brock}, {Hinz}, {Bohn}, {Kenworthy}, {Haffert}, {Snik},
  {Ertel}, {Leisenring}, {Woodward}, \& {Skrutskie}}]{2022AJ....163..217D}
{Doelman}, D.~S., {Stone}, J.~M., {Briesemeister}, Z.~W., {et~al.} 2022, \aj,
  163, 217

\bibitem[{{Emsenhuber} {et~al.}(2021{\natexlab{a}}){Emsenhuber}, {Mordasini},
  {Burn}, {Alibert}, {Benz}, \& {Asphaug}}]{2021A&A...656A..69E}
{Emsenhuber}, A., {Mordasini}, C., {Burn}, R., {et~al.} 2021{\natexlab{a}},
  \aap, 656, A69

\bibitem[{{Emsenhuber} {et~al.}(2021{\natexlab{b}}){Emsenhuber}, {Mordasini},
  {Burn}, {Alibert}, {Benz}, \& {Asphaug}}]{2021A&A...656A..70E}
{Emsenhuber}, A., {Mordasini}, C., {Burn}, R., {et~al.} 2021{\natexlab{b}},
  \aap, 656, A70

\bibitem[{{Gaia Collaboration} {et~al.}(2023){Gaia Collaboration}, {Vallenari},
  {Brown}, {Prusti}, {de Bruijne}, {Arenou}, {Babusiaux}, {Biermann},
  {Creevey}, {Ducourant}, {Evans}, {Eyer}, {Guerra}, {Hutton}, {Jordi},
  {Klioner}, {Lammers}, {Lindegren}, {Luri}, {Mignard}, {Panem}, {Pourbaix},
  {Randich}, {Sartoretti}, {Soubiran}, {Tanga}, {Walton}, {Bailer-Jones},
  {Bastian}, {Drimmel}, {Jansen}, {Katz}, {Lattanzi}, {van Leeuwen}, {Bakker},
  {Cacciari}, {Casta{\~n}eda}, {De Angeli}, {Fabricius}, {Fouesneau},
  {Fr{\'e}mat}, {Galluccio}, {Guerrier}, {Heiter}, {Masana}, {Messineo},
  {Mowlavi}, {Nicolas}, {Nienartowicz}, {Pailler}, {Panuzzo}, {Riclet}, {Roux},
  {Seabroke}, {Sordo}, {Th{\'e}venin}, {Gracia-Abril}, {Portell}, {Teyssier},
  {Altmann}, {Andrae}, {Audard}, {Bellas-Velidis}, {Benson}, {Berthier},
  {Blomme}, {Burgess}, {Busonero}, {Busso}, {C{\'a}novas}, {Carry}, {Cellino},
  {Cheek}, {Clementini}, {Damerdji}, {Davidson}, {de Teodoro}, {Nu{\~n}ez
  Campos}, {Delchambre}, {Dell'Oro}, {Esquej}, {Fern{\'a}ndez-Hern{\'a}ndez},
  {Fraile}, {Garabato}, {Garc{\'\i}a-Lario}, {Gosset}, {Haigron}, {Halbwachs},
  {Hambly}, {Harrison}, {Hern{\'a}ndez}, {Hestroffer}, {Hodgkin}, {Holl},
  {Jan{\ss}en}, {Jevardat de Fombelle}, {Jordan}, {Krone-Martins}, {Lanzafame},
  {L{\"o}ffler}, {Marchal}, {Marrese}, {Moitinho}, {Muinonen}, {Osborne},
  {Pancino}, {Pauwels}, {Recio-Blanco}, {Reyl{\'e}}, {Riello}, {Rimoldini},
  {Roegiers}, {Rybizki}, {Sarro}, {Siopis}, {Smith}, {Sozzetti}, {Utrilla},
  {van Leeuwen}, {Abbas}, {{\'A}brah{\'a}m}, {Abreu Aramburu}, {Aerts},
  {Aguado}, {Ajaj}, {Aldea-Montero}, {Altavilla}, {{\'A}lvarez}, {Alves},
  {Anders}, {Anderson}, {Anglada Varela}, {Antoja}, {Baines}, {Baker},
  {Balaguer-N{\'u}{\~n}ez}, {Balbinot}, {Balog}, {Barache}, {Barbato},
  {Barros}, {Barstow}, {Bartolom{\'e}}, {Bassilana}, {Bauchet}, {Becciani},
  {Bellazzini}, {Berihuete}, {Bernet}, {Bertone}, {Bianchi}, {Binnenfeld},
  {Blanco-Cuaresma}, {Blazere}, {Boch}, {Bombrun}, {Bossini}, {Bouquillon},
  {Bragaglia}, {Bramante}, {Breedt}, {Bressan}, {Brouillet}, {Brugaletta},
  {Bucciarelli}, {Burlacu}, {Butkevich}, {Buzzi}, {Caffau}, {Cancelliere},
  {Cantat-Gaudin}, {Carballo}, {Carlucci}, {Carnerero}, {Carrasco},
  {Casamiquela}, {Castellani}, {Castro-Ginard}, {Chaoul}, {Charlot}, {Chemin},
  {Chiaramida}, {Chiavassa}, {Chornay}, {Comoretto}, {Contursi}, {Cooper},
  {Cornez}, {Cowell}, {Crifo}, {Cropper}, {Crosta}, {Crowley}, {Dafonte},
  {Dapergolas}, {David}, {David}, {de Laverny}, {De Luise}, \& {De
  March}}]{gaia_collaboration_gaia_2023}
{Gaia Collaboration}, {Vallenari}, A., {Brown}, A.~G.~A., {et~al.} 2023, \aap,
  674, A1

\bibitem[{{Gandhi} {et~al.}(2023){Gandhi}, {de Regt}, {Snellen}, {Zhang},
  {Rugers}, {van Leur}, \& {Bosschaart}}]{2023ApJ...957L..36G}
{Gandhi}, S., {de Regt}, S., {Snellen}, I., {et~al.} 2023, \apjl, 957, L36

\bibitem[{{Gautier} \& {Owen}(1989)}]{1989oeps.book..487G}
{Gautier}, D. \& {Owen}, T. 1989, in Origin and Evolution of Planetary and
  Satellite Atmospheres, ed. S.~K. {Atreya}, J.~B. {Pollack}, \& M.~S.
  {Matthews}, 487--512

\bibitem[{{Gontcharov}(2006)}]{gontcharov_pulkovo_2006}
{Gontcharov}, G.~A. 2006, Astronomy Letters, 32, 759

\bibitem[{{GRAVITY Collaboration} {et~al.}(2017){GRAVITY Collaboration},
  {Abuter}, {Accardo}, {Amorim}, {Anugu}, {{\'A}vila}, {Azouaoui}, {Benisty},
  {Berger}, {Blind}, {Bonnet}, {Bourget}, {Brandner}, {Brast}, {Buron},
  {Burtscher}, {Cassaing}, {Chapron}, {Choquet}, {Cl{\'e}net}, {Collin},
  {Coud{\'e} Du Foresto}, {de Wit}, {de Zeeuw}, {Deen},
  {Delplancke-Str{\"o}bele}, {Dembet}, {Derie}, {Dexter}, {Duvert}, {Ebert},
  {Eckart}, {Eisenhauer}, {Esselborn}, {F{\'e}dou}, {Finger}, {Garcia}, {Garcia
  Dabo}, {Garcia Lopez}, {Gendron}, {Genzel}, {Gillessen}, {Gonte}, {Gordo},
  {Grould}, {Gr{\"o}zinger}, {Guieu}, {Haguenauer}, {Hans}, {Haubois}, {Haug},
  {Haussmann}, {Henning}, {Hippler}, {Horrobin}, {Huber}, {Hubert}, {Hubin},
  {Hummel}, {Jakob}, {Janssen}, {Jochum}, {Jocou}, {Kaufer}, {Kellner},
  {Kendrew}, {Kern}, {Kervella}, {Kiekebusch}, {Klein}, {Kok}, {Kolb}, {Kulas},
  {Lacour}, {Lapeyr{\`e}re}, {Lazareff}, {Le Bouquin}, {L{\`e}na}, {Lenzen},
  {L{\'e}v{\^e}que}, {Lippa}, {Magnard}, {Mehrgan}, {Mellein}, {M{\'e}rand},
  {Moreno-Ventas}, {Moulin}, {M{\"u}ller}, {M{\"u}ller}, {Neumann}, {Oberti},
  {Ott}, {Pallanca}, {Panduro}, {Pasquini}, {Paumard}, {Percheron}, {Perraut},
  {Perrin}, {Pfl{\"u}ger}, {Pfuhl}, {Phan Duc}, {Plewa}, {Popovic}, {Rabien},
  {Ram{\'\i}rez}, {Ramos}, {Rau}, {Riquelme}, {Rohloff}, {Rousset},
  {Sanchez-Bermudez}, {Scheithauer}, {Sch{\"o}ller}, {Schuhler}, {Spyromilio},
  {Straubmeier}, {Sturm}, {Suarez}, {Tristram}, {Ventura}, {Vincent},
  {Waisberg}, {Wank}, {Weber}, {Wieprecht}, {Wiest}, {Wiezorrek}, {Wittkowski},
  {Woillez}, {Wolff}, {Yazici}, {Ziegler}, \& {Zins}}]{2017A&A...602A..94G}
{GRAVITY Collaboration}, {Abuter}, R., {Accardo}, M., {et~al.} 2017, \aap, 602,
  A94

\bibitem[{{GRAVITY Collaboration} {et~al.}(2019){GRAVITY Collaboration},
  {Lacour}, {Nowak}, {Wang}, {Pfuhl}, {Eisenhauer}, {Abuter}, {Amorim},
  {Anugu}, {Benisty}, {Berger}, {Beust}, {Blind}, {Bonnefoy}, {Bonnet},
  {Bourget}, {Brandner}, {Buron}, {Collin}, {Charnay}, {Chapron}, {Cl{\'e}net},
  {Coud{\'e} Du Foresto}, {de Zeeuw}, {Deen}, {Dembet}, {Dexter}, {Duvert},
  {Eckart}, {F{\"o}rster Schreiber}, {F{\'e}dou}, {Garcia}, {Garcia Lopez},
  {Gao}, {Gendron}, {Genzel}, {Gillessen}, {Gordo}, {Greenbaum}, {Habibi},
  {Haubois}, {Hau{\ss}mann}, {Henning}, {Hippler}, {Horrobin}, {Hubert},
  {Jimenez Rosales}, {Jocou}, {Kendrew}, {Kervella}, {Kolb}, {Lagrange},
  {Lapeyr{\`e}re}, {Le Bouquin}, {L{\'e}na}, {Lippa}, {Lenzen}, {Maire},
  {Molli{\`e}re}, {Ott}, {Paumard}, {Perraut}, {Perrin}, {Pueyo}, {Rabien},
  {Ram{\'\i}rez}, {Rau}, {Rodr{\'\i}guez-Coira}, {Rousset}, {Sanchez-Bermudez},
  {Scheithauer}, {Schuhler}, {Straub}, {Straubmeier}, {Sturm}, {Tacconi},
  {Vincent}, {van Dishoeck}, {von Fellenberg}, {Wank}, {Waisberg}, {Widmann},
  {Wieprecht}, {Wiest}, {Wiezorrek}, {Woillez}, {Yazici}, {Ziegler}, \&
  {Zins}}]{2019A&A...623L..11G}
{GRAVITY Collaboration}, {Lacour}, S., {Nowak}, M., {et~al.} 2019, \aap, 623,
  L11

\bibitem[{{GRAVITY Collaboration} {et~al.}(2020){GRAVITY Collaboration},
  {Nowak}, {Lacour}, {Molli{\`e}re}, {Wang}, {Charnay}, {van Dishoeck},
  {Abuter}, {Amorim}, {Berger}, {Beust}, {Bonnefoy}, {Bonnet}, {Brandner},
  {Buron}, {Cantalloube}, {Collin}, {Chapron}, {Cl{\'e}net}, {Coud{\'e} Du
  Foresto}, {de Zeeuw}, {Dembet}, {Dexter}, {Duvert}, {Eckart}, {Eisenhauer},
  {F{\"o}rster Schreiber}, {F{\'e}dou}, {Garcia Lopez}, {Gao}, {Gendron},
  {Genzel}, {Gillessen}, {Hau{\ss}mann}, {Henning}, {Hippler}, {Hubert},
  {Jocou}, {Kervella}, {Lagrange}, {Lapeyr{\`e}re}, {Le Bouquin}, {L{\'e}na},
  {Maire}, {Ott}, {Paumard}, {Paladini}, {Perraut}, {Perrin}, {Pueyo}, {Pfuhl},
  {Rabien}, {Rau}, {Rodr{\'\i}guez-Coira}, {Rousset}, {Scheithauer},
  {Shangguan}, {Straub}, {Straubmeier}, {Sturm}, {Tacconi}, {Vincent},
  {Widmann}, {Wieprecht}, {Wiezorrek}, {Woillez}, {Yazici}, \&
  {Ziegler}}]{gravity_collaboration_peering_2020}
{GRAVITY Collaboration}, {Nowak}, M., {Lacour}, S., {et~al.} 2020, \aap, 633,
  A110

\bibitem[{{Gray} {et~al.}(2006){Gray}, {Corbally}, {Garrison}, {McFadden},
  {Bubar}, {McGahee}, {O'Donoghue}, \& {Knox}}]{gray_2006_contribution}
{Gray}, R.~O., {Corbally}, C.~J., {Garrison}, R.~F., {et~al.} 2006, \aj, 132,
  161

\bibitem[{{Greenbaum} {et~al.}(2018){Greenbaum}, {Pueyo}, {Ruffio}, {Wang}, {De
  Rosa}, {Aguilar}, {Rameau}, {Barman}, {Marois}, {Marley}, {Konopacky},
  {Rajan}, {Macintosh}, {Ansdell}, {Arriaga}, {Bailey}, {Bulger}, {Burrows},
  {Chilcote}, {Cotten}, {Doyon}, {Duch{\^e}ne}, {Fitzgerald}, {Follette},
  {Gerard}, {Goodsell}, {Graham}, {Hibon}, {Hung}, {Ingraham}, {Kalas},
  {Larkin}, {Maire}, {Marchis}, {Metchev}, {Millar-Blanchaer}, {Nielsen},
  {Norton}, {Oppenheimer}, {Palmer}, {Patience}, {Perrin}, {Poyneer},
  {Rantakyr{\"o}}, {Savransky}, {Schneider}, {Sivaramakrishnan}, {Song},
  {Soummer}, {Thomas}, {Wallace}, {Ward-Duong}, {Wiktorowicz}, \&
  {Wolff}}]{2018AJ....155..226G}
{Greenbaum}, A.~Z., {Pueyo}, L., {Ruffio}, J.-B., {et~al.} 2018, \aj, 155, 226

\bibitem[{{Groff} {et~al.}(2016){Groff}, {Chilcote}, {Kasdin}, {Galvin},
  {Loomis}, {Carr}, {Brandt}, {Knapp}, {Limbach}, {Guyon}, {Jovanovic},
  {McElwain}, {Takato}, \& {Hayashi}}]{2016SPIE.9908E..0OG}
{Groff}, T.~D., {Chilcote}, J., {Kasdin}, N.~J., {et~al.} 2016, in Society of
  Photo-Optical Instrumentation Engineers (SPIE) Conference Series, Vol. 9908,
  Ground-based and Airborne Instrumentation for Astronomy VI, ed. C.~J.
  {Evans}, L.~{Simard}, \& H.~{Takami}, 99080O

\bibitem[{{Hauschildt} {et~al.}(1999){Hauschildt}, {Allard}, \&
  {Baron}}]{hauschildt_nextgen_1999}
{Hauschildt}, P.~H., {Allard}, F., \& {Baron}, E. 1999, \apj, 512, 377

\bibitem[{{Hayoz} {et~al.}(2023){Hayoz}, {Cugno}, {Quanz}, {Patapis}, {Alei},
  {Bonse}, {Dannert}, {Garvin}, {Gebhard}, {Konrad}, \&
  {Sartori}}]{hayoz_crocodile_2023}
{Hayoz}, J., {Cugno}, G., {Quanz}, S.~P., {et~al.} 2023, \aap, 678, A178

\bibitem[{{Hoch} {et~al.}(2023){Hoch}, {Konopacky}, {Theissen}, {Ruffio},
  {Barman}, {Rickman}, {Perrin}, {Macintosh}, \&
  {Marois}}]{2023AJ....166...85H}
{Hoch}, K. K.~W., {Konopacky}, Q.~M., {Theissen}, C.~A., {et~al.} 2023, \aj,
  166, 85

\bibitem[{{Hoeijmakers} {et~al.}(2018){Hoeijmakers}, {Schwarz}, {Snellen}, {de
  Kok}, {Bonnefoy}, {Chauvin}, {Lagrange}, \&
  {Girard}}]{hoeijmakers_medium-resolution_2018}
{Hoeijmakers}, H.~J., {Schwarz}, H., {Snellen}, I.~A.~G., {et~al.} 2018, \aap,
  617, A144

\bibitem[{{Houll{\'e}} {et~al.}(2025){Houll{\'e}}, {Millour}, {Berio},
  {Scigliuto}, {Lacour}, {Lopez}, {Allouche}, {Augereau}, {Blain}, {Bonnefoy},
  {Carbillet}, {Chauvin}, {Leftley}, {Matter}, {Milli}, {Molli{\`e}re},
  {Nasedkin}, {Nowak}, {Palma-Bifani}, {Pantin}, {Priolet}, {Ravet}, {Woillez},
  {Balmer}, {Boley}, {G{\'a}mez Rosas}, {Girard}, {Haubois}, {Hinkley},
  {Hogerheijde}, {Jaffe}, {Kammerer}, {Kreidberg}, {Lai}, {Lagarde}, {Labdon},
  {Le Bouquin}, {Meilland}, {M{\'e}rand}, {Paladini}, {Petrov}, {Rickman},
  {Rivinius}, {Robbe-Dubois}, {van Boekel}, {Varga}, {Vigan}, {Wang}, \&
  {Weigelt}}]{houlle_mathis_2025}
{Houll{\'e}}, M., {Millour}, F., {Berio}, P., {et~al.} 2025, arXiv e-prints,
  arXiv:2508.18366

\bibitem[{{Houll{\'e}} {et~al.}(2021){Houll{\'e}}, {Vigan}, {Carlotti},
  {Choquet}, {Cantalloube}, {Phillips}, {Sauvage}, {Schwartz}, {Otten},
  {Baraffe}, {Emsenhuber}, \& {Mordasini}}]{Houlle_2020}
{Houll{\'e}}, M., {Vigan}, A., {Carlotti}, A., {et~al.} 2021, \aap, 652, A67

\bibitem[{{Hubeny} \& {Burrows}(2007)}]{2007ApJ...669.1248H}
{Hubeny}, I. \& {Burrows}, A. 2007, \apj, 669, 1248

\bibitem[{{Janson} {et~al.}(2021){Janson}, {Brandeker}, {Olofsson}, \&
  {Liseau}}]{2021A&A...646A.132J}
{Janson}, M., {Brandeker}, A., {Olofsson}, G., \& {Liseau}, R. 2021, \aap, 646,
  A132

\bibitem[{{Kammerer} {et~al.}(2024){Kammerer}, {Lawson}, {Perrin}, {Rebollido},
  {Stark}, {Stolker}, {Girard}, {Pueyo}, {Balmer}, {Worthen}, {Chen}, {van der
  Marel}, {Lewis}, {Ward-Duong}, {Valenti}, {Clampin}, \&
  {Mountain}}]{kammerer_2024_JWST}
{Kammerer}, J., {Lawson}, K., {Perrin}, M.~D., {et~al.} 2024, \aj, 168, 51

\bibitem[{{Kawahara} {et~al.}(2022){Kawahara}, {Kawashima}, {Masuda},
  {Crossfield}, {Pannier}, \& {van den Bekerom}}]{kawahara_2025_autodiff}
{Kawahara}, H., {Kawashima}, Y., {Masuda}, K., {et~al.} 2022, \apjs, 258, 31

\bibitem[{{Kiefer} {et~al.}(2024){Kiefer}, {Bonnefoy}, {Charnay}, {Boccaletti},
  {Lagrange}, {Chauvin}, {B{\'e}zard}, \& {M{\^a}lin}}]{kiefer_new_2024}
{Kiefer}, F., {Bonnefoy}, M., {Charnay}, B., {et~al.} 2024, \aap, 685, A120

\bibitem[{{Kipping} \& {Benneke}(2025)}]{kipping_exoplaneteers_2025}
{Kipping}, D. \& {Benneke}, B. 2025, arXiv e-prints, arXiv:2506.05392

\bibitem[{{Knuth} {et~al.}(2015){Knuth}, {Habeck}, {Malakar}, {Mubeen}, \&
  {Placek}}]{knuth_bayesian_2015}
{Knuth}, K.~H., {Habeck}, M., {Malakar}, N.~K., {Mubeen}, A.~M., \& {Placek},
  B. 2015, Digital Signal Processing, 47, 50

\bibitem[{{Konopacky} {et~al.}(2013){Konopacky}, {Barman}, {Macintosh}, \&
  {Marois}}]{2013Sci...339.1398K}
{Konopacky}, Q.~M., {Barman}, T.~S., {Macintosh}, B.~A., \& {Marois}, C. 2013,
  Science, 339, 1398

\bibitem[{{Kotani} {et~al.}(2020){Kotani}, {Kawahara}, {Ishizuka}, {Jovanovic},
  {Vievard}, {Lozi}, {Sahoo}, {Guyon}, {Yoneta}, \&
  {Tamura}}]{2020SPIE11448E..78K}
{Kotani}, T., {Kawahara}, H., {Ishizuka}, M., {et~al.} 2020, in Society of
  Photo-Optical Instrumentation Engineers (SPIE) Conference Series, Vol. 11448,
  Adaptive Optics Systems VII, ed. L.~{Schreiber}, D.~{Schmidt}, \&
  E.~{Vernet}, 1144878

\bibitem[{{Kuzuhara} {et~al.}(2011){Kuzuhara}, {Tamura}, {Ishii}, {Kudo},
  {Nishiyama}, \& {Kandori}}]{2011AJ....141..119K}
{Kuzuhara}, M., {Tamura}, M., {Ishii}, M., {et~al.} 2011, \aj, 141, 119

\bibitem[{{Lacour} {et~al.}(2021){Lacour}, {Wang}, {Rodet}, {Nowak},
  {Shangguan}, {Beust}, {Lagrange}, {Abuter}, {Amorim}, {Asensio-Torres},
  {Benisty}, {Berger}, {Blunt}, {Boccaletti}, {Bohn}, {Bolzer}, {Bonnefoy},
  {Bonnet}, {Bourdarot}, {Brandner}, {Cantalloube}, {Caselli}, {Charnay},
  {Chauvin}, {Choquet}, {Christiaens}, {Cl{\'e}net}, {Coud{\'e} Du Foresto},
  {Cridland}, {Dembet}, {Dexter}, {de Zeeuw}, {Drescher}, {Duvert}, {Eckart},
  {Eisenhauer}, {Gao}, {Garcia}, {Garcia Lopez}, {Gendron}, {Genzel},
  {Gillessen}, {Girard}, {Haubois}, {Hei{\ss}el}, {Henning}, {Hinkley},
  {Hippler}, {Horrobin}, {Houll{\'e}}, {Hubert}, {Jocou}, {Kammerer},
  {Keppler}, {Kervella}, {Kreidberg}, {Lapeyr{\`e}re}, {Le Bouquin},
  {L{\'e}na}, {Lutz}, {Maire}, {M{\'e}rand}, {Molli{\`e}re}, {Monnier},
  {Mouillet}, {Nasedkin}, {Ott}, {Otten}, {Paladini}, {Paumard}, {Perraut},
  {Perrin}, {Pfuhl}, {Rickman}, {Pueyo}, {Rameau}, {Rousset}, {Rustamkulov},
  {Samland}, {Shimizu}, {Sing}, {Stadler}, {Stolker}, {Straub}, {Straubmeier},
  {Sturm}, {Tacconi}, {van Dishoeck}, {Vigan}, {Vincent}, {von Fellenberg},
  {Ward-Duong}, {Widmann}, {Wieprecht}, {Wiezorrek}, {Woillez}, {Yazici},
  {Young}, \& {Gravity Collaboration}}]{lacour_mass_2021}
{Lacour}, S., {Wang}, J.~J., {Rodet}, L., {et~al.} 2021, \aap, 654, L2

\bibitem[{{Lagrange} {et~al.}(2009){Lagrange}, {Gratadour}, {Chauvin}, {Fusco},
  {Ehrenreich}, {Mouillet}, {Rousset}, {Rouan}, {Allard}, {Gendron}, {Charton},
  {Mugnier}, {Rabou}, {Montri}, \& {Lacombe}}]{lagrange_probable_2009}
{Lagrange}, A.~M., {Gratadour}, D., {Chauvin}, G., {et~al.} 2009, \aap, 493,
  L21

\bibitem[{{Lagrange} {et~al.}(2020){Lagrange}, {Rubini}, {Nowak}, {Lacour},
  {Grandjean}, {Boccaletti}, {Langlois}, {Delorme}, {Gratton}, {Wang},
  {Flasseur}, {Galicher}, {Kral}, {Meunier}, {Beust}, {Babusiaux}, {Le
  Coroller}, {Thebault}, {Kervella}, {Zurlo}, {Maire}, {Wahhaj}, {Amorim},
  {Asensio-Torres}, {Benisty}, {Berger}, {Bonnefoy}, {Brandner}, {Cantalloube},
  {Charnay}, {Chauvin}, {Choquet}, {Cl{\'e}net}, {Christiaens}, {Coud{\'e} Du
  Foresto}, {de Zeeuw}, {Desidera}, {Duvert}, {Eckart}, {Eisenhauer},
  {Galland}, {Gao}, {Garcia}, {Garcia Lopez}, {Gendron}, {Genzel}, {Gillessen},
  {Girard}, {Hagelberg}, {Haubois}, {Henning}, {Heissel}, {Hippler},
  {Horrobin}, {Janson}, {Kammerer}, {Kenworthy}, {Keppler}, {Kreidberg},
  {Lapeyr{\`e}re}, {Le Bouquin}, {L{\'e}na}, {M{\'e}rand}, {Messina},
  {Molli{\`e}re}, {Monnier}, {Ott}, {Otten}, {Paumard}, {Paladini}, {Perraut},
  {Perrin}, {Pueyo}, {Pfuhl}, {Rodet}, {Rodriguez-Coira}, {Rousset}, {Samland},
  {Shangguan}, {Schmidt}, {Straub}, {Straubmeier}, {Stolker}, {Vigan},
  {Vincent}, {Widmann}, {Woillez}, \& {GRAVITY
  Collaboration}}]{lagrange_unveiling_2020}
{Lagrange}, A.~M., {Rubini}, P., {Nowak}, M., {et~al.} 2020, \aap, 642, A18

\bibitem[{{Landman} {et~al.}(2024){Landman}, {Stolker}, {Snellen}, {Costes},
  {de Regt}, {Zhang}, {Gandhi}, {Molliere}, {Kesseli}, {Vigan}, \&
  {Sanchez-L{\'o}pez}}]{landman__2024}
{Landman}, R., {Stolker}, T., {Snellen}, I.~A.~G., {et~al.} 2024, \aap, 682,
  A48

\bibitem[{{Lecluse} {et~al.}(1996){Lecluse}, {Robert}, {Gautier}, \&
  {Guiraud}}]{1996P&SS...44.1579L}
{Lecluse}, C., {Robert}, F., {Gautier}, D., \& {Guiraud}, M. 1996, \planss, 44,
  1579

\bibitem[{{Macintosh} {et~al.}(2014){Macintosh}, {Graham}, {Ingraham},
  {Konopacky}, {Marois}, {Perrin}, {Poyneer}, {Bauman}, {Barman}, {Burrows},
  {Cardwell}, {Chilcote}, {De Rosa}, {Dillon}, {Doyon}, {Dunn}, {Erikson},
  {Fitzgerald}, {Gavel}, {Goodsell}, {Hartung}, {Hibon}, {Kalas}, {Larkin},
  {Maire}, {Marchis}, {Marley}, {McBride}, {Millar-Blanchaer}, {Morzinski},
  {Norton}, {Oppenheimer}, {Palmer}, {Patience}, {Pueyo}, {Rantakyro},
  {Sadakuni}, {Saddlemyer}, {Savransky}, {Serio}, {Soummer},
  {Sivaramakrishnan}, {Song}, {Thomas}, {Wallace}, {Wiktorowicz}, \&
  {Wolff}}]{2014PNAS..11112661M}
{Macintosh}, B., {Graham}, J.~R., {Ingraham}, P., {et~al.} 2014, Proceedings of
  the National Academy of Science, 111, 12661

\bibitem[{{Males} {et~al.}(2014){Males}, {Close}, {Morzinski}, {Wahhaj}, {Liu},
  {Skemer}, {Kopon}, {Follette}, {Puglisi}, {Esposito}, {Riccardi}, {Pinna},
  {Xompero}, {Briguglio}, {Biller}, {Nielsen}, {Hinz}, {Rodigas}, {Hayward},
  {Chun}, {Ftaclas}, {Toomey}, \& {Wu}}]{males_magellan_2014}
{Males}, J.~R., {Close}, L.~M., {Morzinski}, K.~M., {et~al.} 2014, \apj, 786,
  32

\bibitem[{{Mamajek}(2009)}]{2009AIPC.1158....3M}
{Mamajek}, E.~E. 2009, in American Institute of Physics Conference Series, Vol.
  1158, Exoplanets and Disks: Their Formation and Diversity, ed. T.~{Usuda},
  M.~{Tamura}, \& M.~{Ishii} (AIP), 3--10

\bibitem[{{Mamajek} \& {Bell}(2014)}]{mamajek_age_2014}
{Mamajek}, E.~E. \& {Bell}, C. P.~M. 2014, \mnras, 445, 2169

\bibitem[{{Marleau} {et~al.}(2019){Marleau}, {Coleman}, {Leleu}, \&
  {Mordasini}}]{2019A&A...624A..20M}
{Marleau}, G.-D., {Coleman}, G. A.~L., {Leleu}, A., \& {Mordasini}, C. 2019,
  \aap, 624, A20

\bibitem[{{Marleau} \& {Cumming}(2014)}]{2014MNRAS.437.1378M}
{Marleau}, G.~D. \& {Cumming}, A. 2014, \mnras, 437, 1378

\bibitem[{{Marois} {et~al.}(2010){Marois}, {Zuckerman}, {Konopacky},
  {Macintosh}, \& {Barman}}]{2010Natur.468.1080M}
{Marois}, C., {Zuckerman}, B., {Konopacky}, Q.~M., {Macintosh}, B., \&
  {Barman}, T. 2010, \nat, 468, 1080

\bibitem[{{Mesa} {et~al.}(2023){Mesa}, {Gratton}, {Kervella}, {Bonavita},
  {Desidera}, {D'Orazi}, {Marino}, {Zurlo}, \&
  {Rigliaco}}]{2023A&A...672A..93M}
{Mesa}, D., {Gratton}, R., {Kervella}, P., {et~al.} 2023, \aap, 672, A93

\bibitem[{{Miles} {et~al.}(2023){Miles}, {Biller}, {Patapis}, {Worthen},
  {Rickman}, {Hoch}, {Skemer}, {Perrin}, {Whiteford}, {Chen}, {Sargent},
  {Mukherjee}, {Morley}, {Moran}, {Bonnefoy}, {Petrus}, {Carter}, {Choquet},
  {Hinkley}, {Ward-Duong}, {Leisenring}, {Millar-Blanchaer}, {Pueyo}, {Ray},
  {Sallum}, {Stapelfeldt}, {Stone}, {Wang}, {Absil}, {Balmer}, {Boccaletti},
  {Bonavita}, {Booth}, {Bowler}, {Chauvin}, {Christiaens}, {Currie},
  {Danielski}, {Fortney}, {Girard}, {Grady}, {Greenbaum}, {Henning}, {Hines},
  {Janson}, {Kalas}, {Kammerer}, {Kennedy}, {Kenworthy}, {Kervella}, {Lagage},
  {Lew}, {Liu}, {Macintosh}, {Marino}, {Marley}, {Marois}, {Matthews},
  {Matthews}, {Mawet}, {McElwain}, {Metchev}, {Meyer}, {Molliere}, {Pantin},
  {Quirrenbach}, {Rebollido}, {Ren}, {Schneider}, {Vasist}, {Wyatt}, {Zhou},
  {Briesemeister}, {Bryan}, {Calissendorff}, {Cantalloube}, {Cugno}, {De
  Furio}, {Dupuy}, {Factor}, {Faherty}, {Fitzgerald}, {Franson}, {Gonzales},
  {Hood}, {Howe}, {Kraus}, {Kuzuhara}, {Lagrange}, {Lawson}, {Lazzoni}, {Liu},
  {Llop-Sayson}, {Lloyd}, {Martinez}, {Mazoyer}, {Quanz}, {Redai}, {Samland},
  {Schlieder}, {Tamura}, {Tan}, {Uyama}, {Vigan}, {Vos}, {Wagner}, {Wolff},
  {Ygouf}, {Zhang}, {Zhang}, \& {Zhang}}]{2023ApJ...946L...6M}
{Miles}, B.~E., {Biller}, B.~A., {Patapis}, P., {et~al.} 2023, \apjl, 946, L6

\bibitem[{{Millar-Blanchaer} {et~al.}(2015){Millar-Blanchaer}, {Graham},
  {Pueyo}, {Kalas}, {Dawson}, {Wang}, {Perrin}, {moon}, {Macintosh}, {Ammons},
  {Barman}, {Cardwell}, {Chen}, {Chiang}, {Chilcote}, {Cotten}, {De Rosa},
  {Draper}, {Dunn}, {Duch{\^e}ne}, {Esposito}, {Fitzgerald}, {Follette},
  {Goodsell}, {Greenbaum}, {Hartung}, {Hibon}, {Hinkley}, {Ingraham},
  {Jensen-Clem}, {Konopacky}, {Larkin}, {Long}, {Maire}, {Marchis}, {Marley},
  {Marois}, {Morzinski}, {Nielsen}, {Palmer}, {Oppenheimer}, {Poyneer},
  {Rajan}, {Rantakyr{\"o}}, {Ruffio}, {Sadakuni}, {Saddlemyer}, {Schneider},
  {Sivaramakrishnan}, {Soummer}, {Thomas}, {Vasisht}, {Vega}, {Wallace},
  {Ward-Duong}, {Wiktorowicz}, \& {Wolff}}]{millar-blanchaer_beta_2015}
{Millar-Blanchaer}, M.~A., {Graham}, J.~R., {Pueyo}, L., {et~al.} 2015, \apj,
  811, 18

\bibitem[{{Molli{\`e}re} {et~al.}(2022){Molli{\`e}re}, {Molyarova}, {Bitsch},
  {Henning}, {Schneider}, {Kreidberg}, {Eistrup}, {Burn}, {Nasedkin},
  {Semenov}, {Mordasini}, {Schlecker}, {Schwarz}, {Lacour}, {Nowak}, \&
  {Schulik}}]{2022ApJ...934...74M}
{Molli{\`e}re}, P., {Molyarova}, T., {Bitsch}, B., {et~al.} 2022, \apj, 934, 74

\bibitem[{{Molli{\`e}re} {et~al.}(2020){Molli{\`e}re}, {Stolker}, {Lacour},
  {Otten}, {Shangguan}, {Charnay}, {Molyarova}, {Nowak}, {Henning}, {Marleau},
  {Semenov}, {van Dishoeck}, {Eisenhauer}, {Garcia}, {Garcia Lopez}, {Girard},
  {Greenbaum}, {Hinkley}, {Kervella}, {Kreidberg}, {Maire}, {Nasedkin},
  {Pueyo}, {Snellen}, {Vigan}, {Wang}, {de Zeeuw}, \&
  {Zurlo}}]{2020A&A...640A.131M}
{Molli{\`e}re}, P., {Stolker}, T., {Lacour}, S., {et~al.} 2020, \aap, 640, A131

\bibitem[{{Molli{\`e}re} {et~al.}(2019){Molli{\`e}re}, {Wardenier}, {van
  Boekel}, {Henning}, {Molaverdikhani}, \& {Snellen}}]{2019A&A...627A..67M}
{Molli{\`e}re}, P., {Wardenier}, J.~P., {van Boekel}, R., {et~al.} 2019, \aap,
  627, A67

\bibitem[{{Mordasini} {et~al.}(2014){Mordasini}, {Klahr}, {Alibert}, {Miller},
  \& {Henning}}]{2014A&A...566A.141M}
{Mordasini}, C., {Klahr}, H., {Alibert}, Y., {Miller}, N., \& {Henning}, T.
  2014, \aap, 566, A141

\bibitem[{Morley {et~al.}(2024)Morley, Mukherjee, Marley, Fortney, Visscher,
  Lupu, Gharib-Nezhad, Thorngren, Freedman, \& Batalha~7}]{morley_sonora_2024}
Morley, C.~V., Mukherjee, S., Marley, M.~S., {et~al.} 2024, The {Sonora}
  {Substellar} {Atmosphere} {Models}. {III}. {Diamondback}: {Atmospheric}
  {Properties}, {Spectra}, and {Evolution} for {Warm} {Cloudy} {Substellar}
  {Objects}, publication Title: arXiv e-prints ADS Bibcode: 2024arXiv240200758M

\bibitem[{{Morley} {et~al.}(2019){Morley}, {Skemer}, {Miles}, {Line}, {Lopez},
  {Brogi}, {Freedman}, \& {Marley}}]{2019ApJ...882L..29M}
{Morley}, C.~V., {Skemer}, A.~J., {Miles}, B.~E., {et~al.} 2019, \apjl, 882,
  L29

\bibitem[{{Morris} {et~al.}(2024){Morris}, {Wang}, {Hsu}, {Ruffio}, {Xuan},
  {Delorme}, {Hood}, {Bryan}, {Martin}, {Pezzato}, {Mawet}, {Skemer}, {Baker},
  {Bartos}, {Calvin}, {Cetre}, {Doppmann}, {Echeverri}, {Finnerty},
  {Fitzgerald}, {Jovanovic}, {Liberman}, {Lopez}, {Sappey}, {Schofield},
  {Wallace}, \& {Wang}}]{2024arXiv240513125M}
{Morris}, E.~C., {Wang}, J.~J., {Hsu}, C.-C., {et~al.} 2024, \aj, 168, 144

\bibitem[{{Morzinski} {et~al.}(2015){Morzinski}, {Males}, {Skemer}, {Close},
  {Hinz}, {Rodigas}, {Puglisi}, {Esposito}, {Riccardi}, {Pinna}, {Xompero},
  {Briguglio}, {Bailey}, {Follette}, {Kopon}, {Weinberger}, \&
  {Wu}}]{morzinski_magellan_2015}
{Morzinski}, K.~M., {Males}, J.~R., {Skemer}, A.~J., {et~al.} 2015, \apj, 815,
  108

\bibitem[{{Moses} {et~al.}(2016){Moses}, {Marley}, {Zahnle}, {Line}, {Fortney},
  {Barman}, {Visscher}, {Lewis}, \& {Wolff}}]{2016ApJ...829...66M}
{Moses}, J.~I., {Marley}, M.~S., {Zahnle}, K., {et~al.} 2016, \apj, 829, 66

\bibitem[{{Nasedkin} {et~al.}(2024){Nasedkin}, {Molli{\`e}re}, {Lacour},
  {Nowak}, {Kreidberg}, {Stolker}, {Wang}, {Balmer}, {Kammerer}, {Shangguan},
  {Abuter}, {Amorim}, {Asensio-Torres}, {Benisty}, {Berger}, {Beust}, {Blunt},
  {Boccaletti}, {Bonnefoy}, {Bonnet}, {Bordoni}, {Bourdarot}, {Brandner},
  {Cantalloube}, {Caselli}, {Charnay}, {Chauvin}, {Chavez}, {Choquet},
  {Christiaens}, {Cl{\'e}net}, {Coud{\'e} Du Foresto}, {Cridland}, {Davies},
  {Dembet}, {Dexter}, {Drescher}, {Duvert}, {Eckart}, {Eisenhauer},
  {F{\"o}rster Schreiber}, {Garcia}, {Garcia Lopez}, {Gendron}, {Genzel},
  {Gillessen}, {Girard}, {Grant}, {Haubois}, {Hei{\ss}el}, {Henning},
  {Hinkley}, {Hippler}, {Houll{\'e}}, {Hubert}, {Jocou}, {Keppler}, {Kervella},
  {Kurtovic}, {Lagrange}, {Lapeyr{\`e}re}, {Le Bouquin}, {Lutz}, {Maire},
  {Mang}, {Marleau}, {M{\'e}rand}, {Monnier}, {Mordasini}, {Ott}, {Otten},
  {Paladini}, {Paumard}, {Perraut}, {Perrin}, {Pfuhl}, {Pourr{\'e}}, {Pueyo},
  {Ribeiro}, {Rickman}, {Ruffio}, {Rustamkulov}, {Shimizu}, {Sing}, {Stadler},
  {Straub}, {Straubmeier}, {Sturm}, {Tacconi}, {van Dishoeck}, {Vigan},
  {Vincent}, {von Fellenberg}, {Widmann}, {Winterhalder}, {Woillez}, {Yazici},
  \& {Gravity Collaboration}}]{2024arXiv240403776N}
{Nasedkin}, E., {Molli{\`e}re}, P., {Lacour}, S., {et~al.} 2024, \aap, 687,
  A298

\bibitem[{{Nissen}(2013)}]{nissen_carbon--oxygen_2013}
{Nissen}, P.~E. 2013, \aap, 552, A73

\bibitem[{Nowak(2019)}]{nowak_2017_2019}
Nowak, M. 2019, phdthesis, Université Paris sciences et lettres

\bibitem[{{Nowak} {et~al.}(2020){Nowak}, {Lacour}, {Lagrange}, {Rubini},
  {Wang}, {Stolker}, {Abuter}, {Amorim}, {Asensio-Torres}, {Baub{\"o}ck},
  {Benisty}, {Berger}, {Beust}, {Blunt}, {Boccaletti}, {Bonnefoy}, {Bonnet},
  {Brandner}, {Cantalloube}, {Charnay}, {Choquet}, {Christiaens}, {Cl{\'e}net},
  {Coud{\'e} Du Foresto}, {Cridland}, {de Zeeuw}, {Dembet}, {Dexter},
  {Drescher}, {Duvert}, {Eckart}, {Eisenhauer}, {Gao}, {Garcia}, {Garcia
  Lopez}, {Gardner}, {Gendron}, {Genzel}, {Gillessen}, {Girard}, {Grandjean},
  {Haubois}, {Hei{\ss}el}, {Henning}, {Hinkley}, {Hippler}, {Horrobin},
  {Houll{\'e}}, {Hubert}, {Jim{\'e}nez-Rosales}, {Jocou}, {Kammerer},
  {Kervella}, {Keppler}, {Kreidberg}, {Kulikauskas}, {Lapeyr{\`e}re}, {Le
  Bouquin}, {L{\'e}na}, {M{\'e}rand}, {Maire}, {Molli{\`e}re}, {Monnier},
  {Mouillet}, {M{\"u}ller}, {Nasedkin}, {Ott}, {Otten}, {Paumard}, {Paladini},
  {Perraut}, {Perrin}, {Pueyo}, {Pfuhl}, {Rameau}, {Rodet},
  {Rodr{\'\i}guez-Coira}, {Rousset}, {Scheithauer}, {Shangguan}, {Stadler},
  {Straub}, {Straubmeier}, {Sturm}, {Tacconi}, {van Dishoeck}, {Vigan},
  {Vincent}, {von Fellenberg}, {Ward-Duong}, {Widmann}, {Wieprecht},
  {Wiezorrek}, {Woillez}, \& {GRAVITY Collaboration}}]{nowak_direct_2020}
{Nowak}, M., {Lacour}, S., {Lagrange}, A.~M., {et~al.} 2020, \aap, 642, L2

\bibitem[{{{\"O}berg} \& {Bergin}(2016)}]{2016ApJ...831L..19O}
{{\"O}berg}, K.~I. \& {Bergin}, E.~A. 2016, \apjl, 831, L19

\bibitem[{{Owen}(1992)}]{1992mars.book..818O}
{Owen}, T. 1992, in Mars, ed. M.~{George}, 818--834

\bibitem[{{Owen} \& {Encrenaz}(2003)}]{2003SSRv..106..121O}
{Owen}, T. \& {Encrenaz}, T. 2003, \ssr, 106, 121

\bibitem[{{Palma-Bifani} {et~al.}(2023){Palma-Bifani}, {Chauvin}, {Bonnefoy},
  {Rojo}, {Petrus}, {Rodet}, {Langlois}, {Allard}, {Charnay}, {Desgrange},
  {Homeier}, {Lagrange}, {Beuzit}, {Baudoz}, {Boccaletti}, {Chomez}, {Delorme},
  {Desidera}, {Feldt}, {Ginski}, {Gratton}, {Maire}, {Meyer}, {Samland},
  {Snellen}, {Vigan}, \& {Zhang}}]{2023A&A...670A..90P}
{Palma-Bifani}, P., {Chauvin}, G., {Bonnefoy}, M., {et~al.} 2023, \aap, 670,
  A90

\bibitem[{{Parker} {et~al.}(2024){Parker}, {Birkby}, {Landman}, {Wardenier},
  {Young}, {Vaughan}, {van Sluijs}, {Brogi}, {Parmentier}, \&
  {Line}}]{parker_into_2024}
{Parker}, L.~T., {Birkby}, J.~L., {Landman}, R., {et~al.} 2024, \mnras, 531,
  2356

\bibitem[{{Petrus} {et~al.}(2020){Petrus}, {Bonnefoy}, {Chauvin}, {Babusiaux},
  {Delorme}, {Lagrange}, {Florent}, {Bayo}, {Janson}, {Biller}, {Manjavacas},
  {Marleau}, \& {Kopytova}}]{petrus_new_2020}
{Petrus}, S., {Bonnefoy}, M., {Chauvin}, G., {et~al.} 2020, \aap, 633, A124

\bibitem[{{Petrus} {et~al.}(2021){Petrus}, {Bonnefoy}, {Chauvin}, {Charnay},
  {Marleau}, {Gratton}, {Lagrange}, {Rameau}, {Mordasini}, {Nowak}, {Delorme},
  {Boccaletti}, {Carlotti}, {Houll{\'e}}, {Vigan}, {Allard}, {Desidera},
  {D'Orazi}, {Hoeijmakers}, {Wyttenbach}, \&
  {Lavie}}]{petrus_medium-resolution_2021}
{Petrus}, S., {Bonnefoy}, M., {Chauvin}, G., {et~al.} 2021, \aap, 648, A59

\bibitem[{{Petrus} {et~al.}(2023){Petrus}, {Chauvin}, {Bonnefoy}, {Tremblin},
  {Charnay}, {Delorme}, {Marleau}, {Bayo}, {Manjavacas}, {Lagrange},
  {Molli{\`e}re}, {Palma-Bifani}, {Biller}, {Jenkins}, {Goyal}, \&
  {Hoch}}]{petrus_x-shyne_2023}
{Petrus}, S., {Chauvin}, G., {Bonnefoy}, M., {et~al.} 2023, \aap, 670, L9

\bibitem[{{Petrus} {et~al.}(2025){Petrus}, {Chauvin}, {Bonnefoy}, {Tremblin},
  {Morley}, {Charnay}, {Suarez}, {Gagn{\'e}}, {Palma-Bifani}, {Denis}, {Ravet},
  {Bayo}, {B{\'e}zard}, {Biller}, {Delorme}, {Faherty}, {Goyal}, {Hoch}, {Hoy},
  {Jenkins}, {Lagrange}, {Lavie}, {Liu}, {Manjavacas}, {Marleau}, {McElwain},
  {Molli{\`e}re}, {Mordasini}, {Phillips}, {Rojo}, {Zhang}, \&
  {Zurlo}}]{petrus_fulllib_2025}
{Petrus}, S., {Chauvin}, G., {Bonnefoy}, M., {et~al.} 2025, \aap, 701, A208

\bibitem[{{Petrus} {et~al.}(2024){Petrus}, {Whiteford}, {Patapis}, {Biller},
  {Skemer}, {Hinkley}, {Su{\'a}rez}, {Palma-Bifani}, {Morley}, {Tremblin},
  {Charnay}, {Vos}, {Wang}, {Stone}, {Bonnefoy}, {Chauvin}, {Miles}, {Carter},
  {Lueber}, {Helling}, {Sutlieff}, {Janson}, {Gonzales}, {Hoch}, {Absil},
  {Balmer}, {Boccaletti}, {Bonavita}, {Booth}, {Bowler}, {Briesemeister},
  {Bryan}, {Calissendorff}, {Cantalloube}, {Chen}, {Choquet}, {Christiaens},
  {Cugno}, {Currie}, {Danielski}, {De Furio}, {Dupuy}, {Factor}, {Faherty},
  {Fitzgerald}, {Fortney}, {Franson}, {Girard}, {Grady}, {Henning}, {Hines},
  {Hood}, {Howe}, {Kalas}, {Kammerer}, {Kennedy}, {Kenworthy}, {Kervella},
  {Kim}, {Kitzmann}, {Kraus}, {Kuzuhara}, {Lagage}, {Lagrange}, {Lawson},
  {Lazzoni}, {Leisenring}, {Lew}, {Liu}, {Liu}, {Llop-Sayson}, {Lloyd},
  {Macintosh}, {M{\^a}lin}, {Manjavacas}, {Marino}, {Marley}, {Marois},
  {Martinez}, {Matthews}, {Matthews}, {Mawet}, {Mazoyer}, {McElwain},
  {Metchev}, {Meyer}, {Millar-Blanchaer}, {Molli{\`e}re}, {Moran}, {Mukherjee},
  {Pantin}, {Perrin}, {Pueyo}, {Quanz}, {Quirrenbach}, {Ray}, {Rebollido},
  {Adams Redai}, {Ren}, {Rickman}, {Sallum}, {Samland}, {Sargent}, {Schlieder},
  {Stapelfeldt}, {Tamura}, {Tan}, {Theissen}, {Uyama}, {Vasist}, {Vigan},
  {Wagner}, {Ward-Duong}, {Wolff}, {Worthen}, {Wyatt}, {Ygouf}, {Zurlo},
  {Zhang}, {Zhang}, {Zhang}, \& {Zhou}}]{petrus_jwst_2024}
{Petrus}, S., {Whiteford}, N., {Patapis}, P., {et~al.} 2024, \apjl, 966, L11

\bibitem[{{Phillips} {et~al.}(2020){Phillips}, {Tremblin}, {Baraffe},
  {Chabrier}, {Allard}, {Spiegelman}, {Goyal}, {Drummond}, \&
  {H{\'e}brard}}]{2020A&A...637A..38P}
{Phillips}, M.~W., {Tremblin}, P., {Baraffe}, I., {et~al.} 2020, \aap, 637, A38

\bibitem[{{Quanz} {et~al.}(2010){Quanz}, {Meyer}, {Kenworthy}, {Girard},
  {Kasper}, {Lagrange}, {Apai}, {Boccaletti}, {Bonnefoy}, {Chauvin}, {Hinz}, \&
  {Lenzen}}]{quanz_first_2010}
{Quanz}, S.~P., {Meyer}, M.~R., {Kenworthy}, M.~A., {et~al.} 2010, \apjl, 722,
  L49

\bibitem[{{Rajan} {et~al.}(2017){Rajan}, {Rameau}, {De Rosa}, {Marley},
  {Graham}, {Macintosh}, {Marois}, {Morley}, {Patience}, {Pueyo}, {Saumon},
  {Ward-Duong}, {Ammons}, {Arriaga}, {Bailey}, {Barman}, {Bulger}, {Burrows},
  {Chilcote}, {Cotten}, {Czekala}, {Doyon}, {Duch{\^e}ne}, {Esposito},
  {Fitzgerald}, {Follette}, {Fortney}, {Goodsell}, {Greenbaum}, {Hibon},
  {Hung}, {Ingraham}, {Johnson-Groh}, {Kalas}, {Konopacky}, {Lafreni{\`e}re},
  {Larkin}, {Maire}, {Marchis}, {Metchev}, {Millar-Blanchaer}, {Morzinski},
  {Nielsen}, {Oppenheimer}, {Palmer}, {Patel}, {Perrin}, {Poyneer},
  {Rantakyr{\"o}}, {Ruffio}, {Savransky}, {Schneider}, {Sivaramakrishnan},
  {Song}, {Soummer}, {Thomas}, {Vasisht}, {Wallace}, {Wang}, {Wiktorowicz}, \&
  {Wolff}}]{2017AJ....154...10R}
{Rajan}, A., {Rameau}, J., {De Rosa}, R.~J., {et~al.} 2017, \aj, 154, 10

\bibitem[{{Rotman} {et~al.}(2025){Rotman}, {Welbanks}, {Line}, {McGill},
  {Radica}, \& {Nixon}}]{rotman_2025_enable}
{Rotman}, Y., {Welbanks}, L., {Line}, M.~R., {et~al.} 2025, \apj, 989, 201

\bibitem[{{Ruffio} {et~al.}(2019){Ruffio}, {Macintosh}, {Konopacky}, {Barman},
  {De Rosa}, {Wang}, {Wilcomb}, {Czekala}, \& {Marois}}]{ruffio_radial_2019}
{Ruffio}, J.-B., {Macintosh}, B., {Konopacky}, Q.~M., {et~al.} 2019, \aj, 158,
  200

\bibitem[{{Samland} {et~al.}(2017){Samland}, {Molli{\`e}re}, {Bonnefoy},
  {Maire}, {Cantalloube}, {Cheetham}, {Mesa}, {Gratton}, {Biller}, {Wahhaj},
  {Bouwman}, {Brandner}, {Melnick}, {Carson}, {Janson}, {Henning}, {Homeier},
  {Mordasini}, {Langlois}, {Quanz}, {van Boekel}, {Zurlo}, {Schlieder},
  {Avenhaus}, {Beuzit}, {Boccaletti}, {Bonavita}, {Chauvin}, {Claudi}, {Cudel},
  {Desidera}, {Feldt}, {Fusco}, {Galicher}, {Kopytova}, {Lagrange}, {Le
  Coroller}, {Martinez}, {Moeller-Nilsson}, {Mouillet}, {Mugnier}, {Perrot},
  {Sevin}, {Sissa}, {Vigan}, \& {Weber}}]{samland_spectral_2017}
{Samland}, M., {Molli{\`e}re}, P., {Bonnefoy}, M., {et~al.} 2017, \aap, 603,
  A57

\bibitem[{{Schneider} \& {Bitsch}(2021{\natexlab{a}})}]{2021A&A...654A..71S}
{Schneider}, A.~D. \& {Bitsch}, B. 2021{\natexlab{a}}, \aap, 654, A71

\bibitem[{{Schneider} \& {Bitsch}(2021{\natexlab{b}})}]{2021A&A...654A..72S}
{Schneider}, A.~D. \& {Bitsch}, B. 2021{\natexlab{b}}, \aap, 654, A72

\bibitem[{{Skemer} {et~al.}(2015){Skemer}, {Hinz}, {Montoya}, {Skrutskie},
  {Leisenring}, {Durney}, {Woodward}, {Wilson}, {Nelson}, {Bailey}, {Defrere},
  \& {Stone}}]{2015SPIE.9605E..1DS}
{Skemer}, A.~J., {Hinz}, P., {Montoya}, M., {et~al.} 2015, in Society of
  Photo-Optical Instrumentation Engineers (SPIE) Conference Series, Vol. 9605,
  Techniques and Instrumentation for Detection of Exoplanets VII, ed.
  S.~{Shaklan}, 96051D

\bibitem[{{Skemer} {et~al.}(2018){Skemer}, {Hinz}, {Stone}, {Skrutskie},
  {Woodward}, {Leisenring}, \& {Briesemeister}}]{2018SPIE10702E..0CS}
{Skemer}, A.~J., {Hinz}, P., {Stone}, J., {et~al.} 2018, in Society of
  Photo-Optical Instrumentation Engineers (SPIE) Conference Series, Vol. 10702,
  Ground-based and Airborne Instrumentation for Astronomy VII, ed. C.~J.
  {Evans}, L.~{Simard}, \& H.~{Takami}, 107020C

\bibitem[{{Skilling}(2004)}]{skilling_nested_2004}
{Skilling}, J. 2004, in American Institute of Physics Conference Series, Vol.
  735, Bayesian Inference and Maximum Entropy Methods in Science and
  Engineering: 24th International Workshop on Bayesian Inference and Maximum
  Entropy Methods in Science and Engineering, ed. R.~{Fischer}, R.~{Preuss}, \&
  U.~V. {Toussaint} (AIP), 395--405

\bibitem[{{Smith} \& {Terrile}(1984)}]{smith_circumstellar_1984}
{Smith}, B.~A. \& {Terrile}, R.~J. 1984, Science, 226, 1421

\bibitem[{{Snellen} {et~al.}(2014{\natexlab{a}}){Snellen}, {Brandl}, {de Kok},
  {Brogi}, {Birkby}, \& {Schwarz}}]{snellen_fast_2014}
{Snellen}, I., {Brandl}, B., {de Kok}, R., {et~al.} 2014{\natexlab{a}}, arXiv
  e-prints, arXiv:1404.7506

\bibitem[{{Snellen} {et~al.}(2014{\natexlab{b}}){Snellen}, {Brandl}, {de Kok},
  {Brogi}, {Birkby}, \& {Schwarz}}]{2014Natur.509...63S}
{Snellen}, I. A.~G., {Brandl}, B.~R., {de Kok}, R.~J., {et~al.}
  2014{\natexlab{b}}, \nat, 509, 63

\bibitem[{{Spiegel} \& {Burrows}(2012)}]{2012ApJ...745..174S}
{Spiegel}, D.~S. \& {Burrows}, A. 2012, \apj, 745, 174

\bibitem[{{Squicciarini} {et~al.}(2022){Squicciarini}, {Gratton}, {Janson},
  {Mamajek}, {Chauvin}, {Delorme}, {Langlois}, {Vigan}, {Ringqvist}, {Meeus},
  {Reffert}, {Kenworthy}, {Meyer}, {Bonnefoy}, {Bonavita}, {Mesa}, {Samland},
  {Desidera}, {D'Orazi}, {Engler}, {Alecian}, {Miglio}, {Henning}, {Quanz},
  {Mayer}, {Flasseur}, \& {Marleau}}]{2022A&A...664A...9S}
{Squicciarini}, V., {Gratton}, R., {Janson}, M., {et~al.} 2022, \aap, 664, A9

\bibitem[{{Stolker} {et~al.}(2020){Stolker}, {Quanz}, {Todorov}, {K{\"u}hn},
  {Molli{\`e}re}, {Meyer}, {Currie}, {Daemgen}, \&
  {Lavie}}]{2020A&A...635A.182S}
{Stolker}, T., {Quanz}, S.~P., {Todorov}, K.~O., {et~al.} 2020, \aap, 635, A182

\bibitem[{{Swastik} {et~al.}(2021){Swastik}, {Banyal}, {Narang}, {Manoj},
  {Sivarani}, {Reddy}, \& {Rajaguru}}]{swastik_host_2021}
{Swastik}, C., {Banyal}, R.~K., {Narang}, M., {et~al.} 2021, \aj, 161, 114

\bibitem[{{Tremblin} {et~al.}(2015){Tremblin}, {Amundsen}, {Mourier},
  {Baraffe}, {Chabrier}, {Drummond}, {Homeier}, \&
  {Venot}}]{tremblin_fingering_2015}
{Tremblin}, P., {Amundsen}, D.~S., {Mourier}, P., {et~al.} 2015, \apjl, 804,
  L17

\bibitem[{{Vigan} {et~al.}(2024){Vigan}, {El Morsy}, {Lopez}, {Otten},
  {Garcia}, {Costes}, {Muslimov}, {Viret}, {Charles}, {Zins}, {Murray},
  {Costille}, {Paufique}, {Seemann}, {Houll{\'e}}, {Anwand-Heerwart},
  {Phillips}, {Abinanti}, {Balard}, {Baraffe}, {Benedetti}, {Blanchard},
  {Blanco}, {Beuzit}, {Choquet}, {Cristofari}, {Desidera}, {Dohlen}, {Dorn},
  {Ely}, {Fuenteseca}, {Garcia}, {Jaquet}, {Jaubert}, {Kasper}, {Le Merrer},
  {Maire}, {N'Diaye}, {Pallanca}, {Popovic}, {Pourcelot}, {Reiners}, {Rochat},
  {Sehim}, {Schmutzer}, {Smette}, {Tchoubaklian}, {Tomlinson}, \& {Valenzuela
  Soto}}]{2024A&A...682A..16V}
{Vigan}, A., {El Morsy}, M., {Lopez}, M., {et~al.} 2024, \aap, 682, A16

\bibitem[{{Wang} {et~al.}(2021){Wang}, {Kulikauskas}, \&
  {Blunt}}]{wang_asclnet_whereistheplanet}
{Wang}, J.~J., {Kulikauskas}, M., \& {Blunt}, S. 2021, {whereistheplanet:
  Predicting positions of directly imaged companions}, Astrophysics Source Code
  Library, record ascl:2101.003

\bibitem[{{Worthen} {et~al.}(2024){Worthen}, {Chen}, {Law}, {Lu}, {Hoch},
  {Chai}, {Sloan}, {Sargent}, {Kammerer}, {Hines}, {Rebollido}, {Balmer},
  {Perrin}, {Watson}, {Pueyo}, {Girard}, {Lisse}, \&
  {Stark}}]{worthen_miri_2024}
{Worthen}, K., {Chen}, C.~H., {Law}, D.~R., {et~al.} 2024, \apj, 964, 168

\bibitem[{{Xuan} {et~al.}(2024){Xuan}, {Hsu}, {Finnerty}, {Wang}, {Ruffio},
  {Zhang}, {Knutson}, {Mawet}, {Mamajek}, {Inglis}, {Wallack}, {Bryan},
  {Blake}, {Molli{\`e}re}, {Hejazi}, {Baker}, {Bartos}, {Calvin}, {Cetre},
  {Delorme}, {Doppmann}, {Echeverri}, {Fitzgerald}, {Jovanovic}, {Liberman},
  {L{\'o}pez}, {Morris}, {Pezzato}, {Sappey}, {Schofield}, {Skemer}, {Wallace},
  {Wang}, {Agrawal}, \& {Horstman}}]{2024ApJ...970...71X}
{Xuan}, J.~W., {Hsu}, C.-C., {Finnerty}, L., {et~al.} 2024, \apj, 970, 71

\bibitem[{{Zhang} {et~al.}(2021){Zhang}, {Snellen}, {Bohn}, {Molli{\`e}re},
  {Ginski}, {Hoeijmakers}, {Kenworthy}, {Mamajek}, {Meshkat}, {Reggiani}, \&
  {Snik}}]{zhang_13co-rich_2021}
{Zhang}, Y., {Snellen}, I. A.~G., {Bohn}, A.~J., {et~al.} 2021, \nat, 595, 370

\bibitem[{{Zhang} {et~al.}(2024){Zhang}, {Xuan}, {Mawet}, {Wang}, {Hsu},
  {Ruffio}, {Knutson}, {Inglis}, {Blake}, {Chachan}, {Horstman}, {Baker},
  {Bartos}, {Calvin}, {Cetre}, {Delorme}, {Doppmann}, {Echeverri}, {Finnerty},
  {Fitzgerald}, {Jovanovic}, {Liberman}, {L{\'o}pez}, {Morris}, {Pezzato},
  {Sappey}, {Schofield}, {Skemer}, {Wallace}, {Wang}, \& {Do
  {\'O}}}]{2024arXiv240720952Z}
{Zhang}, Y., {Xuan}, J.~W., {Mawet}, D., {et~al.} 2024, \aj, 168, 131

\bibitem[{{Zhang}(2024)}]{2024RNAAS...8..114Z}
{Zhang}, Z. 2024, Research Notes of the American Astronomical Society, 8, 114

\bibitem[{{Zucker}(2003)}]{zucker_cross-correlation_2003}
{Zucker}, S. 2003, \mnras, 342, 1291

\bibitem[{{Zwintz} {et~al.}(2019){Zwintz}, {Reese}, {Neiner}, {Pigulski},
  {Kuschnig}, {M{\"u}llner}, {Zieba}, {Abe}, {Guillot}, {Handler}, {Kenworthy},
  {Stuik}, {Moffat}, {Popowicz}, {Rucinski}, {Wade}, {Weiss}, {Bailey},
  {Crawford}, {Ireland}, {Lomberg}, {Mamajek}, {Mellon}, \&
  {Talens}}]{zwintz_revisiting_2019}
{Zwintz}, K., {Reese}, D.~R., {Neiner}, C., {et~al.} 2019, \aap, 627, A28

\end{thebibliography}

\begin{appendix}
\label{sec8}

\section[Simulation of 13CO detection]{Simulation of $^{13}$CO detection}\label{sec8.2}

This section summarizes the tests we performed using synthetic spectra to constrained both the value and detection limits of the isotopologue $^{13}$CO using GRAVITY. We generated two classes synthetic spectra using the Exo-REM grid with similar parameters as the ones predicted for $\beta$~Pic~b (namely \cite{chilcote_1-24_2017}) both with and without $^{13}$CO and at the resolution, binning and wavelength coverage of GRAVITY. We then created our data samples divided by signal-to-noise: low (S/N~$\sim10$), medium (S/N~$\sim20$, similar to the GRAVITY data) and high (S/N~$\sim30$). This noise was effectively applied to the data $d$ using this formula,

\begin{equation}
    d_{noisy}=d+G\left(0, \frac{d}{\text{S/N}}\right)
,\end{equation}

where $d$ is the original synthetic spectrum, $G\left(0,\sigma\right)$ a Gaussian distribution centered on 0 with a standard deviation of $\sigma$ and $d_{noisy}$ final noised spectrum. This effectively scale the error bars with flux to obtain an overall constant S/N across the K band.

\begin{figure}[ht!]
\centering
    \includegraphics[scale=0.6]{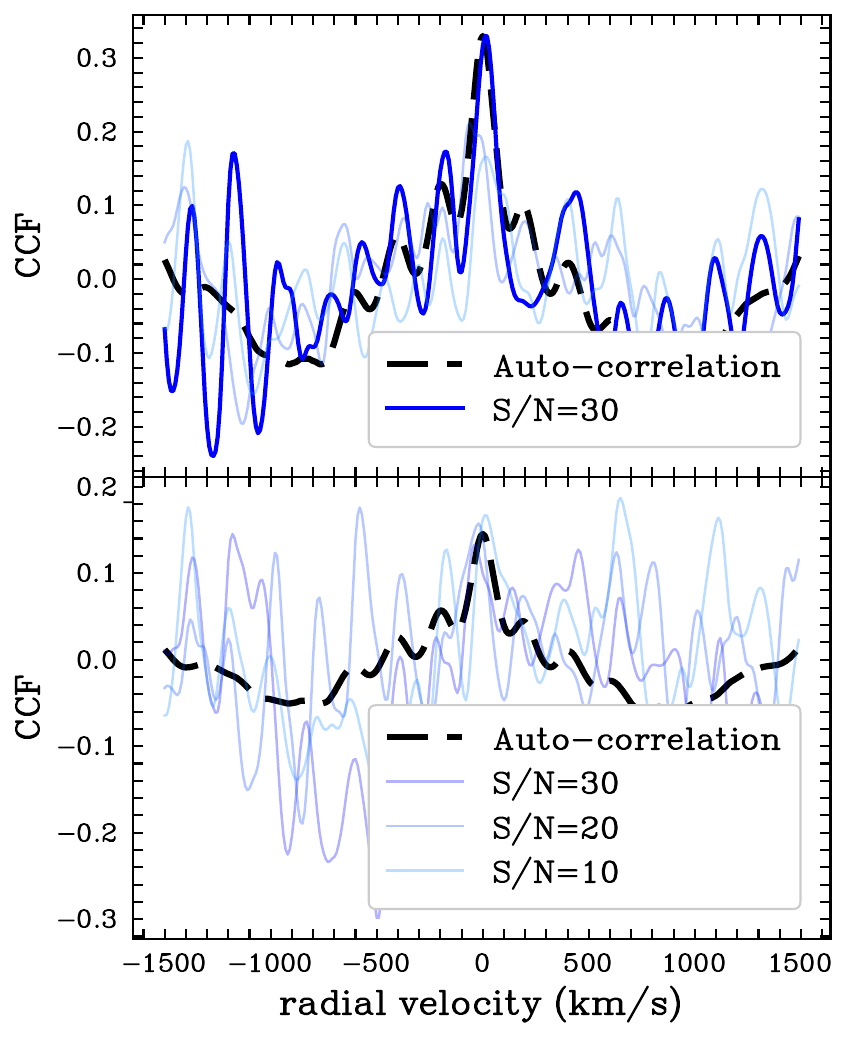}
    \caption{Cross-correlation functions between synthetic spectra with added Gaussian noise and containing $^{13}$CO generated using Exo-REM and their associated model spectrum. Gaussian noise at different levels (S/N = 10, 20, and 30) was applied to each synthetic spectrum. \textit{Top:} CCFs without atmospheric noise. \textit{Bottom:} CCFs with atmospheric noise. The signal is clearly detected for the spectrum with S/N = 30 without atmospheric noise, with a detection significance of S/N$_{CCF}=3.85$.}
    \label{CCF_synth}
\end{figure}

We then perform the CCF analysis presented in Sect.~\ref{sec3.2} on our different synthetic spectra. Results are plotted on the top panel of Fig.~\ref{CCF_synth} showing the CCFs obtained using synthetic spectra with $^{13}$CO. We see that a synthetic spectrum with S/N~$\sim20$ and containing $^{13}$CO is almost sufficient to retrieve a signal (S/N~$_{CCF}=2.48$) with relatively clear wings.

Obviously, this is an idealized case where the synthetic observation is only affected by Gaussian (photon) noise. To better mimic the effect of the atmospheric residuals seen on the GRAVITY spectrum, we modulated the synthetic flux with an observed transmission spectrum (T) taken from \href{https://www.eso.org/observing/etc/skycalc/skycalc.htm}{ESO/SkyCalc}. This modulation was applied using the following formula,

\begin{equation}
\begin{split}
    d_{noisy+atm} & = d_{noisy}+G\left(0,(1-\text{T})*k\right)\\
    & = d + G\left(0,\frac{d}{\text{S/N}}\right) + G\left(0,(1-\text{T})*k\right)
\end{split}
,\end{equation}

where $k$ is a scaling parameter, set to $k=2\times10^{-15}$ hereafter. The retrieved CCFs are shown on the bottom panel of Fig.~\ref{CCF_synth}. We observe that the detection signals are strongly affected and no detection is visible.

Obtaining a detection signal using the GRAVITY spectrum (S/N~$\sim$~20 and telluric residuals) is very unlikely in this setup. We thus argue that the faint signals we obtain in Sect.~\ref{sec3.2} and \ref{sec4} are probably due to residual telluric lines polluting the $^{13}$CO absorptions.

\section{Figures and tables}\label{sec8.3}

\begin{figure*}[t]
    \centering
    \includegraphics[scale=0.57]{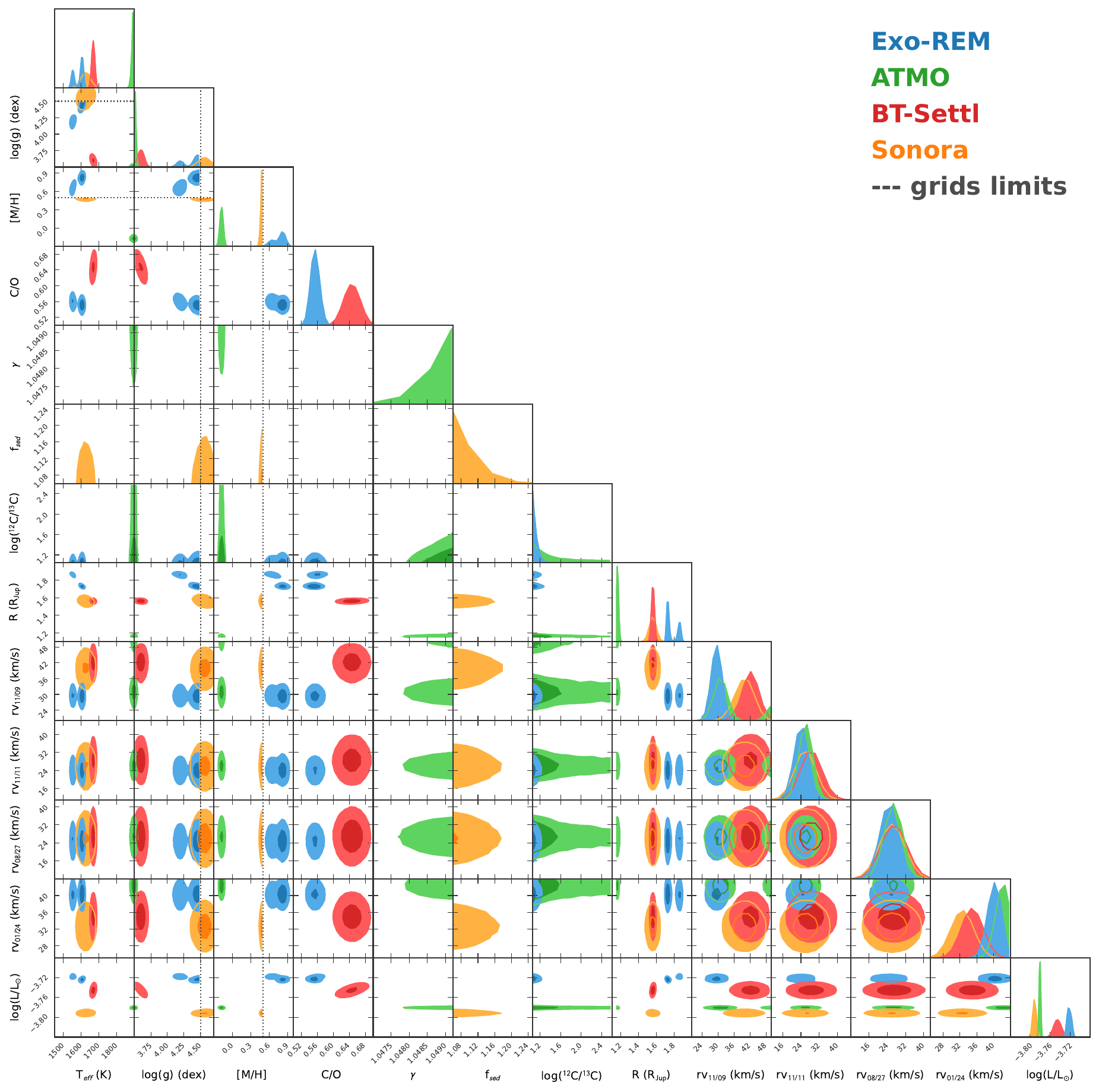}
    \caption{Corner plot of grid and extra-grid parameters for the inversions presented in Sect.~\ref{sec3} (using GRAVITY only). Each color represents a different model: blue for Exo-REM, green for ATMO, red for BT-Settl, and orange for Sonora. Dotted lines indicate the boundaries of the respective model grids when encountered during the inversion.}
    \label{GRAVITY_post_comp}
\end{figure*}

\begin{figure*}[ht!]
    \centering
    \includegraphics[scale=0.42]{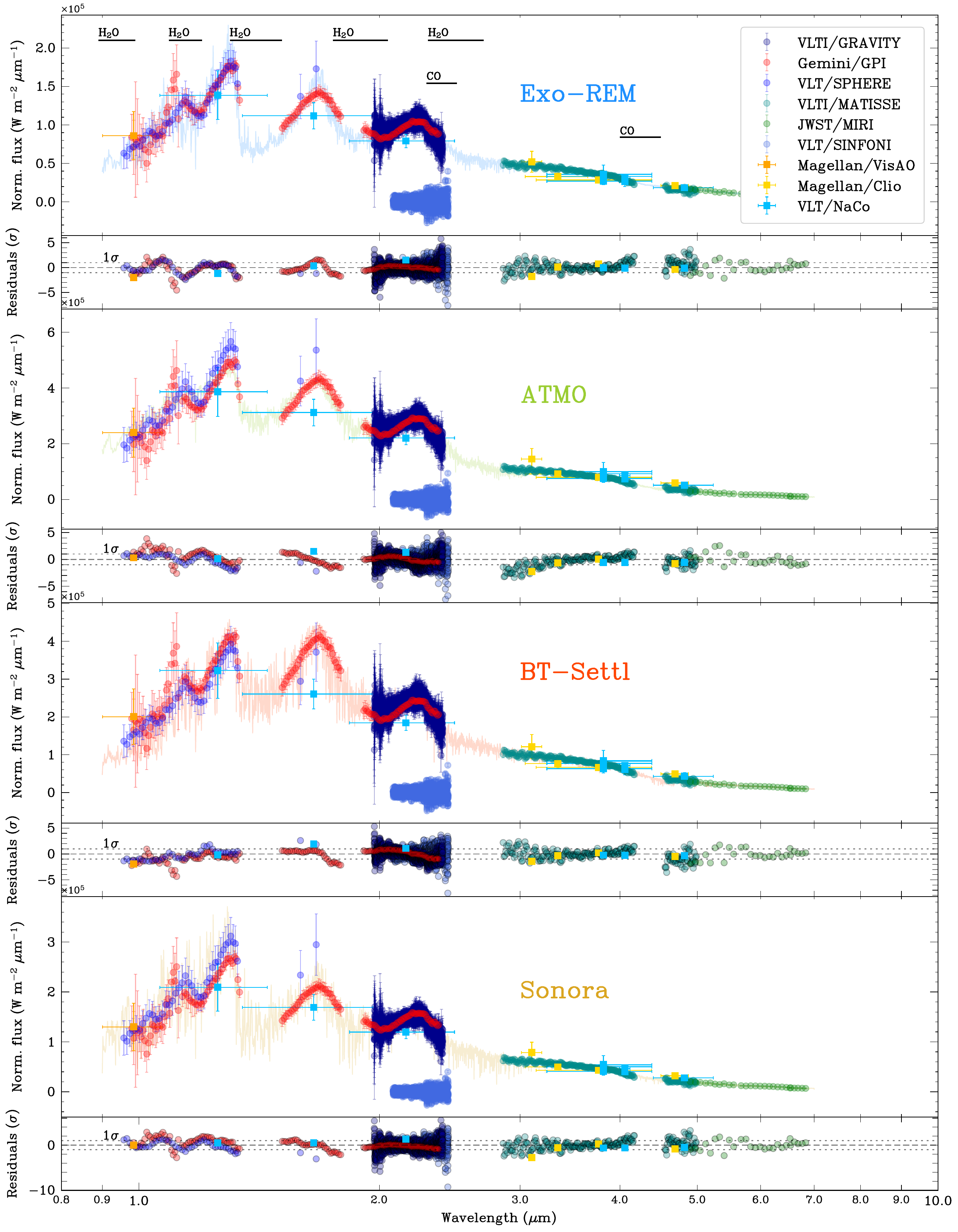}
    \caption{Results of the forward modeling of $\beta$~Pic~b using the most recent spectro/photometric observations of the planet (except CRIRES$_+$). From top to bottom: Exo-REM, ATMO, BT-Settl and Sonora. \textit{Top sub-panels:} spectra (points) and photometry (squares) of $\beta$~Pic~b alongside a R$_{\lambda}\sim$ 4,000 spectrum extracted using the best fit parameters for the inversion on all datasets. Observations have been re-normalized using the analytic scaling factors computed during the inversion. \textit{Bottom sub-panel:} residuals for each fit. Dotted lines represents the $\pm\sigma$ (68$\%$) confidence interval.}
    \label{GRAVITY_MOSAIC_fullspec}
\end{figure*}

\begin{figure}[ht!]
\centering
    \includegraphics[scale=0.75]{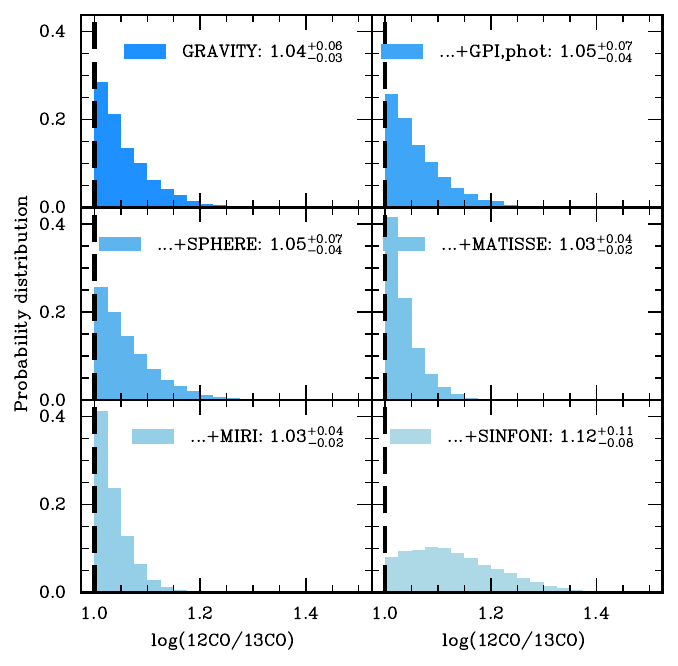}
    \caption{Comparison of the log($^{12}$C/$^{13}$C) when adding observations iteratively from dark (only GRAVITY) to light blue (all except CRIRES$_+$). These values where obtain for the "Free" approach.}
    \label{Free_MOSAIC_13CO}
\end{figure}

\begin{figure}[ht!]
\centering
    \includegraphics[scale=0.69]{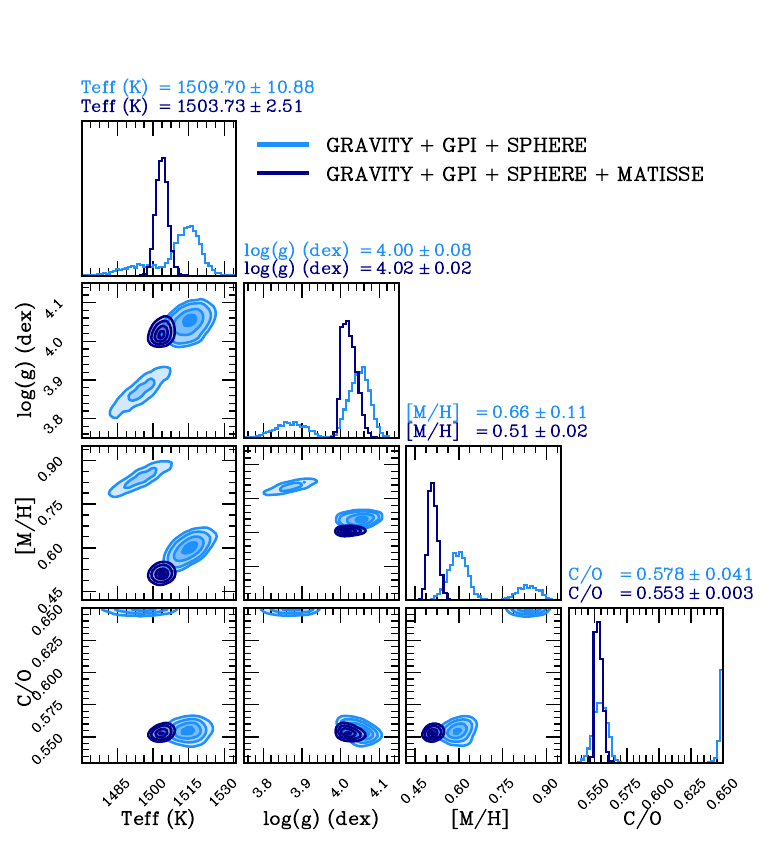}
    \caption{Corner plot comparing Exo-REM inversions without $^{13}$CO, including MATISSE data (dark blue) or excluding it (light blue). The addition of MATISSE breaks the bimodal solution.}
    \label{bimodal_exemple}
\end{figure}

\setlength{\tabcolsep}{7pt}
\renewcommand{\arraystretch}{1.5}
\begin{table*}[ht!]
\tiny
    \centering
    \caption{Grid priors and posteriors for the "free" approach.}
        \begin{tabular}{llllllll}
    \hline
    \hline
    Parameter & \Teff & log(g) & [M/H] & C/O & $\gamma$ & f$_\text{sed}$ & log($^{12}$C/$^{13}$C) \\
    & (K) & (dex) & & & & & \\
    \hline
    Exo-REM priors & $U(1200, 2000)$ & $U(3.5, 4.5)$ & $U(-0.5, 1)$ & $U(0.4, 0.65)$ & & & $U(1.0, 2.6)$ \\
    Exo-REM posteriors & & & & & & & \\
    GRAVITY & $B(1553.23, 1601.08)$ & $B(4.19, 4.42)$ & $B(0.66, 0.81)$ & $0.55\pm0.01$ & & & $<1.07$ \\
    ... + GPI, phot & $1486.75^{+9.85}_{-12.71}$ & $3.82^{+0.05}_{-0.06}$ & $0.82\pm0.03$ & $B(0.55, 0.65)$ & & & $<1.08$ \\
    ... + SPHERE & $1487.85^{+9.16}_{-11.75}$ & $3.82^{+0.05}_{-0.06}$ & $0.82^{+0.04}_{-0.05}$ & $>0.648$ & & & $<1.08$ \\
    ... + MATISSE & $1503.21\pm2.55$ & $4.01^{+0.02}_{-0.01}$ & $0.52\pm0.02$ & $0.554\pm0.003$ & & & $<1.04$ \\
    ... + MIRI & $1503.14^{+2.58}_{-2.48}$ & $4.01^{+0.02}_{-0.01}$ & $0.52\pm0.02$ & $0.554\pm0.003$ & & & $<1.04$ \\
    ... + SINFONI & $1503.38^{+2.30}_{-2.31}$ & $4.00\pm0.01$ & $0.51^{+0.02}_{-0.01}$ & $0.552^{+0.003}_{-0.002}$ & & & $1.12^{+0.11}_{-0.08}$ \\
    \hline
    ATMO priors & $U(1400, 1900)$ & $U(3.5, 4.5)$ & $U(-0.6, 0.6)$ & & $U(1.01,1.05)$ & & \\
    ATMO posteriors & & & & & & & \\
    GRAVITY &$>1897.74$ & $3.63\pm0.05$ & $<-0.596$ & & $1.0198\pm0.0002$ & & \\
    ... + GPI, phot & $1841.16^{+8.46}_{-7.99}$ & $<3.53$ & $<-0.596$ & & $1.0194\pm0.0002$ & & \\
    ... + SPHERE & $1838.17^{+8.02}_{-7.84}$ & $<3.53$ & $<-0.596$ & & $1.0194\pm0.0002$ & & \\
    ... + MATISSE & $1893.89^{+3.64}_{-5.52}$ & $<3.505$ & $-0.599\pm0.001$ & & $1.0147\pm0.0001$ & &  \\
    ... + MIRI & $>1894.05$ & $<3.503$ & $<-0.599$ & & $1.0147\pm0.0001$ & \\
    ... + SINFONI & $>1897.74$ & $<3.504$ & $<-0.599$ & & $1.0150\pm0.0001$ & & \\
    \hline
    BT-Settl priors & $U(1400, 2000)$ & $U(3.5, 5.0)$ & & $U(0.2754, 1.096)$ & & & \\
    BT-Settl posteriors & & & & & & & \\
    GRAVITY & $1668.48^{+5.49}_{-5.87}$ & $3.60\pm0.04$ & & $0.65\pm0.02$ & & & \\
    ...+GPI, phot & $1748.93^{+3.87}_{-3.86}$ & $3.67^{+0.03}_{-0.02}$ & & $0.59\pm0.01$ & & & \\
    ...+SPHERE & $1748.50^{+3.86}_{-3.88}$ & $3.68\pm0.02$ & & $0.59\pm0.01$ & & & \\
    ...+MATISSE & $1890.09^{+2.59}_{-2.54}$ & $<3.50$ & & $0.969\pm0.001$ & & & \\
    ...+MIRI & $1890.21^{+2.46}_{-2.56}$ & $<3.50$ & & $0.969\pm0.001$ & & & \\
    ...+SINFONI & $1893.97^{+2.55}_{-2.53}$ & $<3.50$ & & $0.968\pm0.001$ & & & \\
    \hline
    Sonora priors & $U(900, 2400)$ & $U(3.5, 5.5)$ & $U(-0.5, 0.5)$ & & & $U(1, 8)$ & \\
    Sonora posteriors & & & & & & & \\
    GRAVITY & $1625.43^{+11.41}_{-11.48}$ & $4.56^{+0.06}_{-0.07}$ & $>0.487$ & & & $<1.03$ & \\
    ...+GPI, photo & $1521.60^{+7.12}_{-6.41}$ & $4.25\pm0.04$ & $>0.492$ & & & $<1.03$ & \\
    ...+SPHERE & $1521.38^{+6.85}_{-6.62}$ & $4.25\pm0.04$ & $>0.493$ & & & $<1.03$ & \\
    ...+MATISSE & $1588.51^{+6.95}_{-7.82}$ & $4.60\pm0.02$ & $>0.497$ & & & $<1.01$ & \\
    ...+MIRI & $1589.41^{+7.06}_{-7.66}$ & $4.60\pm0.02$ & $>0.496$ & & & $<1.01$ & \\
    ...+SINFONI & $1593.35^{+5.10}_{-6.35}$ & $4.60\pm0.02$ & $>0.497$ & & & $<1.01$ & \\
    \hline
    \end{tabular}
    \label{priors_post_MOSAIC_Free}
\end{table*}

\newpage

\setlength{\tabcolsep}{5pt} 
\renewcommand{\arraystretch}{1.5}
\begin{table*}[ht!]
\tiny
    \centering
    \caption{Grid priors and posteriors for the "physically~informed" approach.}
        \begin{tabular}{llllllllll}
    \hline
    \hline
    Parameter & \Teff & log(g) & [M/H] & C/O & $\gamma$ & f$_\text{sed}$ & log($^{12}$C/$^{13}$C) & log(L/L$_{\odot}$) \\
    & (K) & (dex) & & & & & & (dex) \\
    \hline
    Exo-REM priors & $U(1200, 2000)$ & $G(4.18, 0.13)$ & $U(-0.5, 1)$ & $U(0.4, 0.65)$ & & & $U(1.0, 2.6)$ & \\
    Exo-REM posteriors & & & & & & & & \\
    GRAVITY & $B(1554.90, 1597.54)$ & $B(4.20, 4.39)$ & $B(0.67, 0.81)$ & $0.56\pm0.01$ & & & $<1.07$ & $-3.95\pm0.05$ \\
    ... + GPI, phot & $1514.21^{+4.71}_{-4.74}$ & $4.04\pm0.03$ &  $0.60\pm0.03$ &  $0.556\pm0.005$ & & & $<1.07$ & $-4.00^{+0.04}_{-0.05}$ \\
    ... + SPHERE & $1494.78^{+6.89}_{-7.88}$ & $B(3.86, 4.03)$ & $B(0.60, 0.85)$ & $>0.648$ & & & $<1.08$ & $-4.02^{+0.04}_{-0.05}$ \\
    ... + MATISSE & $1502.49^{+2.41}_{-2.38}$ & $4.01^{+0.02}_{-0.01}$ & $0.51^{+0.02}_{-0.01}$ & $0.553^{+0.003}_{-0.002}$ & & & $<1.04$ & $-4.01^{+0.04}_{-0.05}$ \\
    ... + MIRI & $1502.43^{+2.48}_{-2.34}$ & $4.01^{+0.02}_{-0.01}$ & $0.51^{+0.02}_{-0.01}$ & $0.553^{+0.003}_{-0.002}$ & & & $<1.04$ & $-4.01^{+0.04}_{-0.05}$ \\
    ... + SINFONI & $1502.74^{+2.32}_{-2.14}$ & $4.00\pm0.01$ & $0.50\pm0.01$ & $0.552^{+0.003}_{-0.002}$ & & & $1.12^{+0.11}_{-0.08}$ & $-4.01^{+0.04}_{-0.05}$ \\
    \hline
    ATMO priors & $U(1400, 1900)$ & $G(4.18, 0.13)$ & $U(-0.6, 0.6)$ & & $U(1.01,1.05)$ & & & \\
    ATMO posteriors & & & & & & & & \\
    GRAVITY & $>1897.88$ & $3.70\pm0.04$ & $<-0.596$ & & $1.0194\pm0.0002$ & & & $-3.62\pm0.04$ \\
    ... + GPI, phot & $1840.78^{+8.31}_{-6.99}$ & $3.55^{+0.04}_{-0.05}$ & $<-0.596$ & & $1.0194\pm0.0002$ & & & $-3.68^{+0.05}_{-0.04}$ \\
    ... + SPHERE & $1838.74^{+7.46}_{-8.62}$ & $3.54\pm0.04$ & $<-0.596$ & & $1.0194\pm0.0002$ & & & $-3.69^{+0.05}_{-0.04}$ \\
    ... + MATISSE & $>1893.17$ & $<3.50$ & $<-0.6$ & & $1.0149\pm0.0001$ & & & $-3.65^{+0.05}_{-0.04}$ \\
    ... + MIRI & $>1893.27$ & $<3.50$ & $<-0.6$ & & $1.0149\pm0.0001$ & & & $-3.66\pm0.05$ \\
    ... + SINFONI & $>1897.68$ & $<3.50$ & $<-0.6$ & & $1.0153\pm0.0001$ & & & $-3.67^{+0.06}_{-0.05}$ \\
    \hline
    BT-Settl priors & $U(1400, 2000)$ & $G(4.18, 0.13)$ & & $U(0.2754, 1.096)$ & & & & \\
    BT-Settl posteriors & & & & & & & & \\
    GRAVITY & $1661.71^{+4.23}_{-4.59}$ & $3.66\pm0.03$ & & $0.62\pm0.01$ & & & & $-3.84\pm0.04$ \\
    ...+GPI, phot & $1748.80^{+3.76}_{-3.65}$ & $3.69\pm0.02$ & & $0.59\pm0.01$ & & & & $-3.78\pm0.05$ \\
    ...+SPHERE & $1748.37^{+3.71}_{-3.52}$ & $3.69\pm0.02$ & & $0.59\pm0.01$ & & & & $-3.79\pm0.05$ \\
    ...+MATISSE & $1714.55^{+2.68}_{-2.78}$ & $<3.50$ & & $0.71\pm0.01$ & & & & $-3.82\pm0.05$ \\
    ...+MIRI & $1714.48^{+2.74}_{-2.67}$ & $<3.50$ & & $0.71\pm0.01$ & & & & $-3.83\pm0.05$ \\
    ...+SINFONI & $1714.87^{+2.69}_{-2.63}$ & $<3.50$ & & $0.691\pm0.005$ & & & & $-3.83\pm0.05$ \\
    \hline
    Sonora priors & $U(900, 2400)$ & $G(4.18, 0.13)$ & $U(-0.5, 0.5)$ & & & $U(1, 8)$ & & \\
    Sonora posteriors & & & & & & & & \\
    GRAVITY & $1613.36^{+10.11}_{-9.14}$ & $4.48^{+0.06}_{-0.05}$ & $>0.483$ & & & $<1.03$ & & $-3.90\pm0.04$ \\
    ...+GPI, phot & $1521.11^{+6.61}_{-6.36}$ & $4.24\pm0.04$ & $>0.492$ & & & $<1.03$ & & $-4.00\pm0.04$ \\
    ...+SPHERE & $1520.36^{+6.33}_{-6.06}$ & $4.24\pm0.04$ & $>0.493$ & & & $<1.03$ & & $-4.01\pm0.04$ \\
    ...+MATISSE & $1584.92^{+7.68}_{-7.46}$ & $4.58\pm0.02$ & $>0.497$ & & & $<1.01$ & & $-3.94\pm0.05$ \\
    ...+MIRI & $1585.09^{+7.53}_{-7.85}$ & $4.57\pm0.02$ & $>0.497$ & & & $<1.01$ & & $-3.94\pm0.05$ \\
    ...+SINFONI & $1590.80^{+5.96}_{-6.88}$ & $4.57\pm0.02$ & $>0.496$ & & & $<1.01$ & & $-3.94\pm0.05$ \\
    \hline
    \end{tabular}
    \label{priors_post_MOSAIC_Phys}
\end{table*}

\setlength{\tabcolsep}{15pt}
\renewcommand{\arraystretch}{1.5}
\begin{table*}[ht!]
\tiny
    \centering
    \caption{"Best"extra-grid priors and posteriors for the combined forward modeling using the "Physically~informed" approach.}
        \begin{tabular}{lllllllll}
        \hline
        \hline
        Parameter & R & rv$_\text{SINFONI}$ & rv$_\text{GRAVITY}$ & rv$_\text{GRAVITY}$ & rv$_\text{GRAVITY}$ & rv$_\text{GRAVITY}$ \\
         & (\Rjup) & (km.s$^{-1}$) & (km.s$^{-1}$) & (km.s$^{-1}$) & (km.s$^{-1}$) & (km.s$^{-1}$) \\
         & & 2014/09/10 & 2019/11/09 & 2019/11/11 & 2021/08/27 & 2022/01/24 \\    
        \hline
        Exo-REM priors & $G(1.4,0.1)$ & $U(-50, 50)$ & $U(-50, 50)$ & $U(-50, 50)$ & $U(-50, 50)$ & $U(-50, 50)$ \\
        Exo-REM posteriors & $1.43\pm0.07$ & $17.61^{+0.48}_{-0.50}$ & $B(29.57, 48.97)$ & $24.18\pm1.65$ & $24.86^{+2.23}_{-2.28}$ & $39.97^{+1.10}_{-1.05}$ \\
        \hline
        ATMO priors & $G(1.4,0.1)$ & $U(-50, 50)$ & $U(-50, 50)$ & $U(-50, 50)$ & $U(-50, 50)$ & $U(-50, 50)$ \\
        ATMO posteriors & $1.32^{+0.09}_{-0.07}$ & $17.78^{+0.54}_{-0.56}$ & $B(33.55, 48.97)$ & $22.74^{+2.27}_{-2.26}$ & $26.54^{+3.03}_{-2.90}$ & $40.17^{+1.44}_{-1.47}$ \\
        \hline
        BT-Settl priors & $G(1.4,0.1)$ & $U(-50, 50)$ & $U(-50, 50)$ & $U(-50, 50)$ & $U(-50, 50)$ & $U(-50, 50)$ \\
        BT-Settl posteriors & $1.35^{+0.08}_{-0.07}$ & $13.71^{+0.74}_{-0.75}$ & $42.23^{+1.92}_{-2.00}$ & $28.11^{+2.88}_{-2.84}$ & $26.53^{+3.89}_{-3.85}$ & $35.95^{+1.86}_{-1.89}$ \\
        \hline
        Sonora priors & $G(1.4,0.1)$ & $U(-50, 50)$ & $U(-50, 50)$ & $U(-50, 50)$ & $U(-50, 50)$ & $U(-50, 50)$ \\
        Sonora posteriors & $1.37\pm0.08$ & $11.11^{+0.82}_{-0.84}$ & $40.20^{+2.05}_{-2.45}$ & $26.33^{+3.63}_{-3.51}$ & $26.01^{+4.80}_{-4.68}$ & $32.08\pm2.03$ \\ 
        \hline
    \end{tabular}
    \tablefoot{The radial velocities display here are in the reference frame of the sun (e.g., rv~=~rv$_{\beta~\text{Pic~A}}$~+~rv$_{\beta~\text{Pic~b}}$).}
    \label{priors_post_extra_MOSAIC_All}
\end{table*}

\newpage

\begin{figure}[ht!]
    \centering
    \includegraphics[scale=0.45]{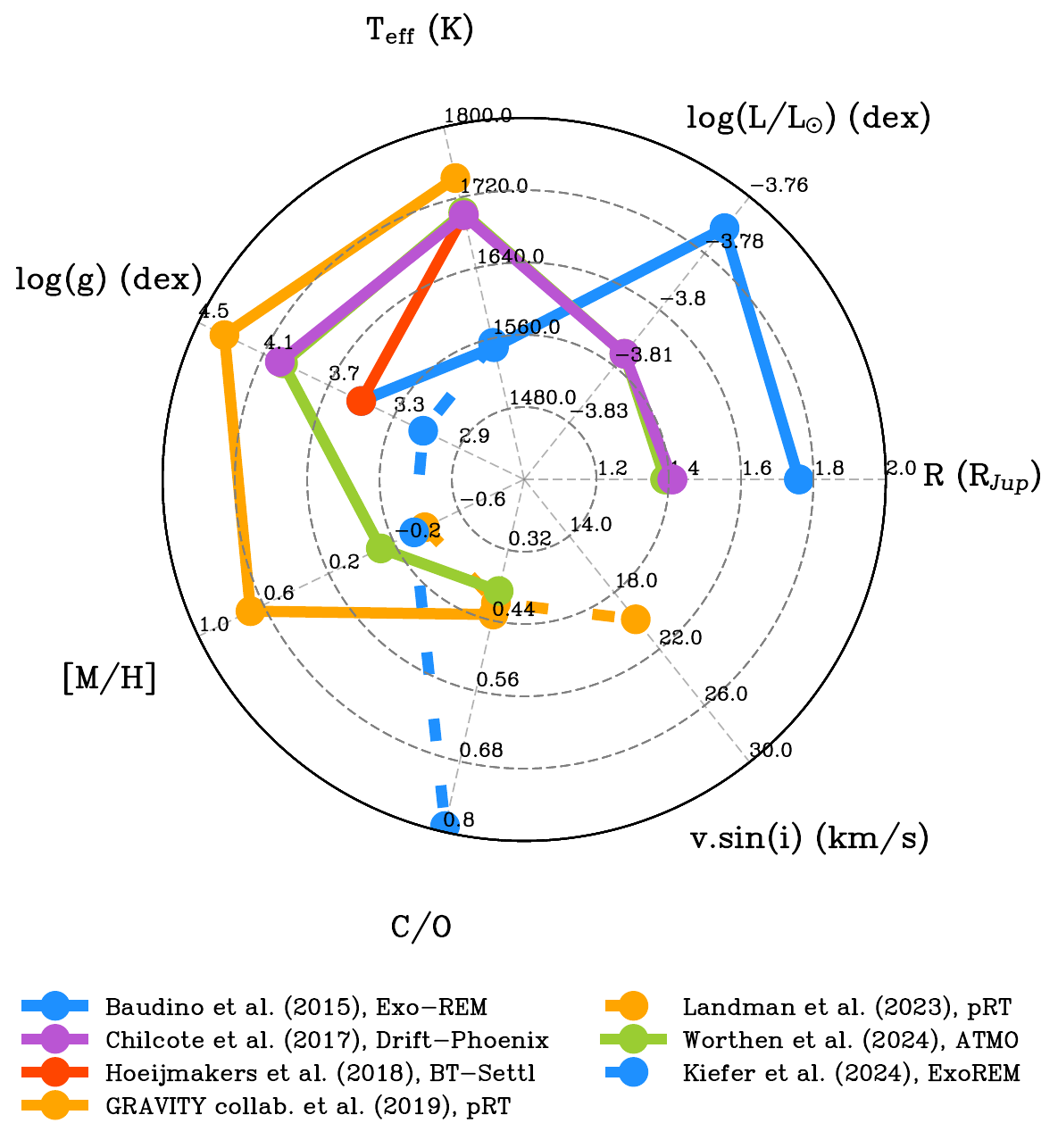}
    \caption{Recent planetary parameters estimates of $\beta$~Pic~b using atmospheric models, showing great variations. Next to each reference, the model used is cited.}
    \label{litt_spider}
\end{figure}

\begingroup

\setlength{\tabcolsep}{1pt}
\renewcommand{\arraystretch}{1.5} 
\begin{table*}[ht!]
\tiny
    \caption{Compilation of the best existing physical parameters estimations for $\beta$~Pic~b.}
    \begin{tabular}{lllllllll}
        \hline
        \hline
        Reference & Data & Model & \Teff & log(g) & [M/H] & C/O & R & M \\
         & & & (K) & (dex) & & & (\Rjup) & (\Mjup) \\
        \hline
        \cite{lagrange_probable_2009} & NaCo & COND & $\sim1600$ &  &  &  &  & $9^{+3}_{-2}$ \\
         & NaCo & DUSTY & $\sim1400$ &  &  &  &  & $8^{+4}_{-2}$ \\
        \hline
        \cite{quanz_first_2010} & NaCo & DUSTY & $\sim1470$ & $\sim4.0$ &  &  &  & $\sim8$ \\
        \hline
        \cite{bonnefoy_high_2011} & NaCo & BT-Settl & $1700\pm300$ &  &  &  &  &  \\
         & NaCo & COND/DUSTY & $1470-1700$ & &  &  &  & $7-11$ \\
        \hline
        \cite{currie_combined_2013} & photo + NICI & DUSTY & $1575-1650$ & $3.8\pm0.2$ &  &  & $1.65\pm0.06$ & $7^{+4}_{-3}$ \\
        \hline
        \cite{bonnefoy_near-infrared_2013} & photo & PHOENIX & $1700\pm100$ & $4.0\pm0.5$ &  &  & $1.4\pm0.2$ & \\
         & photo & COND/DUSTY & $1594^{+105}_{-77}$ & $4.0^{+0.17}_{-0.08}$ &  &  & $1.5^{+0.04}_{-0.10}$ & $9^{+3.4}_{-1.5}$ \\
        \hline
        \cite{males_magellan_2014} & photo + VisAO & COND/DUSTY & $1643\pm32$ & $3.8\pm0.2$ &  &  & $1.43\pm0.02$ & $11.9\pm0.7$ \\
        \hline
        \cite{bonnefoy_physical_2014} & GPI, rv & astrometry & & & & & & $<20$ \\
         & GPI & PHOENIX & $1650\pm150$ & $<4.7$ & & & $1.5\pm0.2$ & \\
        \hline
        \cite{chilcote_first_2015} & GPI & PHOENIX & $1600-1700$ & $3.5-4.5$ & & & & \\
        \hline
        \cite{baudino_interpreting_2015} & photo & Exo-REM & $1550\pm150$ & $3.5\pm1$ & & & $1.76\pm0.24$ & \\
        \hline
        \cite{morzinski_magellan_2015} & photo + Clio & PHOENIX & $1708\pm23$ & $\sim4.2$ & & & $1.45\pm0.02$ & $12.7\pm0.3$ \\
        \hline
        \cite{chilcote_1-24_2017} & GPI, photo & COND/DUSTY & $1724\pm15$ & $4.18\pm0.01$ & & & $1.46\pm0.01$ & $12.9\pm0.2$ \\
         & GPI, photo & Drift-Phoenix & $\sim1700$ & $\sim4.0$ &  &  & $\sim1.41$ & \\
         & GPI, photo & AMES-Dusty & $\sim1800$ & $\sim3.5$ &  &  & $\sim1.17$ & \\
         & GPI, photo & BT-Settl & $\sim1800$ & $\sim3.5$ &  &  & $\sim1.22$ & \\
        \hline
        \cite{hoeijmakers_medium-resolution_2018} & SINFONI & BT-Settl & $\sim1700$ & $\sim3.5$ & & & & \\
        \hline
        \citeauthor{gravity_collaboration_peering_2020}* & GRAVITY & Exo-REM & $1700\pm50$ & $\sim3.5$ & $\sim-0.5$ & $\le0.30$ & & $\sim2.0$ \\
        & GRAVITY, GPI & Exo-REM & $1590\pm20$ & $\sim4.0$ & $\sim0.5$ & $0.43\pm0.05$ & & $\sim12.4$ \\
        & GRAVITY & pRT & $1847\pm55$ & $3.3^{+0.54}_{-0.42}$ & $-0.53^{+0.28}_{-0.34}$ & $0.35^{+0.07}_{-0.09}$ & & $1.4^{+3.94}_{-0.87}$ \\
        & GRAVITY, GPI & pRT & $1742\pm10$ & $4.34^{+0.08}_{-0.09}$ & $0.68^{+0.11}_{-0.08}$ & $0.43^{+0.04}_{-0.03}$ & & $15.43^{+2.91}_{-2.79}$ \\
        \hline
        \cite{lagrange_unveiling_2020} & GRAVITY, SPHERE, rv & astrometry & & & & & & $9.7\pm0.7$ \\
        \hline
        \cite{brandt_precise_2021} & GRAVITY, rv & astrometry & & & & & & $9.3^{+2.6}_{-2.5}$ \\
        \hline
        \cite{lacour_mass_2021} & GRAVITY, rv & astrometry & & & & & & $11.9^{+2.93}_{-3.04}$ \\
        \hline
        \cite{landman__2024} & CRIRES$_+$ & pRT, nominal & & & $-0.39\pm0.16$ & $0.41\pm0.04$ & & \\
        & CRIRES$_+$ & pRT, free & & & $0.49\pm0.25$ & $0.41\pm0.04$ & & \\
        & CRIRES$_+$ & pRT, cloudy & & & -$0.10\pm0.20$ & $0.38\pm0.05$ & & \\
        & CRIRES$_+$ & pRT, * priors & & & $0.72\pm0.06$ & $0.48\pm0.03$ & & \\
        \hline
        \cite{worthen_miri_2024} & MIRI, GRAVITY, GPI, photo & Drift-Phoenix & $1738^{+16}_{-21}$ & $4.03\pm0.07$ & $0.18\pm0.07$ & & $1.40^{+0.04}_{-0.03}$ & \\
        & MIRI, GRAVITY, GPI, photo & ATMO & $1703^{+37}_{-44}$ & $3.98^{+0.08}_{-0.12}$ & $-0.12^{+0.20}_{-0.19}$ & $0.39^{+0.10}_{-0.06}$ & $1.39\pm0.06$ & \\
        & MIRI, GRAVITY, GPI, photo & Exo-REM & $1471^{+20}_{-18}$ & $3.71^{+0.07}_{-0.08}$ & $0.90^{+0.07}_{-0.13}$ & $0.36^{+0.13}_{-0.05}$ & $1.97\pm0.05$ & \\
        \hline
        \cite{kiefer_new_2024} & SINFONI & Exo-REM, free & $1555^{+22}_{-29}$ & $3.12^{+0.12}_{-0.09}$ & $-0.325^{+0.065}_{-0.045}$ & $0.79^{+0.01}_{-0.11}$ & & \\
        & SINFONI & Exo-REM, log(g) prior & $1746^{+4}_{-3}$ & $4.185\pm0.010$ & $-0.235^{+0.014}_{-0.011}$ & $0.551\pm0.002$ & & \\
        & SINFONI & Exo-REM, \Teff prior & $1748^{+3}_{-4}$ & $4.216^{+0.027}_{-0.031}$ & $-0.235^{+0.015}_{-0.013}$ & $0.551\pm0.002$ & \\
        & SINFONI & Exo-REM, both priors & $1745^{+3}_{-2}$ & $4.183\pm0.010$ & $-0.235^{+0.015}_{-0.013}$ & $0.551\pm0.002$ & \\
        \hline
        \cite{houlle_mathis_2025} & MATISSE, GRAVITY fixed & Exo-REM & $1529\pm3$ & $3.84\pm0.03$ & $0.30\pm0.03$ & $0.539^{+0.003}_{-0.002}$ & $1.943\pm0.008$ & $10.5\pm0.7$ \\
         & MATISSE fixed, GRAVITY & Exo-REM & $1529\pm3$ & $3.83\pm0.04$ & $0.30\pm0.03$ & $0.539\pm0.003$ & $2.085^{+0.008}_{-0.009}$ & $11.8\pm0.9$ \\
        \hline
    \end{tabular}
    \tablefoot{"photo" refers to photometry litterature that are iteratively combined (NaCo, NICI, VisAO, Clio).}
    \label{litterature}
\end{table*}

\endgroup

\end{appendix}

\end{document}